\documentclass[preprint2,12pt]{aastex6}

\usepackage{natbib}
\usepackage{amsmath}
\usepackage{amssymb}
\usepackage{bm}

\shorttitle{Atmospheric Retrieval for Gas Giants in Reflected Light I}
\shortauthors{Lupu et al.}

\begin{document}

\title{Developing Atmospheric Retrieval Methods for Direct Imaging Spectroscopy of Gas Giants in Reflected Light I: Methane Abundances and Basic Cloud Properties}

\author{Roxana E. Lupu}
\affil{BAER Institute / NASA Ames Research Center, Moffet Field, CA 94035, USA} \email{Roxana.E.Lupu@nasa.gov}
\author{Mark S. Marley}
\affil{NASA Ames Research Center, Moffet Field, CA 94035, USA}
\author{Nikole Lewis}
\affil{Space Telescope Science Institute, 3700 San Martin Drive Baltimore, MD 21218}
\author{Michael Line}
\affil{Univ. California at Santa Cruz, 1156 High St, Santa Cruz, CA 95064}
\author{Wesley A. Traub}
\affil{Jet Propulsion Laboratory, California Institute of Technology, 4800 Oak Grove Dr, Pasadena, CA 91109}
\author{Kevin Zahnle}
\affil{NASA Ames Research Center, Moffet Field, CA 94035, USA}

\begin{abstract}

Upcoming space-based coronagraphic instruments in the next decade will perform reflected light spectroscopy and photometry of cool, directly imaged extrasolar giant planets. We are developing a new atmospheric retrieval methodology to help assess the science return and inform the instrument design for such future missions, and ultimately interpret the resulting observations. Our retrieval technique employs a geometric albedo model coupled with both a Markov chain Monte Carlo  Ensemble Sampler ({\it emcee}) and a multimodal nested sampling algorithm ({\it MultiNest}) to map the posterior distribution. This combination makes the global evidence calculation more robust for any given model, and highlights possible discrepancies in the likelihood maps. As a proof-of-concept, our current atmospheric model contains 1 or 2 cloud layers, methane as a major absorber, and a H$_2$-He background gas. This 6-to-9 parameter model is appropriate for Jupiter-like planets and can be easily expanded in the future. In addition to deriving the marginal likelihood distribution and confidence intervals for the model parameters, we perform model selection to determine the significance of methane and cloud detection as a function of expected signal-to-noise in the presence of spectral noise correlations. After internal validation, the method is applied to realistic spectra of Jupiter, Saturn, and HD 99492 c, a model observing target. We find that the presence or absence of clouds and methane can be determined with high confidence, while parameter uncertainties are model-dependent and correlated. Such general methods will also be applicable to the interpretation of direct imaging spectra of cloudy terrestrial planets.

\end{abstract}


\section{INTRODUCTION}
\label{intro}

Space-based telescopes equipped with coronagraphic imagers can separate light scattered by orbiting planets from that of their primary stars. The detection of light that penetrates deeply into an atmosphere rather than merely skimming its upper layers, as with transit methods, potentially permits more extensive and informative characterization of  atmospheric gaseous absorbers as well as cloud and haze layers.
However the interpretation of the scattered light signal will in practice be limited by a multitude of uncertainties beyond the basic limitations of data quality. Among these are the uncertain or unknown planetary radii, masses, and cloud layers. Here, in the first of what we plan to be a series of papers, we present the initial development of an atmospheric retrieval methodology that quantifies the resultant uncertainties 
and clarifies the precision with which the planet's gravity, composition, and cloud structure and other parameters can be discerned.

Direct imaging offers the possibility of characterizing planets around nearby stars and at larger orbital distances than is possible for transit observations. Directly imaged planets see less stellar irradiation than traditional transit observation targets and can either be young, warm, and self-luminous, or older and much colder than those studied by transit methods. While a multitude of space coronagraph missions have been studied or proposed over the last two decades, the only mission currently in development by NASA with the capability of imaging cool giant planets in reflected light is {\it WFIRST} \citep{Spergel:2015}. 

\begin {figure}
\centering
\includegraphics*[scale=0.9,angle=0]{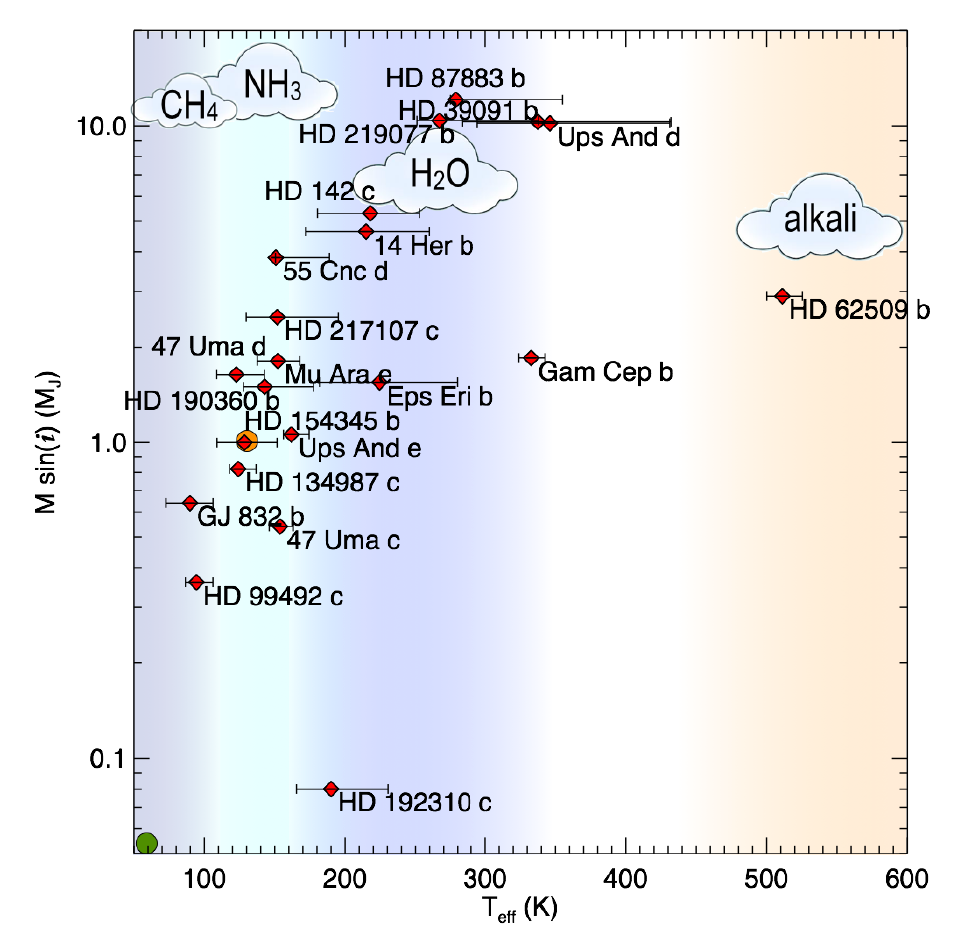}
\caption{$M\mathrm{sin}(i)$ and ranges of estimated effective temperature ($T_{\mathrm{eff}}$) of a selection of
announced RV planets that are favorable for direct imaging. The orange circle represents Jupiter while the green one hints at Uranus which actually falls below the lower axis. Estimated $T_{\mathrm{eff}}$ computed from planet orbits, Jupiter's Bond albedo, and estimated internal heat flows given available constraints on the ages of the primary stars. Bands show major cloud species expected in various ranges of $T_{\mathrm{eff}}$. The existence of two of the planets shown, Ups And e and Eps Eri b, is controversial. \label{fig:rv}}
\end{figure}

Current estimates are that a coronagraph-equipped {\it WFIRST} mission will be able to obtain photometry and spectra for at least a dozen known radial velocity (RV) planets as well as search for lower
mass planets \citep{Traub:2016}. An example of the diversity of the known RV planets favorable for direct imaging is shown in Figure~\ref{fig:rv}. This sample was drawn from the Exoplanet Encyclopedia and will likely increase with future discoveries from RV or {\it WFIRST} surveys. In this figure the known $M\sin i$, measured by RV methods, is plotted against estimated blackbody radiating temperature (or effective temperature) in order to illustrate the phase space of atmospheric conditions that might be expected among these most favorable planets. The effective temperatures have been calculated using an evolution model for the range of masses and the age ranges of the stars, accounting for both internal heat sources and the incident flux \citep{Marley:2014}. The planet's inclination ($i$) will be determined from the direct imaging observations, therefore constraining their approximate
masses and, with the aid of the mass-radius relationship, their surface gravities. Vertical color bands show the approximate ranges over which various atmospheric compounds form clouds. While many Jupiter and Saturn-like worlds with ammonia clouds are expected, some planets with water, alkali, and even methane clouds may also be observed.

The Coronagraph Instrument onboard {\it WIFRST}, in combination with an Integral Field Spectrometer \citep{Traub:2016}, is currently planned to provide us with images (430--970~nm) and low-resolution (spectral resolution $R\sim 70$) reflected light spectra of gaseous planets around nearby Sun-like stars (600-970~nm). Unlike transit spectroscopy that only probes the top of the atmosphere to $\sim 1$~mbar \citep[e.g., ][]{Kreidberg:2014}, reflected light can probe deep into the atmosphere of these gas giants \citep[e.g., ][]{Marley:2014}, and therefore offers a more comprehensive view of composition and cloud layers.

Most planets in Figure~\ref{fig:rv} have effective temperatures of $\sim 150-350\,\rm K$. 
Assuming these worlds are comparable to Solar System gas giants, their $600--970\,\rm nm$ spectra will be dominated by 
cloud decks of water or ammonia and gaseous absorption by methane and possibly water. Photochemical
hazes will doubtless be important as well. There is a long and comprehensive history of 
interpretation of such spectra of Solar System planets dating back to \citet{Sato:1979} and before. For Jupiter-like atmospheres
the continuum scattered flux level at these wavelengths is set by scattering from the bright clouds while Rayleigh scattering is more important at the bluest wavelengths. The bright continuum is punctuated by
gaseous methane absorption features of varying strengths. The relative strengths of the various methane absorption bands, combined with the continuum flux level set by the clouds, together constrain the cloud properties and methane column abundance. Shortward of 600~nm, the photometric measurements will give us information about the shape of the continuum, dominated by Rayleigh, haze, and cloud scattering. If both CH$_4$ and H$_2$O features are present in the spectra, we can constrain the C/O ratio, value related to the place of planet's formation in the circumstellar disk \citep{Bond:2010,Helling:2014,Oberg:2011}.

Extracting such information from low to moderate spectral resolution data at modest signal-to-noise ratios will be a challenge. Cloud properties and location, absorber abundances, planetary radius (and thus gravity), and the atmospheric thermal profile will all be unknown. While forward modeling techniques, such as \citet{Cahoy:2010}, can give insight into the range of possible spectra, extraction of cloud properties and absorber
abundances will require the application of retrieval methods to the available data. 

We aim to develop the necessary theoretical and computational framework to enable such retrievals. As this will be a complex endeavor we approach the problem in steps. Here we present a first step in the development of this framework, focusing on the retrieval of gross cloud properties, surface gravity, and methane mixing ratio. In future papers we will add retrievals for orbital phase, star-planet distance, planet size, additional absorbers and atmospheric thermal profile.

In the remainder of this paper we provide more detailed background on reflected light spectra of giant planets, present the conceptual model and Markov Chain Monte Carlo retrieval method, and the results of this study. The paper is organized as follows: Section~\ref{back} provides more context and background to the problem. Section~\ref{fwd} describes our albedo code and the forward models used in the retrievals; Section~\ref{noise} describes our noise model used to generate the input datasets; Section~\ref{mcmc} contains the Bayesian retrieval scheme, followed by its validation in Section~\ref{results}. Other retrieval results for more realistic spectra of known gas giants are shown in Section~\ref{appl}, and the conclusions are summarized in Section~\ref{sum}.

\begin {figure}
\centering
\includegraphics*[scale=0.4,angle=0]{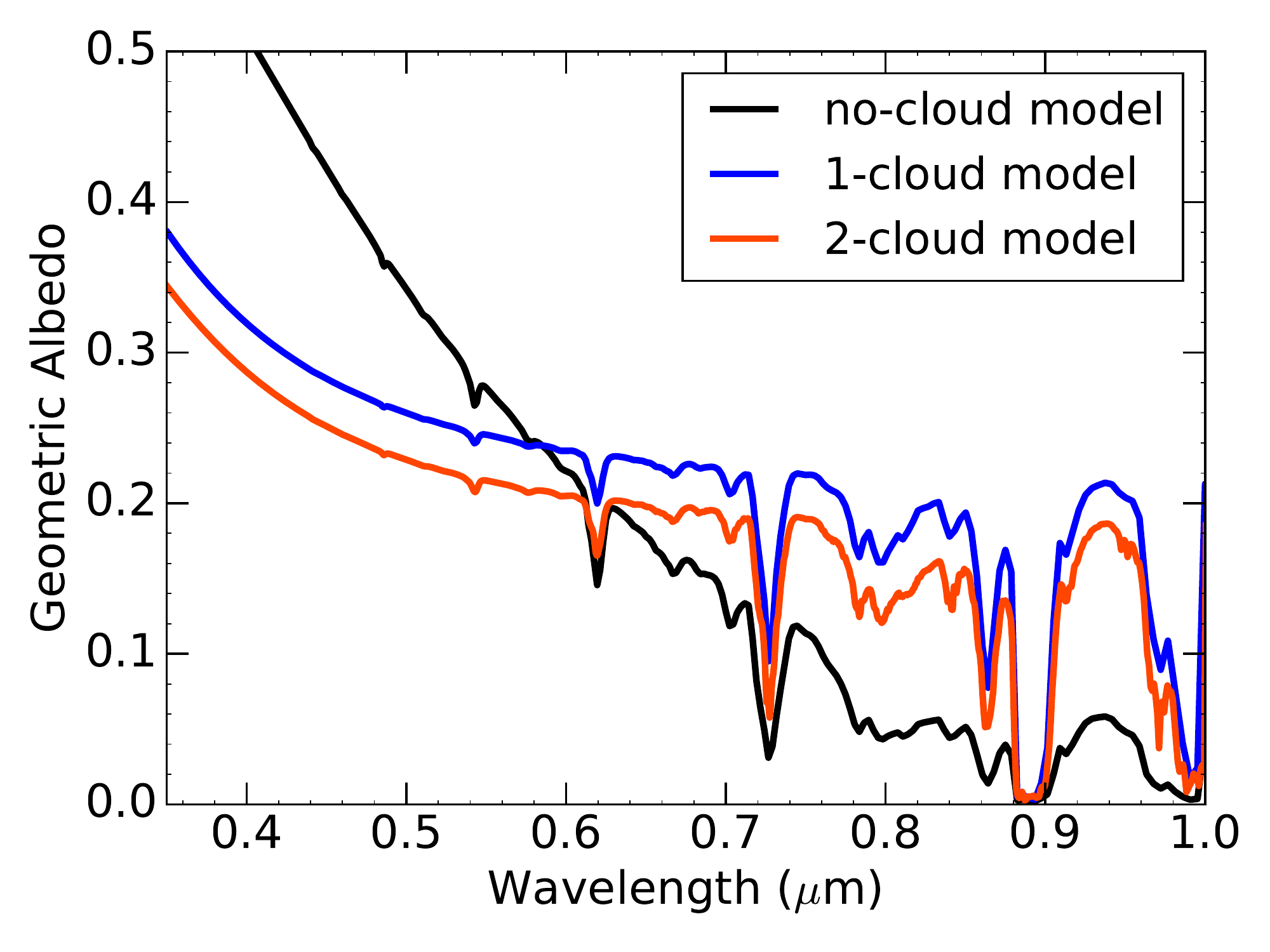}
\caption{Model geometric albedo spectra for three example cases: cloud-free (black), a single optically thick cloud deck (blue), and one cloud deck plus and optically thin haze layer (red). All models assume a CH$_4$ abundance of $10^{-3}$, and a surface gravity of 25 m s$^{-2}$. The cloud deck is at a depth of 1.8 bars in both red and blue examples, and has an albedo of 0.95. The simulated haze layer in the red model has an optical depth of 0.2, an albedo of 0.6, and is occupies the region between 0.2 and 0.5 bar. \label{fig:hd}}
\end{figure}

\section{BACKGROUND}
\label{back}
In this section we provide a brief overview to a few of the key concepts
used throughout the remainder of the paper.

\subsection{Geometric Albedo}
The analysis of extrasolar planet reflection spectra owes much to the
Solar System literature. However this literature also brings its own set
of conventions, not all of which translate smoothly to the exoplanet
context. For expediency we nevertheless choose here to follow these conventions, although we recognize that as exoplanet direct imaging evolves into its own sub-field that this terminology will likely evolve to shed some vestigial structures. 

A foremost concept is the geometric albedo, the ratio of light received from a planet when observed at full phase to that which would be measured from a perfectly reflective Lambert disk of the same size as the planet. 
Because the angular distribution of light scattered by a real atmosphere differs from that scattered by a Lambert disk, the geometric albedo of even a perfectly scattering atmospheres is not unity.  For a conservative, infinitely deep Rayleigh scattering atmosphere the geometric  albedo is 0.750. The fractional reflectivity measured at a star-planet-observer angle differing from $180^\circ$ is given by the product of the geometric albedo and the planetary phase function. Theoretical calculations of reflected light spectra for extrasolar giant planets have been preformed to date by \citet{Marley:1999,Burrows:2004,Burrows:2014,Cahoy:2010,Greco:2015}, showing the wide variations determined by metallicity, effective temperature, cloud presence, and orbital phase angle.

\begin {figure*}
\centering
\includegraphics*[scale=0.55,angle=0]{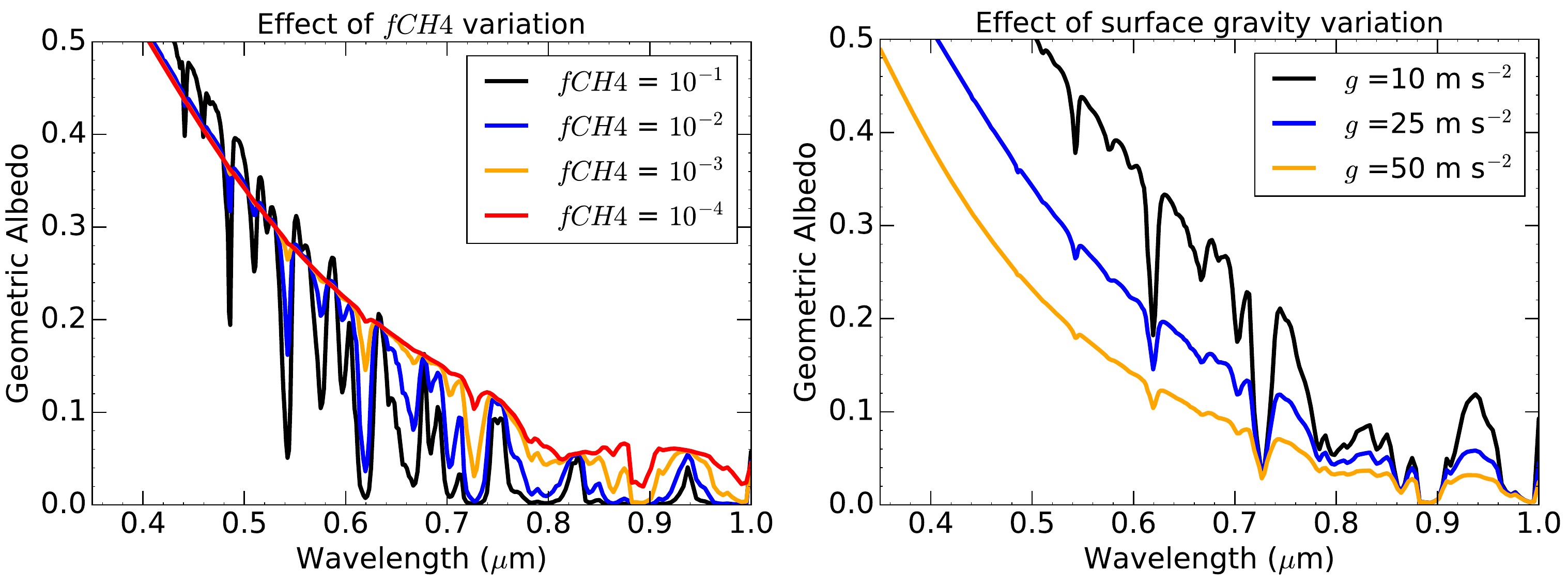}
\caption{Model geometric albedo spectra comparing the effects of increasing methane abundance (left) and surface gravity (right) for a cloud-free planet. In the left plot the surface gravity is kept constant at 25 m s$^{-2}$, while in the right plot the methane abundance is kept constant at 10$^{-3}$. The thermal profile is kept constant in all cases. \label{fig:met0}}
\end{figure*}

\begin {figure*}
\centering
\includegraphics*[scale=0.5,angle=0]{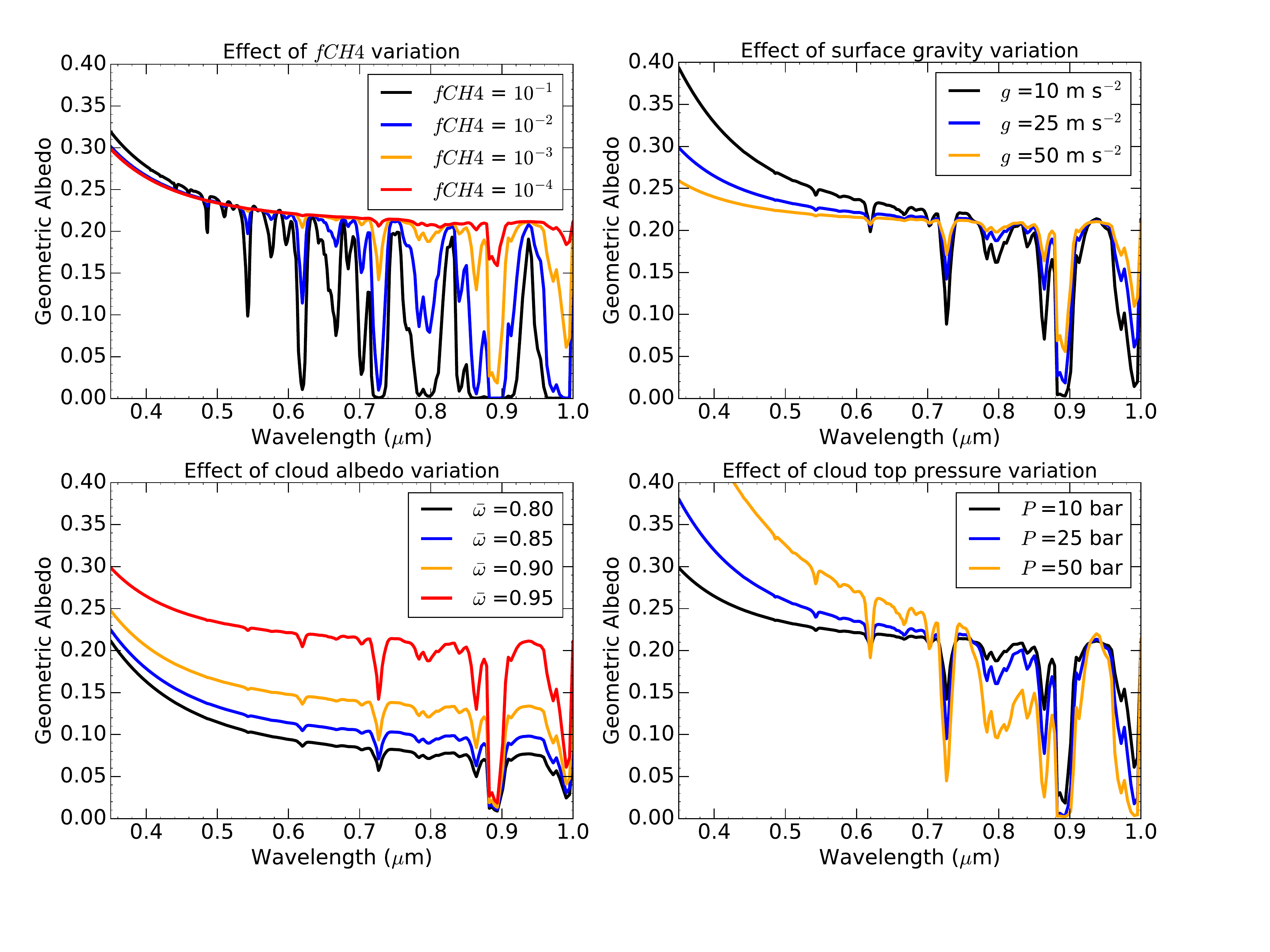}
\caption{Model geometric albedo spectra comparing the effects of increasing methane abundance (top left), surface gravity (top right), cloud albedo (bottom left), and cloud top pressure (bottom right) for a planet with a single cloud deck. When not variable, the model parameters are set to $f_{CH4}=10.^{-3}$, $g=25$m s$^{-2}$, $\bar{\omega}=0.95$, and $P=0.8$ bar. The thermal profile is kept constant in all cases.\label{fig:met1}}
\end{figure*}

There are two important reasons why ``geometric albedo spectra" will not
be directly measured for directly imaged exoplanets. 
First, while transiting planets can be observed at full phase just before they are eclipsed on the ``far" side of their orbits, directly imaged planets will never be observed even close to full phase because they would lie too close to the primary star to be resolved from the star. Second, the radius of a planet will not be directly measured, rather only the product between the planet's area and its reflectivity as a function of wavelength. Thus it is an oversimplification to discuss ``geometric albedo spectra" for directly imaged extrasolar planets. Nevertheless to simplify the model development for this work,  we consider here only the planetary spectrum at full phase, cast as ``geometric albedo spectra". In the second paper of this series (Nayak et al., submitted) we will explore issues arising from the phase dependence of planetary reflectivity (see \citet{Cahoy:2010}) and the unknown planetary radius.  

Figure~\ref{fig:hd} shows model geometric albedo spectra we calculated for three typical planet cases following the methods described in this paper. Depending on the temperature and composition of the planet, certain species can condense forming cloud decks (mostly alkalis, methane, ammonia, and water for the RV planets shown in Figure~\ref{fig:rv}). As known from our Solar System (e.g. Jupiter, Titan), a haze layer can also form in the upper layers of the atmosphere under the action of stellar ultraviolet radiation. The figure compares computed geometric albedo spectra with (blue) and without (black) the expected clouds and haze layer (red). Cloudy giant planets are brighter in reflected light at red wavelengths as incoming photons are scattered before they can be absorbed \citep{Marley:1999}.

Figures~\ref{fig:met0} and \ref{fig:met1} present additional model geometric albedo spectra for varying atmospheric parameters, that can be expected given the diversity of extrasolar planets. These plots emphasize the changes that can be expected in the albedo spectra given variations in methane abundance and surface gravity, as well as cloud albedo and depth in the atmosphere when the atmosphere is not clear of clouds. More spectral variations as a function of mass, orbit, metallicity, and phase are described in detail in \citet{Cahoy:2010} and \citet{Sudarsky:2000}. Distinctive differences diagnostic of important atmospheric processes between the spectra of known planets can clearly be expected. This study explores how well an instrument like the coronagraph on {\it WFIRST} would be able to constrain planet atmospheric composition.

\subsection{Retrieval Approaches}
Our atmospheric retrieval procedure involves combining a well-tested planetary albedo code \citep{McKay:1989,Marley:1999,Cahoy:2010} that can take into account multiple absorbers, cloud and Rayleigh scattering, and arbitrary incident and observed angles, with state-of-the-art Bayesian inference tools, namely the Markov chain Monte Carlo (MCMC) ensemble sample {\it emcee} \citep{Goodman:2010,Foreman-Mackey:2013} and the multimodal nested sampling algorithm {\it MultiNest} \citep{Feroz:2008,Feroz:2009,Feroz:2013} that can be used interchangeably. 

We believe that this is the first time such powerful retrieval techniques have been designed to {\it{simultaneously}} measure molecular abundances and cloud properties and their correlations from scattered light spectra. {\it{NEMESIS}} \citep{Rodgers:2000,Irwin:2008} is the only other existing retrieval method for planetary atmospheres in reflected light that has been applied to exoplanet characterization \citep{Barstow:2014}. By contrast to our Bayesian approach, {\it{NEMESIS}} uses non-linear optimal estimation to derive the best-fit model parameters and their uncertainties, and for exoplanet characterization did not include cloud properties explicitly as free parameters in the retrieval process. Instead, the effect of cloud properties on the retrieval results was investigated separately by calculating the $\chi^2$ goodness-of-fit over a large grid spanning cloud particle size, optical depth, and base pressure \citep{Barstow:2014}. Recently, cloud properties have been introduced in the {\it{NEMESIS}} retrieval scheme to analyze the scattering properties of Uranus \citep{Irwin:2015}. In this new approach the code retrieves the imaginary refractive index spectrum together with a Gamma distribution for particle size, characterized by a mean radius and variance. The extinction cross-section, single scattering albedo and phase function spectra are then calculated using standard Mie theory. Such parameterization allows for a more physical and self-consistent description of cloud and haze layers. Our method goes in the opposite direction, retrieving optical properties (optical depth, scattering albedo, and asymmetry factor) and cloud depth as model parameters, but not linking them to a physical model of cloud composition (such as particle size). As shown later in this paper, the presence of clouds naturally leads to degeneracies between methane abundance, cloud positions, and surface gravity. \citet{Irwin:2015} also highlight this degeneracy and constrain the cloud properties only by using a fixed, previously measured, methane abundance profile.

As shown by \citet{Line:2013,Line:2014a}, the Bayesian inference tools are 
better equipped to handle highly non-gaussian posterior distributions that are expected for future exoplanet observations, given the limited data and complex atmospheric models. Moreover, clouds play a significant role in the atmospheres of both gas giants in our Solar System and the exoplanets considered as future observing targets, given their expected effective temperatures. By including simple cloud properties (optical depth, albedo, depth in the atmosphere, etc.) as model parameters alongside molecular abundances, we can fully explore the degeneracies in the atmospheric structure, given the spectrum.

For our initial retrieval tests we constructed two highly idealized cloud models, one with a single cloud deck of arbitrary opacity, and the other with a scattering haze overlying a completely opaque cloud layer. Such atmospheric models are adequate for the types of planets addressed in this paper, and unquestionably can be improved in future work. Our goal is to determine if consistent results for scientifically interesting quantities (abundances, cloud properties) can be obtained using reflected light spectra from a space based coronagraph, given the likely modest signal-to-noise and spectral resolution.

\section{FORWARD MODEL}
\label{fwd}

Our geometric albedo code for giant planets was originally developed by \citet{Marley:1999} and is based on the methods of \citet{McKay:1989}. This code was subsequently modified and improved by \citet{Cahoy:2010}, who investigated the albedo variations as a function of star-planet distance, metallicity, mass, and phase angle. This original albedo code uses as input parameters the exoplanet's gravity and depth-dependent temperature, pressure, composition, and cloud properties
which are in turn computed by a 1-D radiative-convective
equilibrium model \citep{Marley:1999,Cahoy:2010}. The atmosphere is divided in 60 layers, with the bottom pressure marking the point beyond which photon scattering is negligible. This pressure level is taken from the radiative-convective equilibrium model for HD 99492c, and from the measured pressure-temperature profiles for Jupiter and Saturn \citep{Seiff:1998,Tyler:1982}. In all these cases, this pressure level is below the observable cloud decks. In summary, $P_{bottom}$ is 40 bars for HD 99492 c and the cloud free and 1-cloud validation cases, 10 bars for Jupiter and the 2-cloud validation case, and 251 bars for Saturn. In the full forward model the clouds are parametrized by wavelength-dependent optical depth $\tau_{cld}$, single scattering albedo ($\bar{\omega}_{cld}$), and scattering asymmetry factor ($\bar{g}_{cld}$), obtained from a full Mie scattering treatment of particle sizes
predicted by a cloud model \citep{Ackerman:2001}. The single scattering albedo represents the ratio between the amounts of scattering and total particle extinction, and the asymmetry factor,  $\bar{g}_{cld}$, is a measure of the degree of forward scattering. 

To simulate a spherical planet, we cover the illuminated surface of a sphere with 100 plane--parallel facets \citep{Cahoy:2010}, where each facet may have different incident and observed angles, $\mu_0=\cos \theta_0$ and $\mu_1=\cos \theta_1$, where $\theta_0$ and $\theta_1$ are the angles between the local normal vector and the star and observer, respectively. Although the ability to use different combinations of incident and observed angles allows for arbitrary planet phase angles, we modeled the planet as observed at 0-degree phase angle (face-on), in which case the observer and the source are collinear and $\mu_0=\mu_1$ for every facet. Increasing the number of facets proportionally increases the computing time, and only leads to a modest increase in accuracy. In this case, the albedo code takes about $3\,\rm s$ to run, which is reasonable to use in combination with an MCMC sampler. Although the general case permits $\theta_0 \ne \theta_1$, for the work reported here we set $\theta_0 = \theta_1$ in order
to compute geometric albedo, which by definition is the reflectivity at zero phase angle.
In a future work (Nayak et al, submitted) we will consider observations at arbitrary phase angle.

Following the approach of \citet{Horak:1950} and \citet{Horak:1965}, we use two-dimensional planetary coordinates and Chebyshev-Gauss integration to integrate over the emergent intensities and calculate the albedo spectra. The radiative transfer is performed point by point for each of the points sampling the planetary disk. The scattering source function \citep{Toon:1989,Meador:1980} includes the contributions of both diffuse and direct scattering:
\begin{equation}
\label{eq:source}
\begin{aligned}
S(\tau,\mu_1)=\frac{\bar{\omega}}{4\pi}F_0 p(\mu_1,-\mu_0)e^{-\tau/\mu_0}\\
+\int_{-1}^{1} \frac{\bar{\omega}}{2}I(\tau,\mu')p(\mu_1,\mu') d\mu', 
\end{aligned}
\end{equation}
where $F_0$ is the Solar flux at to top of the atmosphere, normalized to 1, and $p(\mu_1,\mu_2)$ is the scattering phase function. The two terms on the right-hand side represent the single and multiple scattering components, respectively.

We use a two-stream quadrature \citep{Toon:1989} to solve for the diffuse, angle-independent radiation field. This solution is then used as an approximation to the source function, which is then back-propagated to the top of the atmosphere, while adding the angular dependence given by the scattering phase function. This is a completely scalar approach and does not include any polarization effects.

Based on our experience and the results of \citet{Cahoy:2010}, we expect that the most important model parameters for Jupiter-like exoplanets in reflected light will be the methane abundance, surface gravity, and cloud properties. In a future paper
we will consider other gaseous opacity sources. The code uses the opacity for methane in the visible following \citet{Karkoschka:1994}, and the collision-induced absorption (CIA) for H$_2$-H$_2$, H$_2$-He and H$_2$-CH$_4$ as summarized in \citet{Freedman:2008}. 

The total gaseous absorption optical depth is then $\tau_{abs}=\tau_{CH4}+\tau_{CIA}$. In spite of newer methane line lists, difficulties remain in calculating the high-energy transitions of methane and \citet{Karkoschka:1994} is still the best reference for the methane opacity in the visible, and is used to reproduce Solar System measurements. We define $\tau_{total}=\tau_{scat}+\tau_{abs}$, where the total optical depth to scattering is $\tau_{scat}=\tau_{Ray}+\tau_{cloud}$. 

Following \citet{Cahoy:2010}, for the direct scattering (or single scattering term in Equation~\ref{eq:source}) we use a two-term Henyey-Greenstein scattering phase function with high forward scattering and moderate backscattering: 
\begin{equation}
\label{eq:tthg}
p_{TTHG}=\left(1-\frac{\bar{g}^2}{4}\right)p_{HG}(\bar{g},\Theta)+\frac{\bar{g}^2}{4}p_{HG}(-\bar{g}/2,\Theta),
\end{equation}
where
\begin{equation}
\label{eq:hg}
p_{HG}(\bar{g},\Theta)=\frac{1}{4\pi}\frac{1-\bar{g}^2}{(1+\bar{g}^2-2\bar{g}\cos\Theta)^{3/2}}
\end{equation}
and $\Theta$ is the scattering angle, related to the planet's phase angle $\alpha$ by $\alpha=\pi - \Theta$, and $\bar{g}$ is the scattering asymmetry factor associated with the scattering by cloud particles, $\bar{g}=\bar{g}_{cld}\times\tau_{cld}/\tau_{scat}$, since Rayleigh scattering is treated separately. 

For the multiple scattering term in Equation~\ref{eq:source}, the diffuse scattering phase function is written as a Legendre polynomial expansion, assuming azimuthal independence: 

\begin{equation}
p(\mu,\mu')=1+3\bar{g}\mu\mu'+\bar{g}_2(3(\mu\mu')^2-1)/2,
\end{equation}
where $\mu$ and $\mu'$ denote the scattered and incident angle, respectively, and $\bar{g}_2$ contains the Rayleigh scattering contribution $\bar{g}_2=\bar{g}_{Ray}\times\tau_{Ray}/\tau_{scat}$. Here $\mu$ and $\mu'$ are chosen such that the right solution is obtained in the Rayleigh limit. Rayleigh scattering is calculated following \citet{Hansen:1974}, with $\bar{g}_{Ray}=0.5$, and $\bar{\omega}_{Ray}=1$. The total layer single scattering albedo then becomes $(\bar{\omega}_{Ray}\tau_{Ray}+\bar{\omega}_{\rm cld}\tau_{\rm cld})/\tau_{total}$, for every layer in the atmosphere. Further details of the radiative-transfer modeling are described in \citet{Marley:1999,Cahoy:2010}. 

For retrieval purposes, we have preserved the radiative transfer and scattering prescription of the original albedo code, but made large simplifications to the input parameters. The simplified model used in the present study has constant molecular abundances throughout the atmosphere, with H$_2$ and He in primordial solar ratio. The pressure-temperature profile $T(P)$ of the atmosphere is kept fixed since we do not expect that our spectral range of interest ($0.4-1$~ $\mu$m) will contain any information for constraining it (see also \citet{Barstow:2014}). The wavelength dependence of the cloud parameters is also ignored (gray assumption for $\tau_{\rm cld}$, $\bar{g}_{\rm cld}$, and $\bar{\omega}_{\rm cld}$). The depth dependence is limited to parametrizing the cloud height and cloud top pressure, as described below.

In actuality of course the temperature-pressure profile will vary with surface gravity and this will primarily affect the atmospheric scale height. Here our variation of atmospheric gravity, $g$, stands in for variations in both $T(P)$ and $g$. As we add complexity to the model we will explore the sensitivity of retrievals to a varying $T(P)$.

\subsection{Cloud Models}
\begin {figure}
\centering
\includegraphics*[scale=0.5,angle=0]{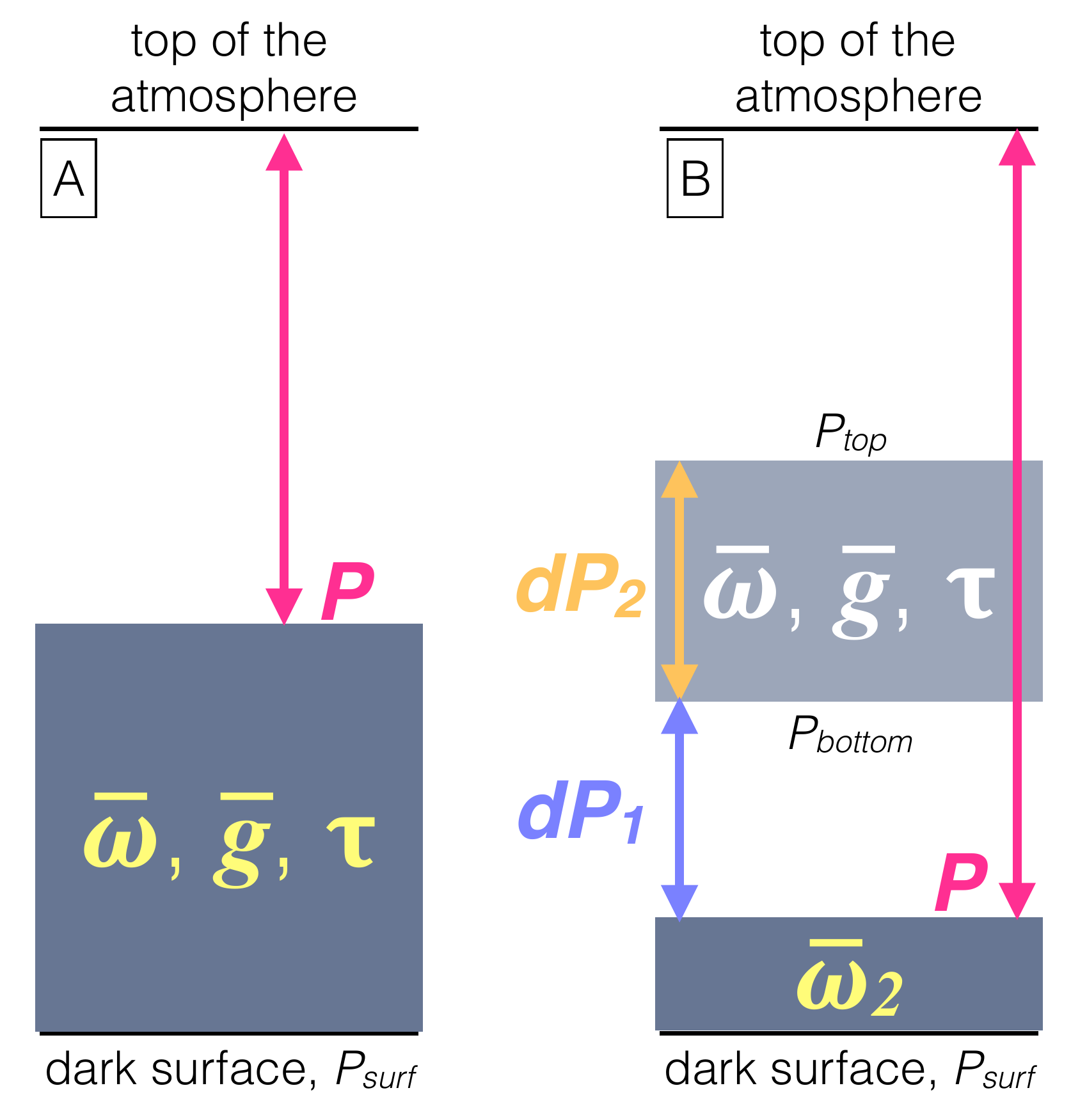}
\caption{Visual representation of our 1-cloud (panel A) and 2-cloud (panel B) models. The definitions of model parameters and their use in the albedo code are given in Sections~\ref{1cld} and \ref{2cld}, respectively.  \label{fig:cld}}
\end{figure}

As commonly employed in solar system giant planet atmosphere retrievals \citep[e.g.,][]{Sato:1979}, for the purposes of atmospheric retrieval we consider two different cloud 
treatments as illustrated in Figure~\ref{fig:cld}. The simpler of the two
models a single cloud layer while the more complex allows for two distinct clouds/hazes.
We describe each model in turn below.

\subsubsection{1-Cloud Model}
\label{1cld}

The one-cloud model is parameterized as a semi-infinite layer with a cloud top at pressure $P$ in the atmosphere and characterized by the single scattering albedo $\bar{\omega}$, scattering asymmetry factor $\bar{g}$, and the gray optical depth $\tau$ of the
layer where the top cloud is found. For simplicity of notation, we have dropped the subscript `cld' from the quantities $\bar{\omega}_{\rm cld}$, $\bar{g}_{\rm cld}$,  $\tau_{\rm cld}$, as defined in the previous section. This structure is shown in panel A of Figure~\ref{fig:cld}.  

The pressure of the cloud top is allowed to vary freely. Our typical input pressure-temperature profile has $N=60$ vertical atmospheric layers. We find the model layer in which the cloud top pressure is located, $j_c$ ($1\le j_c \le N$), and scale the cloud optical depth in this layer by the position of the cloud top pressure relative to the pressure at the bottom of the layer. The next deeper layer ($j=j_c+1$) will have cloud optical depth $\tau_j=\tau_{j_c}\times(P_{j+1}/P_j)$, where the layer number $j$ increases with depth in the atmosphere from 0 to $N$ and $P_j$ denotes the pressure at the top of layer $j$. The cloud optical depths in the following layers all the way to the bottom are calculated iteratively as $\tau_{j+1}=\tau_j\times(P_{j+2}/P_{j+1})$. Thus in this model
$\tau$ is essentially a measure of how opaque the cloud top is, and the optical depth per unit mass is constant over the entire vertical extent of the cloud. Large values of $\tau$ imply a rapid transition from cloudless atmosphere to cloud, whereas small values imply a more gradual increase of cloud opacity. Other cloud profile parameterizations are of course possible and we will explore these in future work.

The cloud single scattering albedo $\bar{\omega}$ and scattering asymmetry factor $\bar{g}$ are kept constant as a function of wavelength and depth in the atmosphere, below the layer containing the top of the cloud, e.g. $\bar{\omega}_j=...=\bar{\omega}_N=\bar{\omega}$ for $j\geq j_c$. This model will be referred in what follows as the ``1-cloud model", and is characterized by 6 parameters: $f_{\rm CH4}$, $g$, $P$, $\bar{\omega}$, $\bar{g}$, and $\tau$, where $g$ is the planet's surface gravity, to be distinguished from $\bar{g}$, and $f_{\rm CH4}$ is the methane abundance.

\subsubsection{2-Cloud Model}
\label{2cld}

Increasing complexity, we created a model appropriate for a cloud deck overlain by a haze
layer with a  very simple 2 layer structure shown in panel B of Figure~\ref{fig:cld}. Such a model is roughly capable of reproducing the structure observed in Solar System planets, and is a slight modification of the model used in the classic analysis of Jupiter's atmosphere by \citet{Sato:1979}. 

The parameters describing the lower cloud are its top pressure $P$ and single scattering albedo ($\bar{\omega}_2$). Following the same approach as in Section~\ref{1cld}, the pressure of the top of the bottom cloud is found in layer $j_c$, the optical depth below this level is scaled in the same way, except now $\tau=1$ in the top cloud layer, and is not variable. Thus this lower cloud has a sharply defined top layer and its total column optical depth is $\gg 1$ in all cases. This ensures that the bottom cloud is always optically thick, and makes it effectively act as a reflective surface, with a reflectivity controlled by $\bar{\omega}_2$, and situated at a variable depth given by $P$. 

The position of the upper cloud (or haze layer) relative to the bottom cloud is parametrized by the pressure difference between the top of the lower cloud and the bottom of the upper cloud ({$dP_1$) and the pressure difference between the top and the bottom of the upper cloud ($dP_2$). For computational convenience, these quantities are defined in log space, and are related to the size and location of the top cloud by $\log P_{\rm bottom}=P-dP_1$ and $\log P_{\rm top}=P-dP_1-dP_2$, where $P_{top}$ and $P_{bottom}$ are the pressures at the top and at the bottom of the upper cloud, respectively (see Panel B, Figure~\ref{fig:cld}).

Similar to the 1-cloud approach, we find the layers in which the top and bottom pressure of the upper cloud are located and the corresponding fractions, or locate the cloud in a single layer, if necessary. For all the layers between the top and the bottom, the optical depth of the upper cloud is scaled as $\tau_j=\tau\times(P_{j+1}-P_j)/(P_{\rm bottom}-P_{\rm top})$, where $\tau$ is the input variable and is wavelength-independent. The single-scattering albedo $\bar{\omega}$ and asymmetry factor $\bar{g}$ are again kept constant as a function of wavelength and for all layers between $P_{top}$ and $P_{bottom}$. This model will be referred in what follows as the ``2-cloud model", and is characterized by 9 parameters: $f_{\rm CH4}$, $g$, $P$, $dP_1$, $dP_2$, $\bar{\omega}$, $\bar{g}$, $\tau$, and $\bar{\omega}_2$.

Note that the haze single scattering albedo is treated as a constant with 
wavelength. Thus hazes that absorb preferentially in the blue, lowering the albedo in the short-wavelength part of the spectrum,
such as are commonly found in Solar System giant planet atmospheres, 
are not taken into account here. These effects become more important below 0.5 $\mu$m, and are unlikely to affect the region of interest for this study ($0.6-1$~$\mu$m). We will address the wavelength dependence of the single scattering albedo in future work, especially when adding photometric points in the blue. 

\begin{turnpage}
\begin{figure*}
\centering
\includegraphics*[scale=0.45,angle=0]{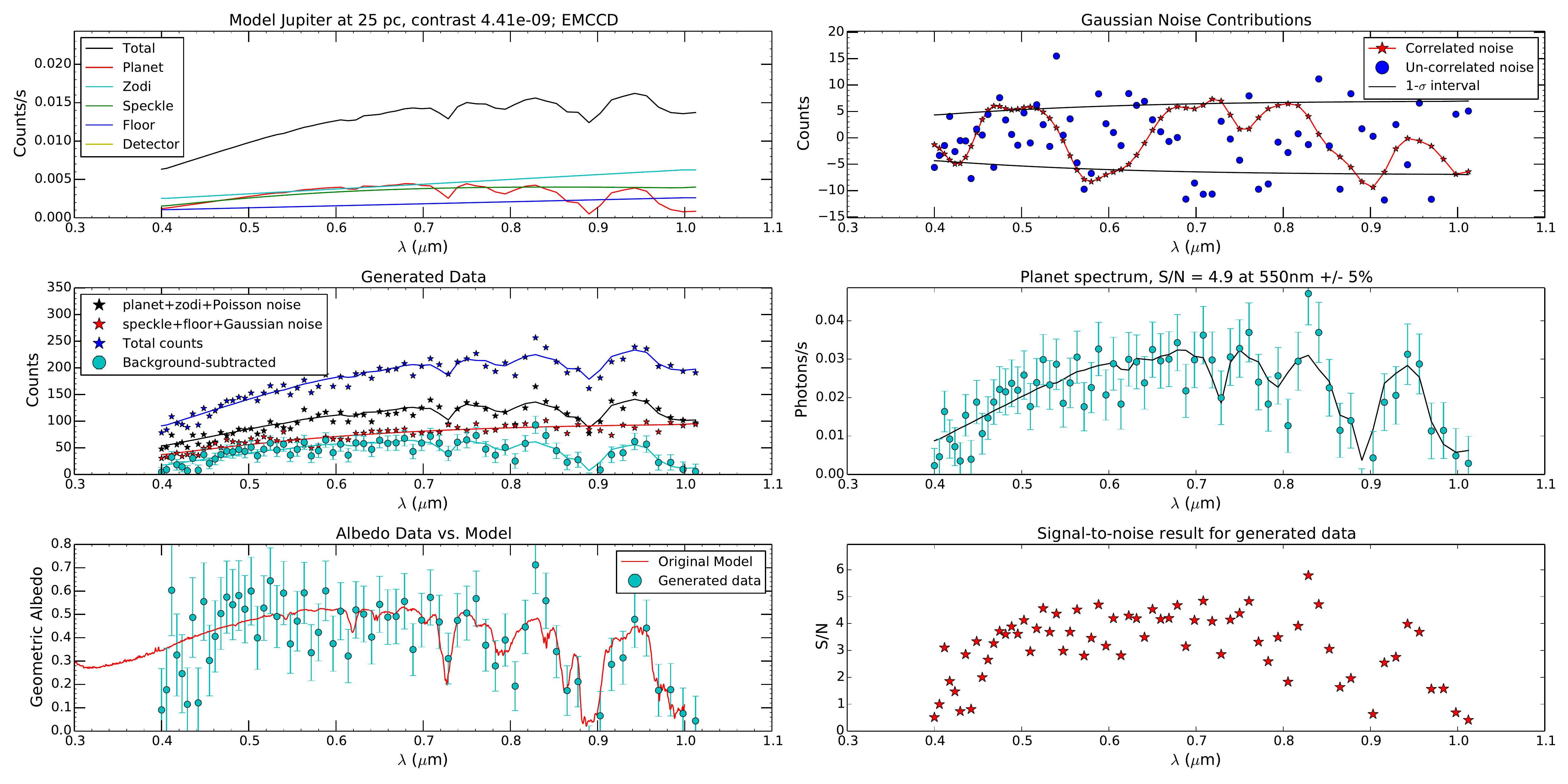}
\caption{In each set, the panels are as follows:
(Top left) Expected count rates from all different sources: planet –red, zodi- cyan, speckle-green, detector noise – blue and yellow (too small to see). The total count rate is shown in black.
(Middle left) Total number of counts after calculating the integration time needed to get a SNR of 5. The model counts are solid lines, and the simulated data are stars.
(Top right) Correlated and un-correlated noise contributions. These are added-in when generating the red star data points in the middle-left panel.
(Middle right) Simulated data converted to photon rate, after background subtraction (cyan), compared to the input model (black).
(Bottom left) Simulated data converted back to geometric albedo, after division by the stellar spectrum (cyan), vs actual albedo of Jupiter (red).
(Bottom right) SNR of the simulated data in each wavelength bin. The nominal SNR (5) corresponds to a 10\% band around 550 nm. \label{fig:noise}}
\end{figure*}
\end{turnpage}

\section{SIMULATED DATA}
\label{noise}

To simulate the direct imaging observations, we use a generic prescription for the total signal and associated noise expected in the planet's point spread function (PSF). This model is sufficient for investigating the effect of data quality (as quantified by the signal-to-noise ratio, SNR) on the size of uncertainties associated with the atmospheric parameters and on the significance of methane and cloud detection. We consider this to be a sufficiently general synthetic data model, that will be improved upon as more a detailed instrument simulator for the {\it WFIRST} coronagraph becomes available \citep[e.g.][]{Robinson:2016}. The plots in Figure~\ref{fig:noise} exemplify our simulated data for a Jupiter-like planet around a Sun-like star, at a distance of 25 pc from our Solar System, using the method detailed below.

Let the total number of counts on the detector, within the planet's PSF, be the sum of planet counts $n_{\rm pl}$, raw speckle counts $n_{\rm spec}$, zodiacal light $n_{\rm zodi}$, and the total detector background counts from all other sources. The spectral bins are chosen such that the resolving power $R=70$ is constant across the $0.4-1.0$~$\mu$m bandpass. For each spectral bin, we define  
\begin{equation}
\label{eq:sn}
\begin{split}
signal (\mathrm{e}^-) & = n_{\rm pl}\times t,\\
noise (\mathrm{e}^-) & = [n_{\rm total}\times t +(f_{pp}\times n_{\rm raw\ speckle}\times t)^2]^{1/2},
\end{split}
\end{equation}
where
\begin{equation}
\begin{split}
n_{\rm total} (\mathrm{e}^-/\mathrm{s}) &= [n_{\rm pl} + n_{\rm zodi} + n_{\rm raw\ speckle}  \\
&+D_c\times m_{\rm pix} + CIC\times m_{\rm pix}/t_{\rm frame} ]\times ENF^2 \\
&+ (N_R/G)^2\times m_{\rm pix}/t_{\rm frame}, 
\end{split}
\end{equation}
$n_{\rm total}$ is the total number of counts within the planet's PSF, $t$ is the total integration time, and the other quantities characterize the detector background noise, with ``typical values" for an electron multiplying (EM) CCD detector: $m_{\rm pix}=5$~pixels, $D_{C}=0.001$~e$^-$(pixel s)$^{-1}$, $N_{R}=3$~RMS~e$^-$(pixel frame)$^{-1}$, $t_{\rm frame}=300$~s, $CIC=0.001$~e$^-$(pixel frame)$^{-1}$, $ENF=1.414$, $G=1000$, and $t=14000$~s. These estimated count rates are generic values, and will vary with the type of planet and wavelength. However, they are a good starting point for our study in SNR space, to scale the relative contributions of different noise sources. The factor $f_{pp}$ quantifies the  speckle reduction efficiency that is expected in post-processing, and can take values roughly between 1/10 and 1/30 \citep{Traub:2016}. We use the generally adopted value $f_{pp}=1/20$ in this paper. 

Assuming the stellar spectrum to be a blackbody at 6000~K, and using the model geometric albedo of the planet, we have calculated the expected number of photons in each spectral bin. This number was converted to a count rate, using estimated count rates of $n_{\rm pl}=0.012$~e$^-$/s, $n_{\rm zodi}=0.012$~e$^-$/s, $n_{\rm spec}=0.010$~e$^-$/s, which contain information about the expected quantum efficiency. It should be noted that here we are making the simplest assumptions on the noise model and in general $n_{\rm pl}$ depends on wavelength and planet type. A more sophisticated noise model for the {\it WFIRST} coronagraph instrument has recently been made available \citep{Robinson:2016} and will be used in future work. The number counts coming from all contributions to the total signal are shown in Figure~\ref{fig:noise}, top left panel. The observed spectrum is simulated assuming that the planet and zodi counts have a Poisson distribution (per channel), while the speckle and detector noise counts have a Gaussian distribution (Figure~\ref{fig:noise} center left). In other words, the simulated data points are drawn from their respective distributions. 

In addition, we consider the possibility of noise correlations among different spectral regions. Since the speckle positions relative to the central star change with wavelength, we expect that at the position of the planet in the observed image certain wavelengths will be more affected by speckle noise than others. In our model, we assume that this will affect only the Gaussian-distributed counts, which are dominated by speckle counts, and not Poisson-distributed ones, which consist of planet and zodi counts. Therefore, the total noise contribution of the Gaussian-distributed counts (their distribution around the mean) was split into 2 components, one spectrally correlated, and one spectrally uncorrelated. The correlated noise component was generated as a Gaussian random process with a squared-exponential kernel and correlation length scale of either 25 or 100 nm. These length scales are appropriate for our chosen spectral range and expected spatial resolution, and the choice of a random process reflects the existing uncertainty in the exact behavior of the speckle noise correlation. Furthermore, we assumed that both correlated and uncorrelated components have equal contributions to the total scatter in the data points, and therefore their distributions will have mean zero and equal variance. This combination of spectrally correlated and uncorrelated noise is shown in the top right panel of Figure~\ref{fig:noise}.

We define the signal-to-noise reference value (SNR$_0$=signal/noise, from Equation~\ref{eq:sn}) as corresponding to the integrated number of counts in a 6\%-wide bandpass centered at 450~nm. Therefore, the integration time needed to achieve a given SNR$_0$ can be calculated as 
\begin{equation}
t (\mathrm{s}) = \frac{SNR_0^2 \times n_{0\rm total}}{n_{0\rm pl}^2-(SNR_0\times f_{pp}\times n_{0\rm raw\  speckle})^2},
\end{equation}
where the index $0$ denotes the fact that these values are calculated for the 550~nm reference bandpass. We calculate the integration time $t_0$ necessary to obtain a SNR$_0$ of 5, 10, or 20, respectively, which is then used to calculate the expected number of counts and scale the signal and noise across the entire bandpass. The final error bars are computed individually for each simulated data point using Equation~\ref{eq:sn}. As shown in Figure~\ref{fig:noise}, the resulting spectrum will have a SNR$<$SNR$_0$ on average, but we will take the SNR$_0$ as the reference value in what follows. The values for SNR$_0$ and speckle noise correlation length as defined above serve as a parametrization of the data space over which we perform our retrievals. The combination of the three SNR values and two possible speckle noise correlation lengths result in 6 simulated datasets for each planet model. 

Lacking more detailed information about the instrument, in the above we have assumed that the entire bandpass is observed simultaneously and the quantum efficiency (detector response) is constant across the bandpass. Although these conditions will not be satisfied in a real observation, they amount to assuming that we can achieve the final SNR distribution with wavelength shown in the bottom right panel of Figure~\ref{fig:noise}. This is just one of the many possible realizations of SNR variation over the bandpass, and this is likely to be unique to each dataset, which will likely be a combination of different observing modes. It is to be expected that the best fit parameter values from our retrievals will depend on the noise distribution with wavelength, as well as on the individual random point generation for each simulated dataset. A complete instrument simulator will be needed to estimate the actual science return from a future mission.

\section{ATMOSPHERIC RETRIEVAL SCHEME}
\label{mcmc}

The allowed ranges and best fit values for the forward model parameters, given the data, are determined using two Bayesian posterior sampling algorithms, namely the affine invariant ensemble Markov chain Monte Carlo sampler, $emcee$ \citep{Goodman:2010, Foreman-Mackey:2013}, and the multimodal nested sampling algorithm $MultiNest$ \citep{Feroz:2008,Feroz:2009,Feroz:2013}. These approaches permit efficient sampling of highly correlated, non-gaussian, and high-dimensional parameter spaces, and are very readily scaleable to multi-processor computing. 

The different approaches taken by the two algorithms in sampling the posterior parameter space can help us avoid the pitfalls of either one. While $emcee$ starts with a first guess and can become trapped in a local minimum, MultiNest starts with a grid of points covering the entire prior parameter space and proceeds by narrowing down the maximum likelihood regions. On the other hand, $MultiNest$ could favor highly-peaked, multi-modal, Gaussian-like distributions, while $emcee$ is more agnostic to the shape of the posterior and can reveal additional tails and correlations. The total evidence for any given model (the integral over the posterior distribution) is automatically calculated by $MultiNest$ as a part of the algorithm, but requires extra steps and can be tricky to compute for $emcee$. Ideally, the two methods will converge to the same solution. 

Overall, we consider the two approaches complementary, and offer greater confidence in avoiding potential biases. Recently, \citet{Allison:2014} have compared in detail these sampling techniques and found that nested sampling is more time-efficient while still providing good accuracy, and the affine-invariant MCMC sampler can be competitive when massively parallelized. They both outperform by far traditional Metropolis-Hastings algorithms. For completeness, we provide a brief description of the two posterior sampling algorithms in the Appendix.

A second component of the retrieval process consists of model comparison, with the purpose of quantifying not only the uncertainties in the model parameters, but also the evidence in support of a chosen model. In this step we can assess whether the 1-cloud or 2-cloud model presented in Section~\ref{1cld} and \ref{2cld} offer a better representation of the data and calculate the significance associated with the cloud or methane detection. The choice between two competing models $\mathcal{M}_X$ and $\mathcal{M}_Y$ then comes down to comparing their probabilities by constructing the {\it Bayes factor} 
\begin{equation}
\label{eq:bdef}
B_{XY}=\frac{\mathcal{P}(\mathcal{M}_X\mid \mathcal{D})}{\mathcal{P}(\mathcal{M}_Y\mid \mathcal{D})}=\frac{\mathcal{Z}_X}{\mathcal{Z}_Y}\frac{\mathcal{P}(\mathcal{M}_X)}{\mathcal{P}(\mathcal{M}_Y)},
\end{equation}
where $\mathcal{Z}$ is the Bayesian evidence defined in the Appendix. Usually the last term in Equation~\ref{eq:bdef} is 1 (both models have the same probability). We use the guidelines provided by \citet{Jeffreys:1961,Raftery:1996} for assessing the evidence in support of model $\mathcal{M}_X$ vs  $\mathcal{M}_Y$ in terms of Bayes factor:

\begin{equation}
\label{eq:bf}
\begin{split}
2\log B_{XY} < 0 & \mbox{:  Negative (supports }\mathcal{M}_Y),\\
0 < 2\log B_{XY} < 2 & \mbox{:  Inconclusive},\\
2 < 2\log B_{XY} < 5 & \mbox{:  Positive},\\
5 < 2\log B_{XY} < 10 & \mbox{:  Moderate},\\
2\log B_{XY} > 10 & \mbox{:  Very Strong (supports }\mathcal{M}_X).
\end{split}
\end{equation}
This ranking system is equally applicable when the evidence supports model $Y$, in which case we simply calculate $B_{YX}$.

Since the posterior distribution in general does not have an analytic form, the difficulty arises when attempting to compute $\mathcal{Z}$ for each model under consideration. In general, the evaluation of Bayesian evidence from an existing MCMC posterior is limited by the poor sampling of regions of low likelihood. This problem can be overcome using thermodynamic integration, at computational costs $10-100\times$ higher than a regular MCMC  \citep[e.g. ][]{Trotta:2008,Calderhead:2009}.  However, as long as the Bayes factor is found within the ranges in Equation~\ref{eq:bf}, the precise value of $B_{XY}$ is not important. In general, some rough assumptions are made on the functional shape of the prior and posterior distributions to be able to approximate the value of this integral. While these approximations are not very accurate, \citet{Cornish:2007} show that for high signal-to noise data (SNR $\gtrsim 9$) all methods converge toward the same values. In this paper we estimate $\mathcal{Z}$ using three different methods: the Schwarz-Bayes information criterion \citep[BIC, ][]{Schwarz:1978}, the Laplace approximation \citep{Lopes:2004, Cornish:2007}, and the Numerical Lebesgue Algorithm (NLA) described by \citet{Weinberg:2012}. We refer the reader to the Appendix for a summary of these methods and relevant definitions. The scatter among the results given by these three methods are indicative of the reliability of these approximations for various models and SNR regimes. In general, we observe that the values converge when the evidence for a given model is very strong. Further, these results obtained from the MCMC samples are validated by comparison with the evidence values calculated by default with the nested sampling algorithm.

\subsection{Priors}
\label{prio}

\begin{deluxetable*}{lcccccc}
\tabletypesize{\scriptsize}
\tablecolumns{8}
\tablecaption{Model parameters and priors for the 1-cloud forward model.\label{tab:param1c}}
\tablehead{
\colhead{Planet} & \colhead{$\log(f_{\rm CH4})$} & \colhead{$\log(g)$} & \colhead{$\log(P)$} & \colhead{$\bar{\omega}$} & \colhead{$\bar{g}$} & \colhead{$\log(\tau_{\rm top})$\tablenotemark{a}} \\
 & & (m s$^{-2}$) & (bar) & & & }
\startdata
Cloud-free case & [-8.,0.] & [-1.,2.] & [-4.4,1.6] & [0.01,0.9999] & [0.01,0.9999] & [-10.,2.]\\
1-cloud case & [-8.,0.] & [-1.,2.]  & [-4.4,1.6] & [0.01,0.9999] & [0.01,0.9999] & [-5.,3.]\\
2-cloud case & [-8.,0.] & [0.,2.] & [-4.4,0.9999] & [0.01,0.9999] & [0.01,0.9999] & [-5.,3.]\\
HD 99492 c & [-8.,0.] & [-1.,2.] & [-4.4,1.6] & [0.01,0.9999] & [0.01,0.9999] & [-4.,3.] \\
Jupiter & [-8.,0.] & [0.,2.] & [-4.4,0.9999] & [0.01,0.9999] & [0.01,0.9999] & [-5.,3.] \\
Saturn & [-8.,0.] & [0.,3.] & [-5.9,2.39] & [0.01,0.9999] & [0.01,0.9999] & [-5.,3.]\\
\enddata
\tablenotetext{a}{For clarity, here the cloud optical depth parameterization is written as $\tau_{\rm top}$, to show the difference between the two forward models (see Sections \ref{1cld} and \ref{2cld}).}
\end{deluxetable*}

\begin{deluxetable*}{lcccccccccc}
\tabletypesize{\scriptsize}
\tablecolumns{8}
\tablecaption{Model parameters and priors for the 2-cloud forward model.\label{tab:param2c}}
\tablehead{
\colhead{Planet} & \colhead{$\log(f_{\rm CH4})$} & \colhead{$\log(g)$} & \colhead{$\log(P)$\tablenotemark{a}} & \colhead{$dP_1$\tablenotemark{a}} & \colhead{$dP_2$\tablenotemark{a}} & \colhead{$10.^{P-dP_1-dP_2}$\tablenotemark{b}} & \colhead{$\bar{\omega}$} & \colhead{$\bar{g}$} & \colhead{$\log(\tau_{\rm total})$\tablenotemark{c}} & \colhead{$\bar{\omega}_2$} \\
 & & (m s$^{-2}$) & (bar) & & & (bar) & & & & }
\startdata
Cloud-free case & [-8.,0.] & [-1.,2.] & [-4.4,1.6] & $>0$ & $>0$ & 4.e-5 & [0.01,0.9999] & [0.01,0.9999] & [0.01,0.9999] & [-10.,3.]\\
1-cloud case & [-8.,0.] & [-1.,2.] & [-4.4,1.6] & $>0$ & $>0$ & 4.e-5 & [0.01,0.9999] & [0.01,0.9999] & [0.01,0.9999] & [-4.,3.]\\
2-cloud case & [-8.,0.] & [0.,2.] & [-5.3,0.9999] & $>0$ & $>0$ & 4.e-5 & [0.01,0.9999] & [0.01,0.9999] & [0.01,0.9999] & [-3.,2.]\\
HD 99492 c & [-8.,0.] & [-1.,2.] & [-4.4,1.6] & $>0$ & $>0$ & 4.e-5 & [0.01,0.9999] & [0.01,0.9999] & [0.01,0.9999] & [-4.,3.]\\
Jupiter & [-8.,0.] & [0.,3.] & [-5.3,0.9999] & $>0$ & $>0$ & 4.e-5 & [0.01,0.9999] & [0.01,0.9999] & [0.01,0.9999] & [-3.,3.]\\
Saturn & [-8.,0.] & [0.,3.] & [-5.9,2.39] & $>0$ & $>0$ & 1.2e-6 & [0.01,0.9999] & [0.01,0.9999] & [0.01,0.9999] & [-3.,3.]\\
\enddata
\tablenotetext{a}{For correspondence with $P_{top}$ and $P_{bottom}$ in Figure~\ref{fig:cld}, $P$, $dP_1$ and $dP_2$ are defined such that $\log P_{\rm bottom}=P-dP_1$ and $\log P_{\rm top}=P-dP_1-dP_2$.}
\tablenotetext{b}{Extra prior for the 2-cloud model ensuring that the sum of the layers does not exceed the height of the atmosphere.}
\tablenotetext{c}{For clarity, here the cloud optical depth parameterization is written as $\tau_{\rm total}$, to show the difference between the two forward models (see Sections \ref{1cld} and \ref{2cld}).}
\end{deluxetable*}

The parameters retrieved for each of the cloud models are described in Sections~\ref{1cld} and \ref{2cld}. In addition to the cloud properties, we are retrieving the methane abundance and surface gravity. For each retrieval case, the priors on the parameteres for the 1-cloud and 2-cloud models are shown in Tables~\ref{tab:param1c} and \ref{tab:param2c}, respectively. Water and alkali abundances will be included as model parameters in future work; however, for the applications considered in this paper (e.g. Jupiter, Saturn), methane is the main absorber. We define the atmospheric methane mixing ratio, {\it{fCH4}}, as the volume mixing ratio of methane. Since in a giant planet atmosphere 98\% of the atmospheric constituents are H$_2$ and He, this uniquely defines the atmospheric methane content. Such an approach would not be possible for a terrestrial planet of course. 

We allowed gravity to vary because in the realistic case neither the size of the planet nor the planetary mass will be known precisely. We allowed an exceptionally large range of gravities to be tested by the retrievals. In a realistic case the planet mass (for RV planets) will be known to substantially better than a factor two by the orbital astrometry solution. From the mass-radius relationship for gas giant planets and albedo scaling arguments  the radius will likely be known to within 50\%, which dominates the gravity uncertainty. Thus for a Jupiter twin the gravity ($g = 25$~m~s$^{-2}$) would plausibly be known to be $<100$~m~s$^{-2}$, not $<1000$~m~s$^{-2}$ as is the constraint placed in most of the results shown here. This turned out to be very important as, all else being equal, a large methane mixing ratio is required at high gravity to produce equivalent
absorption band depths as a lower abundance at lower gravity. 

We recognize the degeneracies that will be introduced by the unknown planet radius and phase angle. In an extension of this work (Nayak et al., submitted) we are explicitly separating the mass and radius and introduce the phase angle as a new parameter. In the current work, the stellar flux is normalized to 1, such that the planet radius does not factor in directly. However, in a realistic case the radius of the planet will act as an overall scaling factor, and we expect to see degeneracies between the radius, phase angle, and planet reflectivity (here $\bar{\omega}$ and/or $\bar{\omega}_2$). These correlations will add to the uncertainties, and have to be seen as a caveat in the present work.

The only restriction on the vertical cloud structure ($P$, $dP_1$, and $dP_2$) is that it does not exceed the total vertical extent of the atmosphere. The cloud albedos and asymmetry factor are allowed to take any value between 0 and 1, while the optical depth of the upper cloud varies between $10^{-3}$ and $10^3$. This optical depth is also varied in the 1-cloud model, but the lower cloud in the 2-cloud model is assumed optically thick (see Section~\ref{2cld}).

The pressure-temperature profile of the atmosphere is kept constant, since there is no information in the spectra at these wavelengths ($0.4-1.0$~$\mu$m) to constrain it. We are considering replacing this fixed profile by a parametrized one, to better account for the effect of surface gravity \citep{Line:2013}. 

\begin {figure*}
\centering
\hspace*{-0.4cm}\includegraphics*[scale=0.62,angle=0]{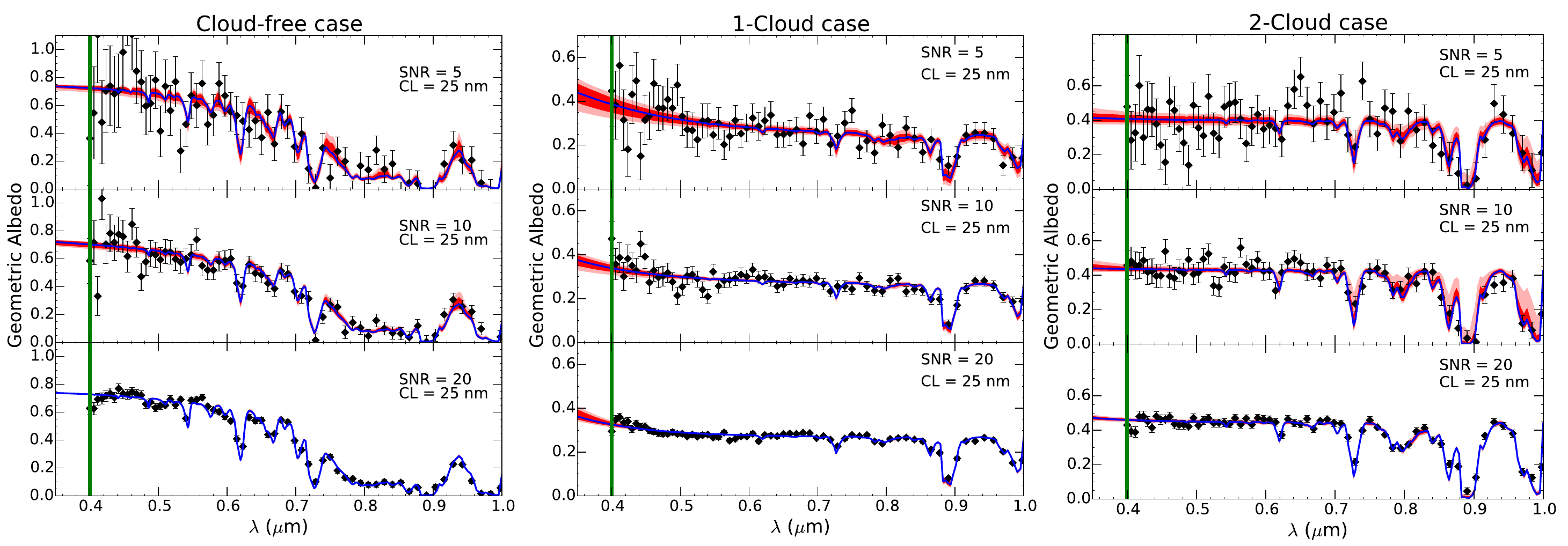}
\caption{Simulated data and best fit spectra for the cloud free case in Section~\ref{v0c} (left) and the single cloud case in Section~\ref{v1c} (middle), using the {\bf 1c} forward model, and for the for the 2-cloud case in Section~\ref{v2c} (right), using the {\bf 2c} forward model. The data correspond to SNR=5, 10, 20, from top to bottom and a spectral correlation noise of 25 nm. The results for a correlation length of 100 nm are similar. The solid and semi-transparent red regions represent $1-\sigma$ and $2-\sigma$ intervals, respectively. These intervals represent the standard deviation a set of 500 spectra generated using random samples from the converged MCMC distribution. The blue line represents the median of this set. The retrieval was performed over the $0.4-1.0$ $\mu$m region, as indicated by the green vertical line. \label{fig:vspec}}
\end{figure*}

\subsection{Implementation}
\label{comp}

The forward models described in Sections~\ref{1cld} and \ref{2cld} have been coded in Fortran and converted into a Python-callable library using f2py (now part of the NumPy package). The retrieval scheme integrates this library with either emcee or PyMultiNest, alternatively. Both MCMC and nested sampling implementations are easily scalable to run from a laptop to a computer cluster. The Fortran code is also parallelizable, but this does not provide a significant increase in speed as long as the MCMC is parallelized. Our retrievals were run on the {\it NASA} {\it Pleiades} Supercomputer, where we highly optimized the code for the forward models, and took advantage of the parallel nature of the algorithms to run on up to 216 processors at the same time (one 24-core node per model parameter). The MultiNest algorithm is found to converge rapidly even when run on just 1-2 nodes.

We have quantified the methane and cloud detections by calculating the ratios of their respective Bayes factors, as described in Section~\ref{mcmc}. For each case (SNR and spectral correlation length combination), a set of four different forward models was used: the 2-cloud model with 9 parameters (Section~\ref{2cld}), the 1-cloud model with 6 parameters (Section~\ref{1cld}), a model without clouds (the cloud subroutines are turned off in the previous models), and a model without methane (the methane abundance is set to $10^{-20}$ in the previous models). Therefore, for each planet example, we ran a set of 24 retrievals using $emcee$. In addition, we performed the same retrievals using $MultiNest$ for the models with a spectral correlation length of 25~nm mainly to cross-check the Bayesian evidence values calculated from the MCMC chains. In cases of good convergence, $MultiNest$ also provided parameter constraints in agreement with $emcee$ at a lower computational costs.

\section{RETRIEVAL VALIDATION}
\label{results}

In order to validate our retrieval procedure, we generate albedo spectra using the 1-cloud and 2-cloud models presented in Section~\ref{1cld} and \ref{2cld}, respectively.  We use the 1-cloud forward model to generate 2 types of spectra: one for an optically thin cloud very deep in the atmosphere, equivalent to a cloud-free atmosphere; and one for an optically thick cloud at moderate height. The third case is generated with the 2-cloud model. The model spectra are then converted to simulated observations using the noise prescription described in Section~\ref{noise}. For each of these three cases we investigate the ability to retrieve the input model parameters, as a function of SNR and noise correlation length. For each of the three cases we ran retrievals using the full 1-cloud and 2-cloud models, a forward model with the clouds turned off (referred to as ``no clouds"; defaults to 0 for all $\bar{g}$, $\bar{\omega}$, and $\tau$'s), and a forward model with negligible methane abundance (referred to as ``no methane", {\it fCH4$=10^{-20}$}). For convenience of notation, we will refer to these four model retrievals as {\bf 1c}, {\bf 2c}, {\bf -c}, and {\bf -m}, where a {\bf 2c-m} notation for example would stand for ``2-cloud forward model without methane". Each SNR and spectral noise correlation length combination was run through the retrieval procedure four times to enable model comparison and assess the significance of methane and cloud detection. Tables~\ref{tab:vres1} and \ref{tab:vres2} summarize the input parameter values for each of the simulated spectra, and the confidence intervals for each parameter obtained after running the retrieval procedure.

\begin {figure*}
\centering
\includegraphics*[scale=0.3,angle=0]{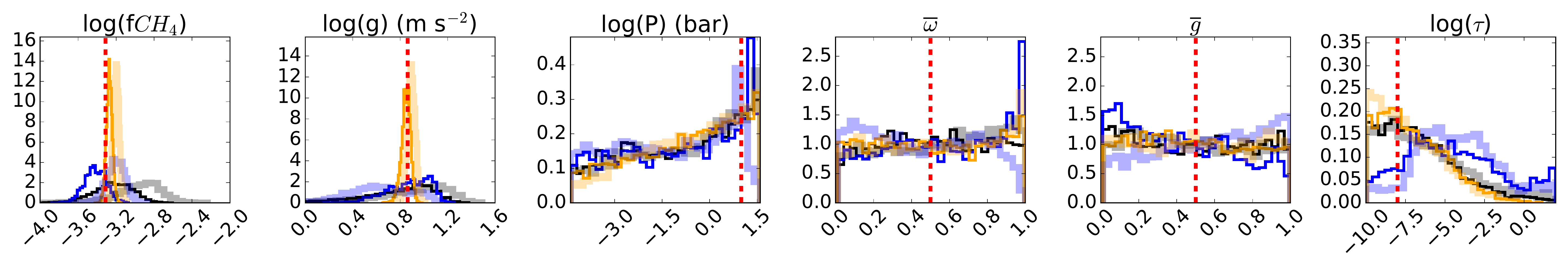}
\includegraphics*[scale=0.3,angle=0]{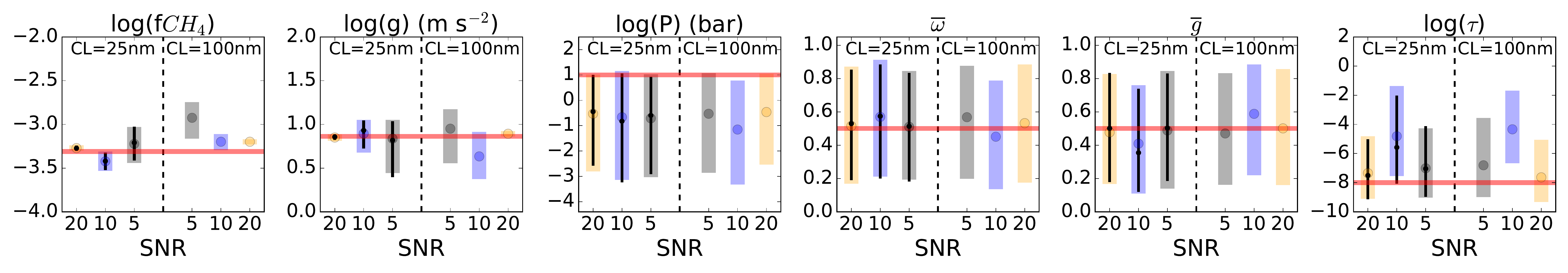}
\caption{Upper: 1-D marginal likelihood distributions for the six parameters in the 1-cloud model ({\bf 1c}) for the cloud-free case in Section~\ref{v0c}. The SNR values are color-coded, with black, blue, and orange for SNR 5, 10, and 20, respectively. The thin solid histograms show the distributions corresponding to a noise correlation length of 25~nm, and the thick semi-transparent ones for a noise correlation length of 100~nm. Lower: Confidence intervals for the model parameters retrieved using MCMC. The color coding matches the upper panel, the black lines show the $1\sigma$ intervals from the nested sampling retrievals, and the red horizontal line shows the input parameter value in the original albedo model. The two spectral correlation lengths are labeled in the left/right parts of the plots. These values are also summarized in Table~\ref{tab:vres1}. Note that the confidence intervals are calculated from the distribution quantiles, and do not reflect possible upper/lower limits or unconstrained parameters that can be seen in the histograms. \label{fig:c0fit}}
\end{figure*}

\begin {figure*}
\centering
\includegraphics*[scale=0.55,angle=0]{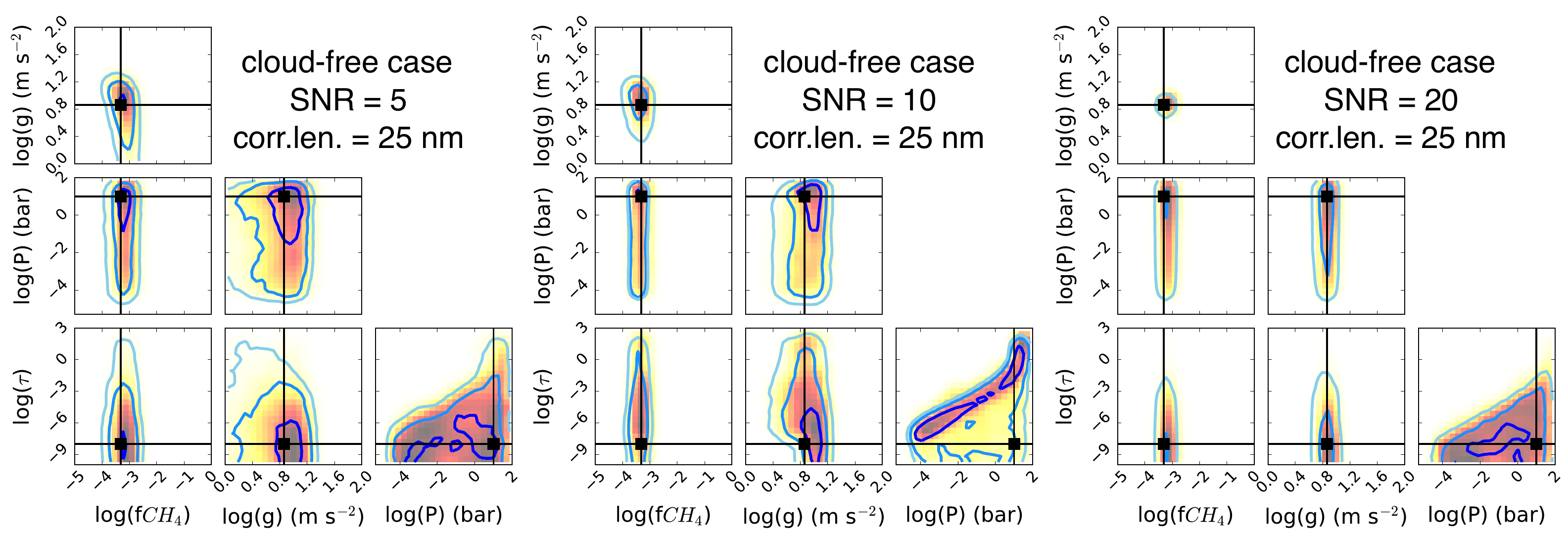}
\caption{2-D marginal posterior probability distributions for SNR=5, 10 and 20, and spectral noise correlation length of 25 nm, for the cloud free case in Section~\ref{v0c}, using the {\bf 1c} forward model. Since the $\bar g$ and $\bar{\omega}$ parameters are unconstrained in this case, we only plot the remaining ones. The red color map corresponds to distributions obtained using the MCMC algorithm, and the blue contours to nested sampling. The black lines show the real solution. \label{fig:c0dd}}
\end{figure*}

\begin {figure}
\centering
\includegraphics*[scale=0.38,angle=0]{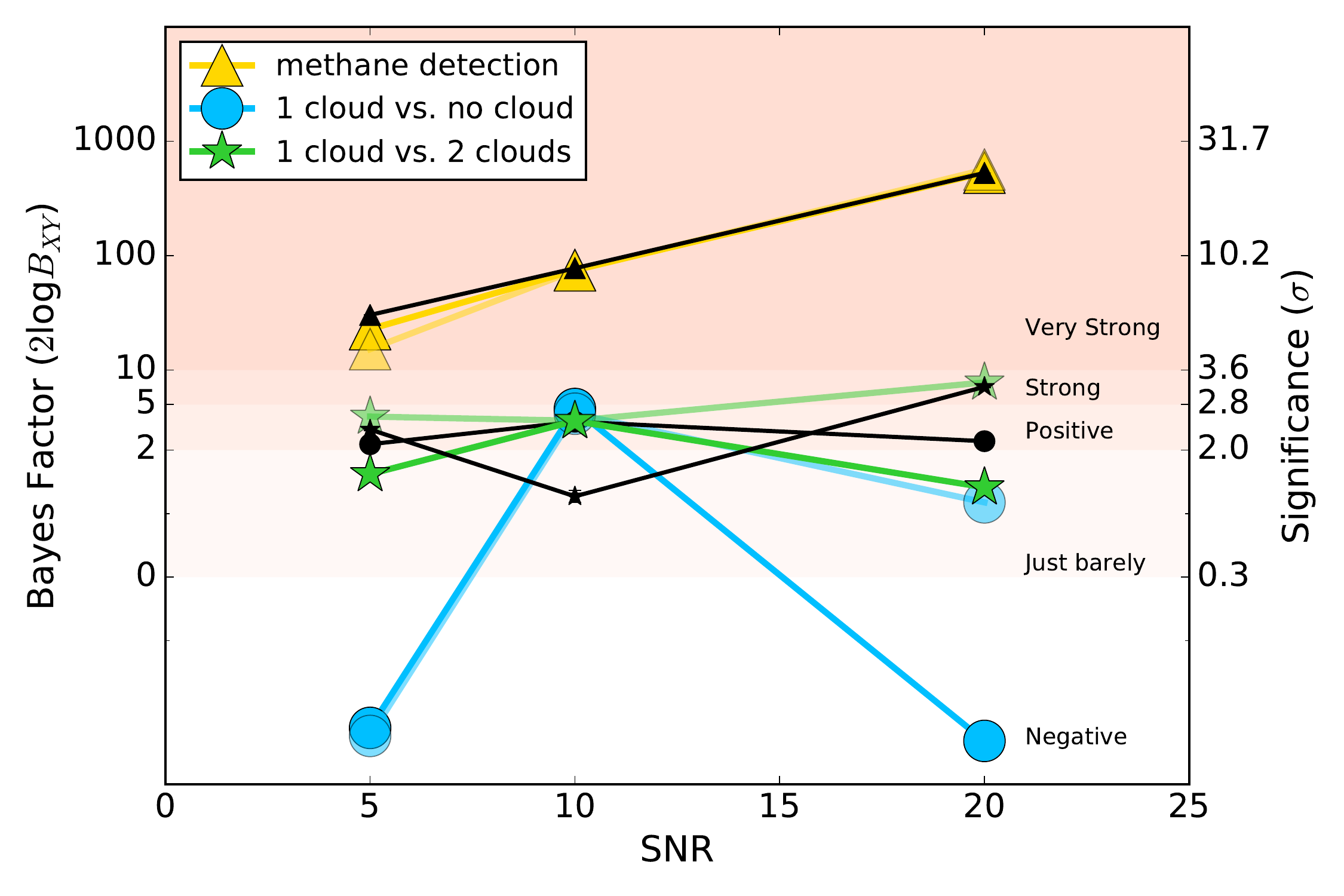}
\caption{Bayes factors and associated significance levels, as defined in Section~\ref{evidence}, for the cloud free case in Section~\ref{v0c}. The vertical shading grades follow the intervals defined in Equation~\ref{eq:bf}. The yellow triangles correspond to the ratios $\mathcal{Z}_{1c}/\mathcal{Z}_{1c-m}$, the blue circles to $\mathcal{Z}_{1c}/\mathcal{Z}_{1c-c}$, and the green stars to $\mathcal{Z}_{1c}/\mathcal{Z}_{2c}$. The colored symbols represent the results derived from the MCMC samples, with the solid color corresponding to a noise correlation length of 25~nm, and the semi-transparent to a noise correlation length of 100~nm. For comparison, the black symbols use the evidence values provided by the nested sampling algorithm for the cases with a noise correlation length of 25~nm. The symbols correspond to the same Bayes factors shown in color. The values calculated using nested sampling have associated error bars, but too small in general to see on this plot. \label{fig:cld0ev}}
\end{figure}

\subsection{Cloud-free Case}
\label{v0c}

We construct the albedo of a cloud-free planet using the 1-cloud model in Section~\ref{1cld}, where the optical depth $\tau$ is set to $10^{-8}$ and the top pressure of the cloud to 10 bar. The other parameters used to generate the model spectrum are listed in Table~\ref{tab:vres1}. Using the noise prescription in Section~\ref{noise}, we generate simulated datasets for SNR values of 5, 10, and 20, and spectral noise correlation lengths of 25 and 100~nm. The data realizations can be seen in the left panel of Figure~\ref{fig:vspec}. The retrieval is performed over the wavelength range 0.4-1.0~$\mu$m, indicated by the green line in Figure~\ref{fig:vspec}. Figures~\ref{fig:c0fit} and \ref{fig:c0dd} show the retrieval results. The marginal probability distributions for the model parameters are shown in the top panel in Figure~\ref{fig:c0fit}. The associated confidence intervals are bounded by the 16\% and 84\% quantiles of the cumulative probability distributions and are shown in the bottom panel of the same figure. These confidence intervals are also listed in Table~\ref{tab:vres1}. 

We find that for a cloud-free planet both the methane abundance {\it fCH4} and surface gravity $g$ are well constrained. The methane abundance is constrained to within a factor of $\sim2.6$ at a SNR of 5 and within a factor of $\sim1.15$ at a SNR of 20. The surface gravity is constrained to within a factor of $\sim4$ at a SNR of 5 and within a factor of $\sim1.2$ at a SNR of 20. As expected, the cloud albedo $\bar{\omega}$ and scattering asymmetry factor $\bar{g}$ are not constrained, since they do not contribute to the observed spectrum. 

The 2-dimensional posterior probability distributions shown in Figure~\ref{fig:c0dd} trace the changes in the parameter constraints as the SNR increases from 5 to 20. This is also reflected by the decrease in the size of confidence intervals shown in the bottom panel of Figure~\ref{fig:c0fit}. The distributions clearly become narrower and more peaked as the SNR increases. This projection also shows that the pressure of the top of the cloud deck in the model is partly correlated with the optical depth $\tau$. A larger top cloud pressure (deeper cloud) allows for a larger range of optical depths. This can be intuitively understood since a deep cloud will have little effect on the observed spectrum even when its optical depth is larger. The range of spectra obtained using parameters drawn from the posterior probability distributions are shown by the red contours in Figure~\ref{fig:vspec}. We also note the excellent agreement between the MCMC and nested sampling methods, where the nested sampling results are shown by the blue contours in Figure~\ref{fig:c0dd}, and by the black lines in Figure~\ref{fig:c0fit}.

The posterior constraints on the cloud parameters $P$, $\tau$, $\bar{\omega}$, and $\bar{g}$ already indicate that the spectrum does not support the presence of an observable cloud. This is further confirmed by the Bayesian evidence analysis. We sample the posterior probability distributions for a set of 4 models: {\bf 1c}, {\bf 1c-m}, {\bf 1c-c}, and {\bf 2c}, as defined above. The pairwise Bayes factors for these models are shown in Figure~\ref{fig:cld0ev}. Clearly, methane is detected with a high significance even when the spectral SNR is 5 (yellow triangles). 
However, the presence of a cloud is not supported. The models containing 2-clouds, 1-cloud, or no clouds are equally able of describing the data, since even in a multiple cloud model the optical depth of the clouds can be very low, effectively acting as a no-cloud model. No preference for a given cloud model in this case means that the presence of a cloud is not necessary to explain the observed spectrum. In this sense, the Bayesian evidence for all these models should be approximately equal, and the scatter in the Bayes factors in Figure~\ref{fig:cld0ev} shows the poor performance of the evidence approximations when the significance is low. A large scatter in the Bayesian evidence calculations by different methods has also been observed by \citet{Cornish:2007} when SNR$\lesssim 7$. When the support for a certain model is low, we also note a lack of correlation between the model significance and the SNR (e.g. green and blue lines in Figure~\ref{fig:cld0ev}). This shows that the retrieval results in such cases are dominated by the particular noise realization. The black symbols in Figure~\ref{fig:cld0ev} show the Bayes factors obtained using the evidence calculated by the nested sampling algorithm. The agreement is excellent for the high-significance methane detection, but lays within the large scatter for the cloud-model comparison.  

\begin {figure*}
\centering
\includegraphics*[scale=0.3,angle=0]{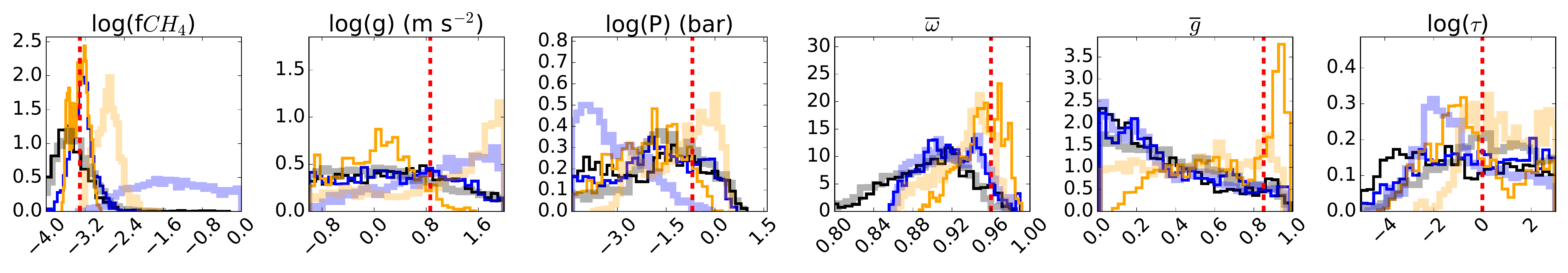}
\includegraphics*[scale=0.3,angle=0]{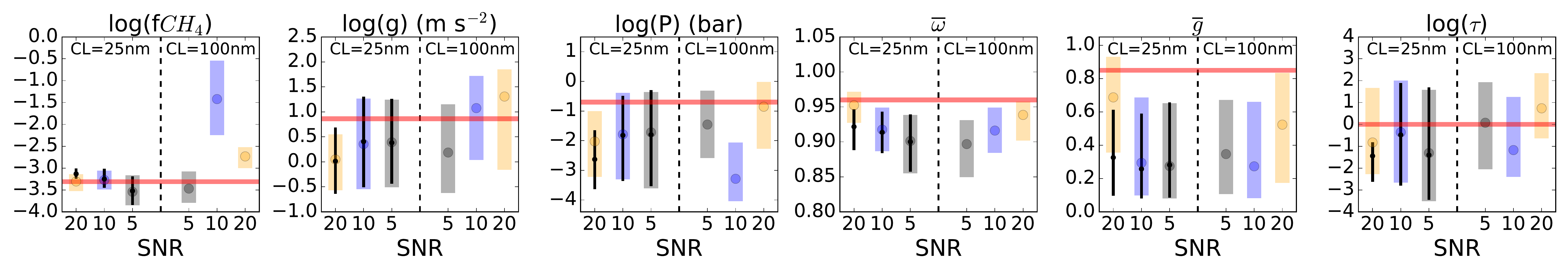}
\caption{Same as Figure~\ref{fig:c0fit}, for the 1-cloud case in Section~\ref{v1c}.  \label{fig:c1fit}}
\end{figure*}

\begin {figure}
\centering
\includegraphics*[scale=0.45,angle=0]{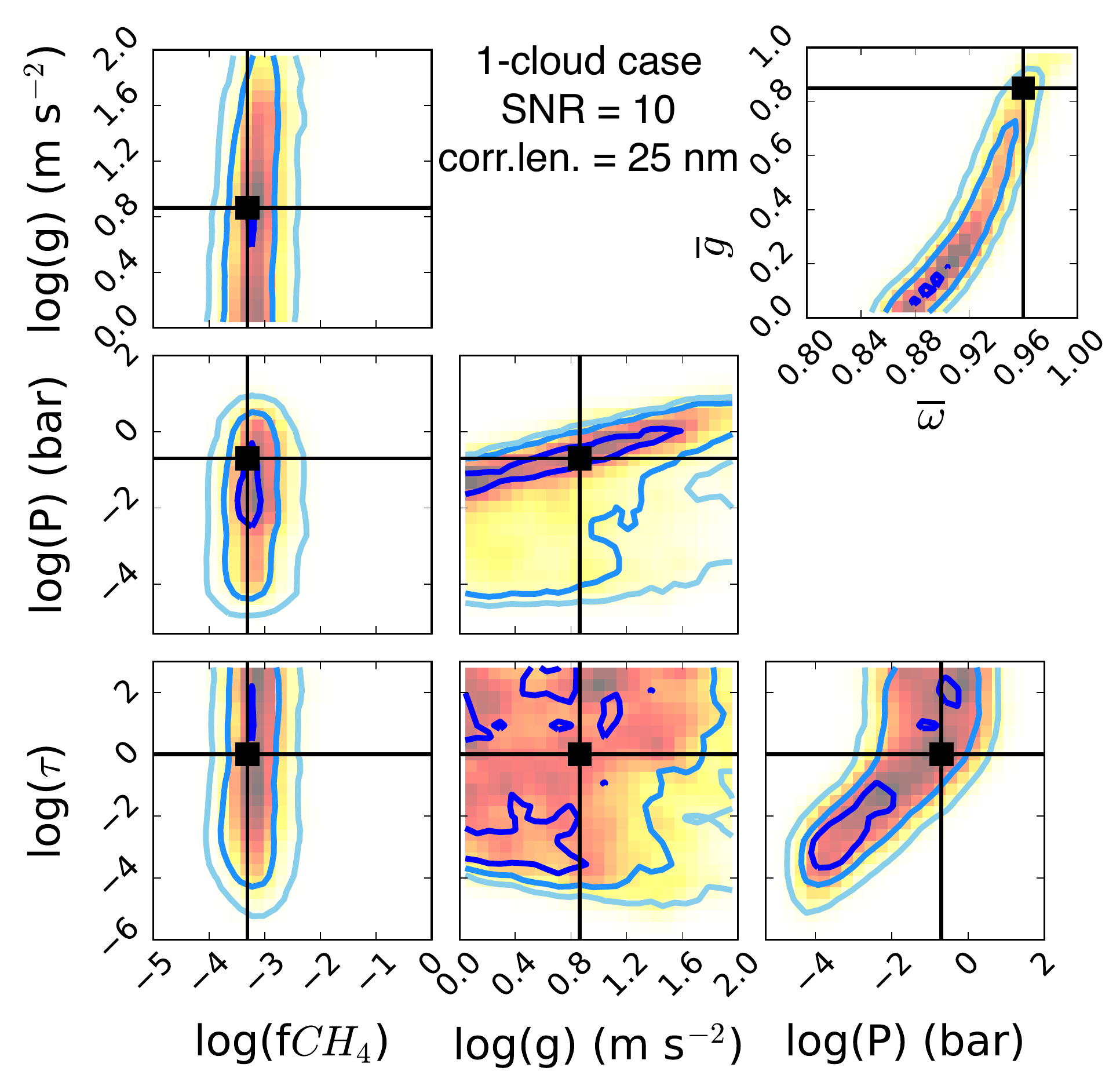}
\caption{Sample 2-D marginal posterior probability distributions for SNR=10 and spectral noise correlation length of 25 nm, for the single cloud case in Section~\ref{v1c}, using the {\bf 1c} forward model. The red color map corresponds to distributions obtained using the MCMC algorithm, and the blue contours to nested sampling. The black lines show the real solution. \label{fig:c1dd}}
\end{figure}

\begin {figure}
\centering
\includegraphics*[scale=0.38,angle=0]{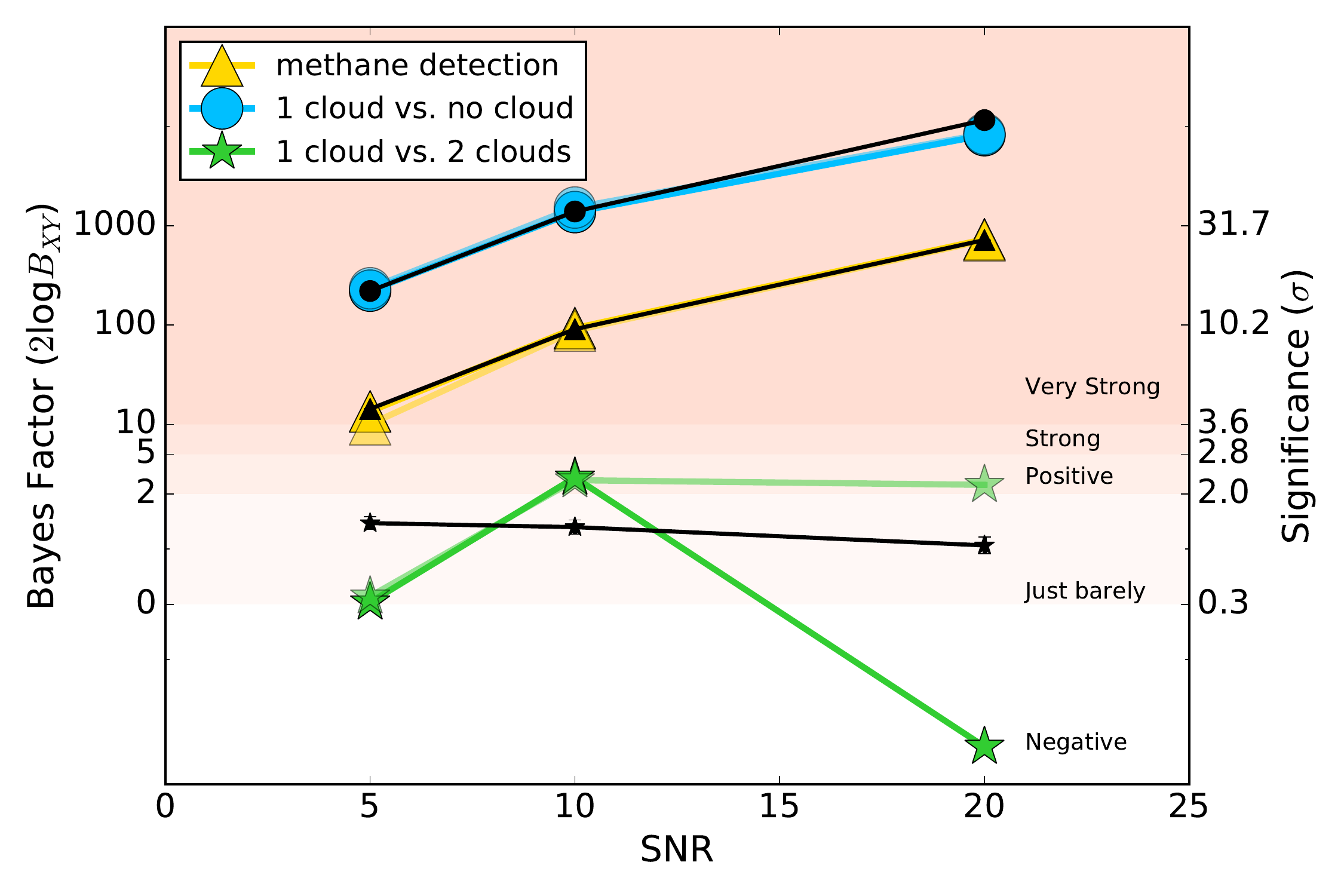}
\caption{Same as Figure~\ref{fig:cld0ev}, for the 1-cloud case in Section~\ref{v1c}. In this case, there is no ambiguity in model selection with a cloud clearly detected at $\sim 20\sigma$ significance even when the SNR of the input data is only 5. \label{fig:cld1ev}}
\end{figure}

\subsection{Single-cloud Case}
\label{v1c}

By raising the optical depth $\tau$ to 1, and the cloud top pressure to 0.2 bar, we can use the 1-cloud model to generate the albedo spectrum of a planet with an observable cloud deck. The simulated observations of such a planet are shown in the middle panel of Figure~\ref{fig:vspec}. The results of this retrieval are shown in Figures~\ref{fig:c1fit} and \ref{fig:c1dd}, and in the bottom half of Table~\ref{tab:vres1}. In this case the methane abundance is still well constrained, although within a wider range than for the no-cloud case, namely within a factor of $\sim 5$ for a SNR of 5 up to within a factor of $\sim 3$ for a SNR of 20. The original abundance value is well within the predicted ranges, where the SNR=10 case with a correlation length of 100~nm seems to be an outlier. 

The surface gravity of the planet is no longer constrained in this case, but is found instead to correlate with the cloud top pressure (Figure~\ref{fig:c1dd}). The power of the posterior sampling lays in discovering such correlations between model parameters. Figure~\ref{fig:c1dd} also shows the correlation between the cloud albedo $\bar{\omega}$ and scattering asymmetry factor $\bar{g}$, and between the top cloud pressure and its optical depth. Essentially, an optically thick cloud also constrains the cloud top pressure between $\sim 0.01$ and 1~bar, while an optically thin cloud would require the cloud top pressure to be very close to the top of the atmosphere. Independent constraints on the surface gravity, such as provided by RV measurements would narrow the allowed range for the cloud top pressure, which in turn would constrain the cloud optical depth. Lacking this information, we obtain a lower limit for the optical depth and an upper limit for the cloud top pressure.

\begin {figure*}
\centering
\includegraphics*[scale=0.3,angle=0]{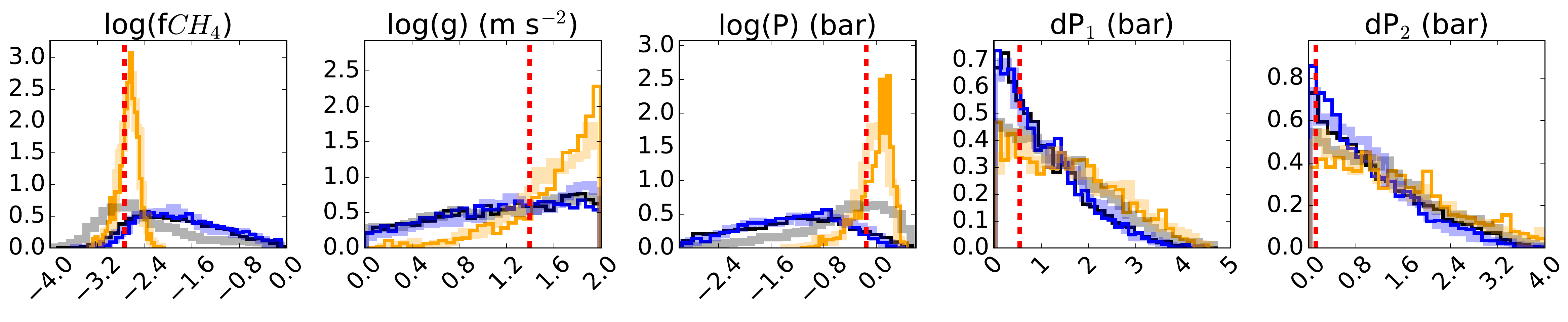}
\includegraphics*[scale=0.3,angle=0]{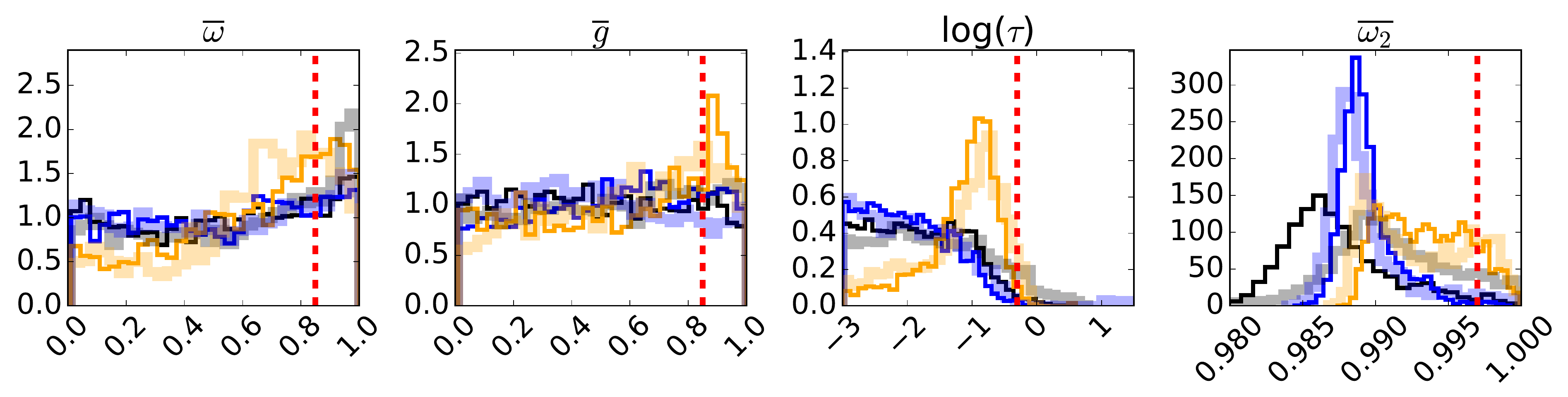}
\includegraphics*[scale=0.3,angle=0]{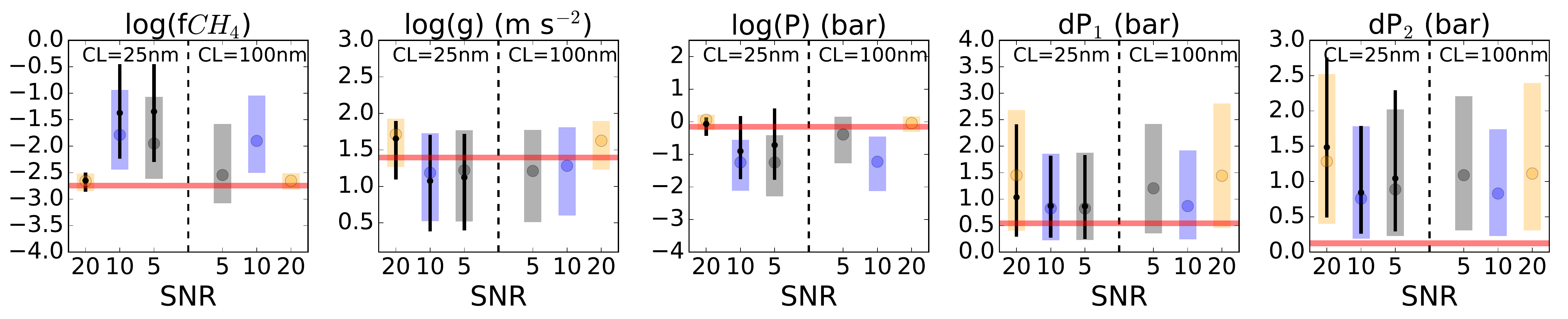}
\includegraphics*[scale=0.3,angle=0]{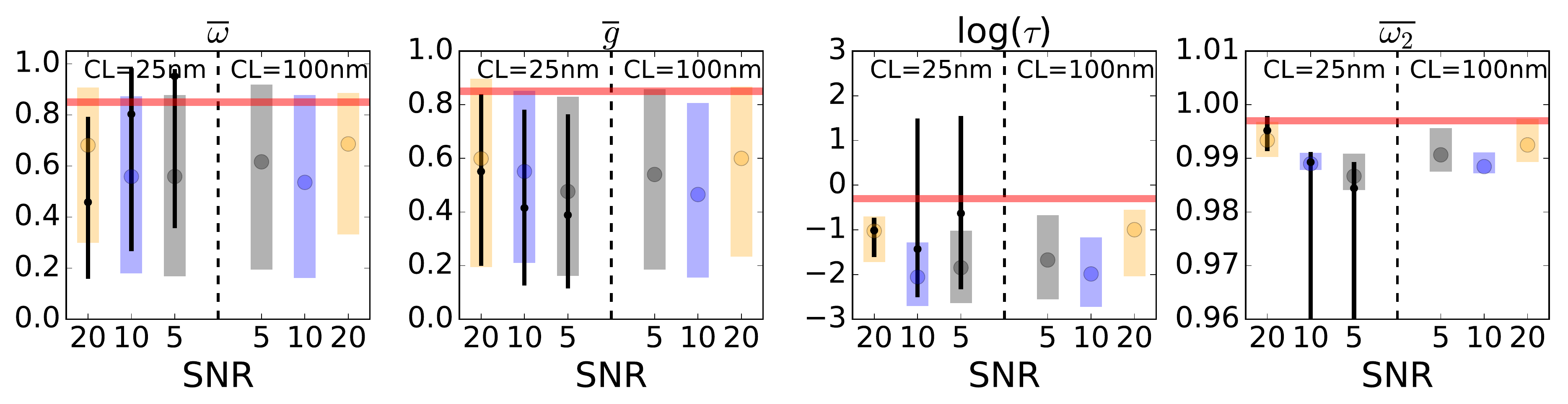}
\caption{Similar to Figure~\ref{fig:c0fit}, for the 2-cloud case in Section~\ref{v2c}. The parameters correspond the the 2-cloud model ({\bf 2c}) in Section~\ref{2cld}. The $1\sigma$ intervals obtained using nested sampling can be affected by possible bi-modal distributions (see also Figure~\ref{fig:c2dd}). \label{fig:c2fit}}
\end{figure*}

\begin {figure*}
\centering
\includegraphics*[scale=0.6,angle=0]{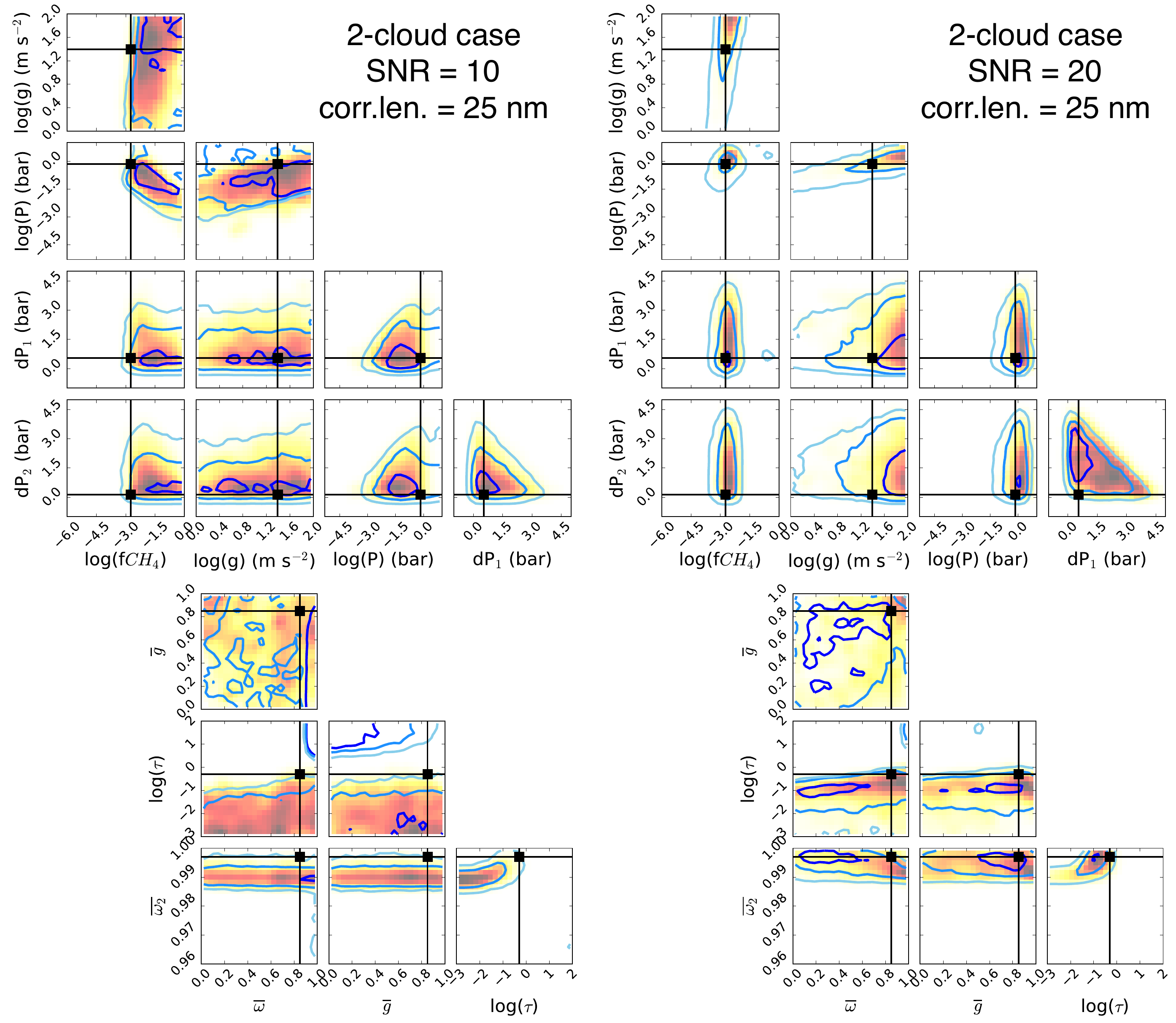}
\caption{Sample 2-D marginal posterior probability distributions for SNR=10 and 20, and spectral noise correlation length of 25 nm, for the 2-cloud case in Section~\ref{v2c}, using the {\bf 2c} forward model. The red color map corresponds to distributions obtained using the MCMC algorithm, and the blue contours to nested sampling. The black lines show the real solution.  \label{fig:c2dd}}
\end{figure*}

\begin {figure}
\centering
\includegraphics*[scale=0.38,angle=0]{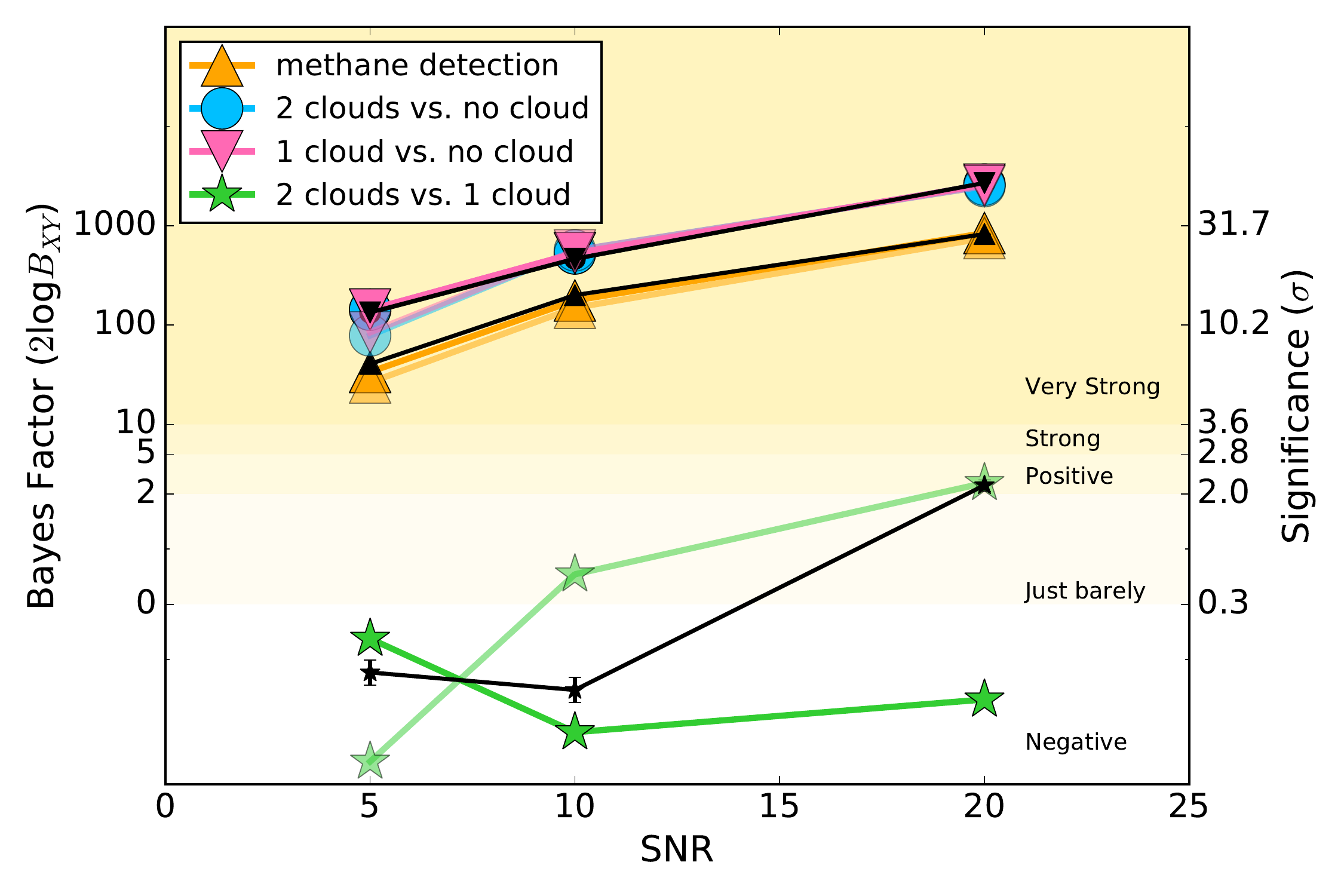}
\caption{Similar plot to Figure~\ref{fig:cld0ev}, for the 2-cloud case in Section~\ref{v2c}.  The color scheme has been modified to emphasize the case where a 2-cloud structure is assumed as default. The orange triangles correspond to the ratios $\mathcal{Z}_{2c}/\mathcal{Z}_{2c-m}$, the blue circles to $\mathcal{Z}_{2c}/\mathcal{Z}_{2c-c}$, the pink triangles to $\mathcal{Z}_{1c}/\mathcal{Z}_{1c-c}$, and the green stars to $\mathcal{Z}_{2c}/\mathcal{Z}_{1c}$. As in the previous examples, the methane and cloud are clearly detected even with a SNR=5 dataset. \label{fig:cld2ev}}
\end{figure}

The other very well constrained parameter is the cloud albedo $\bar{\omega}$. The confidence intervals on this parameter are only of the order $\pm 5$\% to 2\% depending on the SNR and particular noise realization. The correlation with the scattering asymmetry factor leads to a slight asymmetry in these confidence intervals, but the range of allowed values is still remarkably narrow. On the other hand, the scattering asymmetry factor $\bar{g}$ is virtually unconstrained. Similarly to the no-cloud case, there is excellent agreement between the MCMC and nested sampling results.

The high-significance cloud detection is revealed in the Bayes factor plot in Figure~\ref{fig:cld1ev}. The Bayesian evidence is calculated for the posterior distributions corresponding to the models {\bf 1c}, {\bf 1c-m}, {\bf 1c-c}, and {\bf 2c}. The Bayes factors favor the models with clouds relative to the ones without (blue circles), and the model with methane relative to the one without (yellow triangles). The cloud detection significance is $>10 \sigma$ even when the data have a SNR of 5, showing that the cloud deck is required by the observations. The methane detection significance is similar to that in Section~\ref{v0c}. Similarly, the retrieval cannot distinguish between a 1-cloud or a 2-cloud model (green stars), since a 2-cloud model can be reduced to a 1-cloud model as the gap between the 2 cloud decks becomes small and the optical depth of the top cloud becomes large.

\subsection{Two-cloud Case}
\label{v2c}

The final validation case consists of a spectrum generated using the 2-cloud model in Section~\ref{2cld}. The input parameters for the original spectrum are listed in Table~\ref{tab:vres2}, and the simulated datasets are shown in the right panel of Figure~\ref{fig:vspec}. The retrieved marginal probability distributions and confidence intervals are shown in Figure~\ref{fig:c2fit}. In this case, the uncertainty in the methane abundance does not shrink considerably before the SNR reaches a value of 20. The confidence interval for {\it fCH4} extends over a factor of $\sim 30$ ($\sim 60-70$ for nested sampling) when the SNR is 5-10, but drops to a factor of 2 when the SNR reaches 20. Similarly to the 1-cloud case, the surface gravity is not constrained by the data. The multi-dimensional correlation between {\it fCH4}, {\it P}, and {\it g} seen in Figure~\ref{fig:c2dd} (at SNR=10) shows the benefit in reducing the allowed range in {\it g}, via RV and astrometry measurements, which will then propagate into narrowing the allowed ranges in {\it P} and {\it fCH4}. For a SNR=20 dataset, the uncertainties in {\it fCH4} and {\it P} are simultaneously reduced (Figure~\ref{fig:c2dd}). In this case, the pressure at the top of the bottom cloud ($P$) is also constrained to within a factor of $\sim 3$. 

The scattering asymmetry factor $\bar{g}$ of the upper cloud and its albedo $\bar{\omega}$ are both completely unconstrained, while the uncertainty in the albedo of the lower cloud ($\bar{\omega_2}$) is only 1\% even when the data has a SNR of 5. The MCMC algorithm places an upper limit on the optical depth of the upper cloud, which is consistent with the lack of constraints for the other upper cloud parameters, but imposes a very tight constraint on the bottom cloud albedo. Intuitively, as seen in the previous two examples, the parameters of the upper cloud can be constrained as long as this cloud is optically thick, while the properties of the lower cloud (its albedo) can be determined as long as the upper cloud is optically thin. However, especially at lower SNR (see Figure~\ref{fig:c2dd}), the nested sampling algorithm identifies a second set of solutions, with an optically thick upper cloud, associated with a lower methane abundance and a deeper lower cloud. This result suggests that this degeneracy will not be broken unless the scatter in the data points is greatly reduced. Aside from this new mode identified by the nested sampling algorithm, the two Bayesian approaches are again in excellent agreement. The presence of the second mode can be further investigated by starting the MCMC chains in this part of the parameter space.

We have calculated the Bayes factors and compared the models {\bf 2c}, {\bf 1c}, {\bf 2c-c}, and {\bf 2c-m}. Similar to the 1-cloud case, methane and clouds are both detected at very high significance ($\sigma > 4$) even for a dataset with a SNR of 5, as shown in Figure~\ref{fig:cld2ev}. In this case we again cannot distinguish between a 1-cloud and a 2-cloud model, since the first is a special-case limit of the second (green stars). However, both the 1-cloud and the 2-cloud models are equally favored with respect to any cloud free model (blue circles, pink triangles).

\subsection{Importance of SNR and Spectral Noise Correlation Length}

We stress that the quoted significance of the detection itself has no other information on the confidence intervals associated with the model parameters. These confidence intervals, as well as possible correlation and multi-modality, are clearly affected by the SNR of the dataset. The change in the confidence intervals with SNR is shown in Figures~\ref{fig:c0fit}, \ref{fig:c1fit}, and \ref{fig:c2fit}. Overall, while the presence of methane is clearly {\it detected} even at a SNR of 5, its {\it abundance} is well constrained (to within factors of 2-3) only at a SNR of 20. At lower SNR, the uncertainty in the methane abundance  is mainly related to correlations with other models parameters, such as the surface gravity and the position of the cloud deck ({\it P}). This situation is improved in the case of a clear atmosphere, where the methane abundance and surface gravity are simultaneously constrained. However, the {\it presence} of a cloud deck is easy to confirm even at a SNR of 5 (as shown by the Bayes factor plots). This suggests that when the presence of clouds is indicated by early observations, an attempt to  further increase the SNR is justified in order to constrain the methane abundance.

Our results do not indicate any influence of the spectral noise correlation length on the retrieval results. The uncertainties on the model parameters are similar (see Figures~\ref{fig:c0fit}, \ref{fig:c1fit}, and \ref{fig:c2fit}, and Tables~\ref{tab:vres1} and \ref{tab:vres2}). There is a slight bias towards higher values for the retrieved methane abundance in the no-cloud and 1-cloud cases for a spectral noise correlation length of 100~nm, but it is not clear whether this is an effect of the noise correlation length scale or of the particular noise realization in the simulated dataset. Multiple noise realizations for a given correlation length scale would be required to validate this effect.

\begin {figure}
\centering
\hspace*{-0.4cm}\includegraphics*[scale=0.4,angle=0,trim={10 0 0 0},clip]{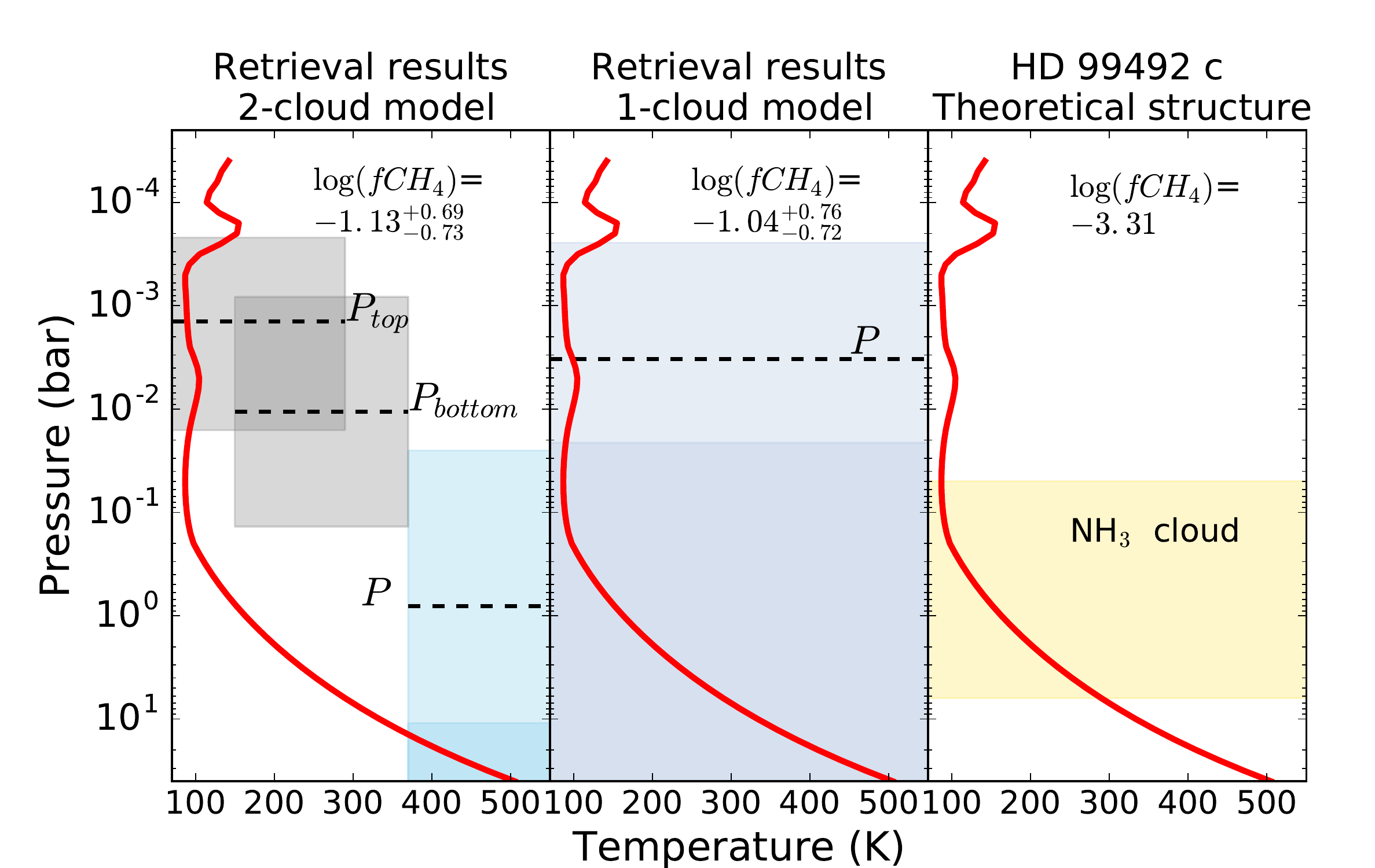}
\caption{Cloud structure for gas giant HD 99492 c, as retrieved using the 2-cloud model (left), and the 1-cloud model (right). The semi-transparent regions are associated with the error bars for the cloud top (bottom) pressures, and the labeling follows the convention in Figure~\ref{fig:cld}. In the left panel, the positions of the cloud layers have been offset for clarity, with the gray regions overlapping to emphasize the fact the both $P_{top}$ and $P_{bottom}$ refer to the same cloud deck, while the blue regions correspond to the second cloud deck defined in Figure~\ref{fig:cld}. The theoretical structure is shown in the right panel, with the region occupied by the cloud calculated using the radiative-convective equilibrium code. The pressure-temperature profile calculated by this code and kept fixed in the retrievals is shown in red in all three panels. The theoretical and retrieved CH$_4$ abundance is shown at the top. \label{fig:hd9cld}}
\end{figure}

\section{REALISTIC TEST CASES}
\label{appl}

For the retrieval tests we used two types of input data, Solar System giants and model planets. We used the Solar System albedo spectra for Jupiter and Saturn from  \citet{Karkoschka:1994}, and a theoretical radiative-convective equilibrium model for HD 99492 c. All of these objects have methane dominated optical reflection spectra. We have applied our albedo retrieval method to a set of 24 cases, comprising 6 combinations of SNR (5, 10, 20) and correlation lengths (25 and 100 nm), the same as for the validation cases. The Solar System-like planets are assumed to be at 25 pc from the Earth, while the distance to the HD 99492 c system is 18 pc. The retrievals use data between 0.6 and 1~$\mu$m to more closely match the projected bandpass of {\it WFIRST} (unlike the validation cases where we used the 0.4-1.0~$\mu$m bandpass). For each case we run the MCMC ensemble sampler with 24 walkers (see Appendix) per parameter, for a total of 3800 steps, and we select the last 400 steps for determining the posterior probability distributions. We also use the nested sampling algorithm for the spectra with noise correlation length of 25~nm. 

\subsection{HD 99492 c}
\label{shd9}

We start by looking at the model planet HD 99492 c, as the real-world example most closely resembling our 1-cloud model. HD 99492 c is thought to be a gas giant with a mass of $0.36\pm0.02$~M$_{Jup}$, and a semimajor axis of $5.4\pm0.1$~AU, orbiting a K2V star. However, its existence has been challenged recently due to high stellar activity \citep{Kane:2016}.

\begin {figure*}
\centering
\includegraphics*[scale=0.3,angle=0]{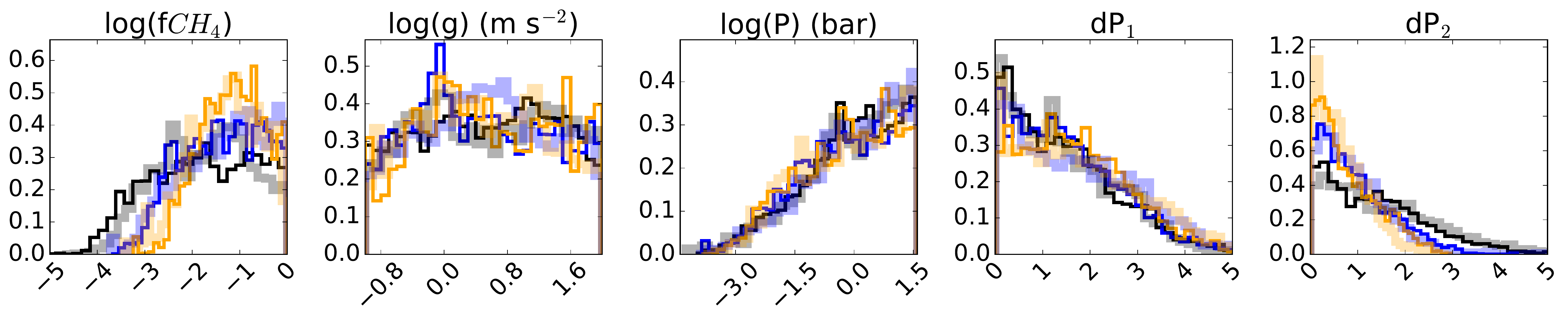}
\includegraphics*[scale=0.3,angle=0]{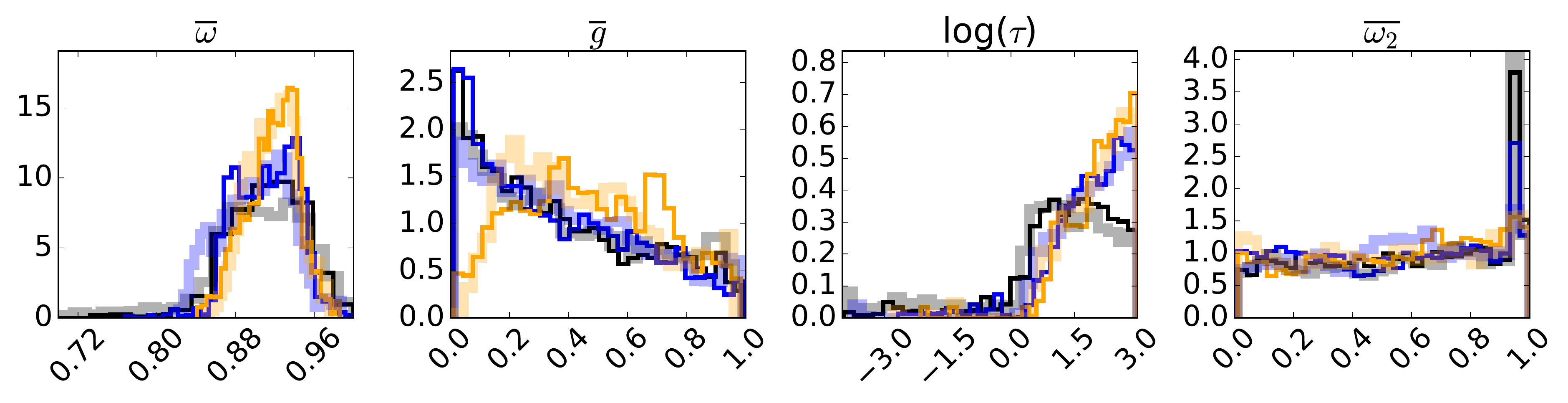}
\includegraphics*[scale=0.3,angle=0]{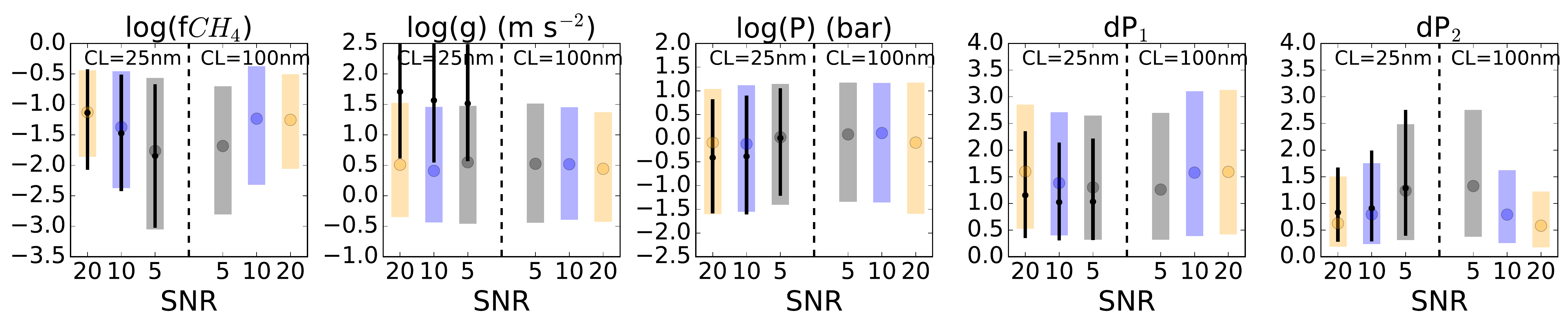}
\includegraphics*[scale=0.3,angle=0]{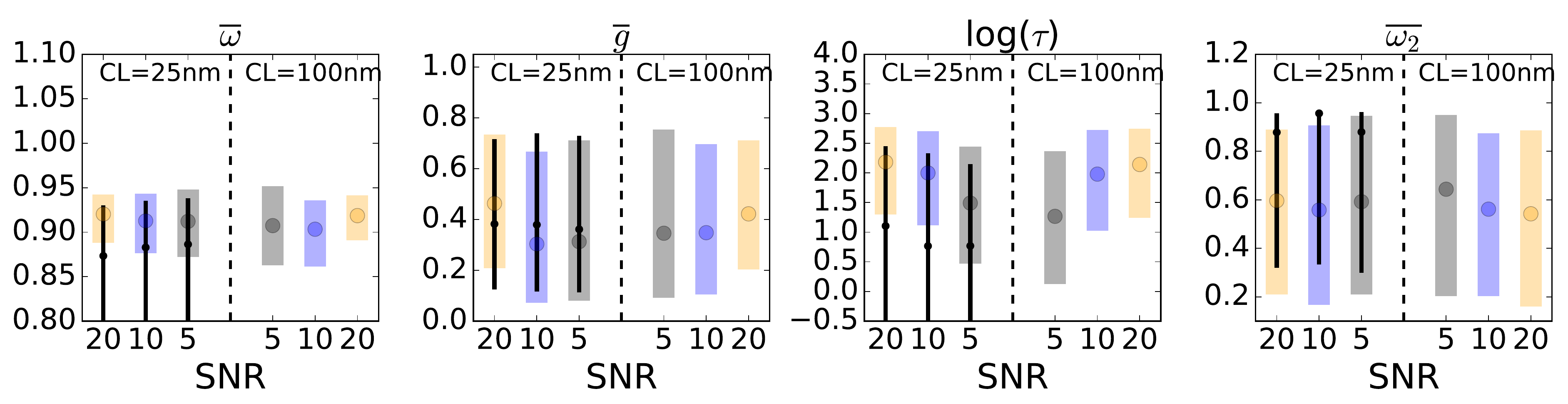}
\caption{Same as Figure~\ref{fig:c2fit}, for the HD 99492 c model in Section~\ref{shd9}. In a realistic scenario, the ``true" parameters values would not be known, and therefore are not shown. \label{fig:hd9fit}}
\end{figure*}

\begin {figure*}
\centering
\includegraphics*[scale=0.6,angle=0]{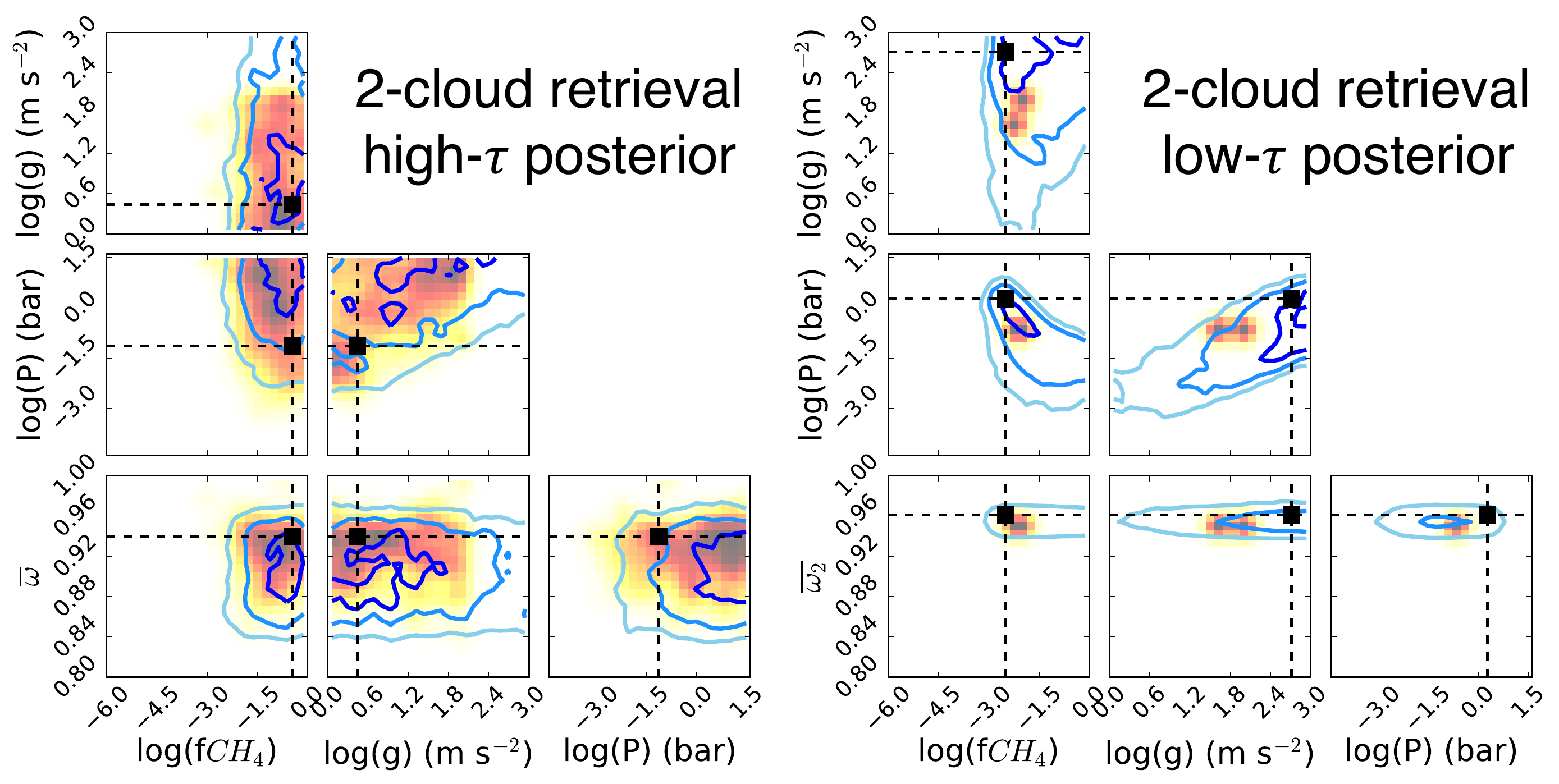}
\caption{ 2-D marginal posterior distributions for HD 99492 c (SNR=20, CL=25~nm), using a 2-cloud model. The full posterior is bi-modal, with a second, low optical depth mode better identified by the nested sampling algorithm (blue contours). For clarity, we plot the two modes separately, the high optical depth on the left, and the low optical depth on the right. The black dashed lines mark the position of the {\it best fit solution} for each mode. \label{fig:hd9tri}}
\end{figure*}

\begin {figure*}
\centering
\includegraphics*[scale=0.6,angle=0]{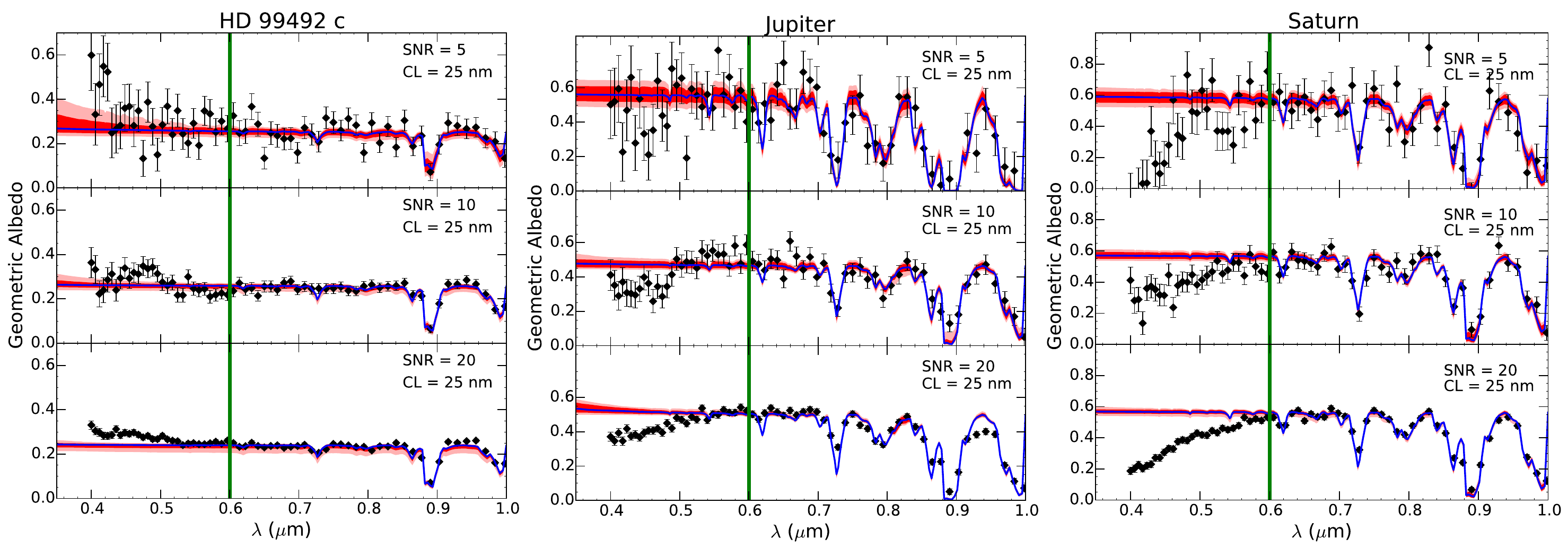}
\caption{Simulated data and best fit spectra for HD 99492 c (left), Jupiter (middle), Saturn (right), using the {\bf 2c} forward model. The data correspond to SNR=5, 10, 20, from top to bottom and a spectral correlation noise of 25 nm. Same conventions as in Figure~\ref{fig:vspec}. The retrieval was performed over the $0.6-1.0$ $\mu$m region, as indicated by the green vertical line. \label{fig:tspec}}
\end{figure*}

 We first determined the pressure-temperature profile for HD 99492 c by computing a 1D radiative-convective equilibrium model following the methods of \citep{Cahoy:2010} while accounting for clouds with the treatment of \citep{Ackerman:2001}. This code computes a self-consistent cloud with vertically varying abundances and particle sizes of each condensible species. This theoretical structure is shown in the right-hand panel in Figure~\ref{fig:hd9cld}. We then input the resulting pressure-temperature profile into a fine-grid albedo code to produce an albedo spectrum comparable to the Solar System data. This high resolution spectrum is then converted to simulated data following the prescription in Section~\ref{noise}, for each chosen combination of SNR and noise correlation length. 
 
Figure \ref{fig:hd9fit} shows the summary of the retrieval results for the gas giant HD 99492 c, with the quantiles listed in Table~\ref{tab:hd9}. An example for the posterior probability distributions for the retrieval using the 2-cloud model is shown in Figure~\ref{fig:hd9tri}. In the 2-cloud scenario, the posterior is bimodal, similar to that found in Section~\ref{v2c}, and we show the most important parameters for the two modes separately in the two panels. The notable difference is that for the mode with a {\it low} optical depth for the top cloud ($\tau$), the albedo of the bottom cloud ($\bar{\omega}_2$) is very well constrained, while for the mode with a {\it high} optical depth for the top cloud, the albedo of the top cloud ($\bar{\omega}$) is very well-constrained, to within $\sim 6\%$. This is easily understood, since in the case of low optical depth we can ``see through" the top cloud, and the albedo of the bottom cloud surface is what determines the spectrum, while the opposite is true when the top cloud is optically thick. 

\begin {figure*}[ht]
\centering
\includegraphics*[scale=0.5,angle=0]{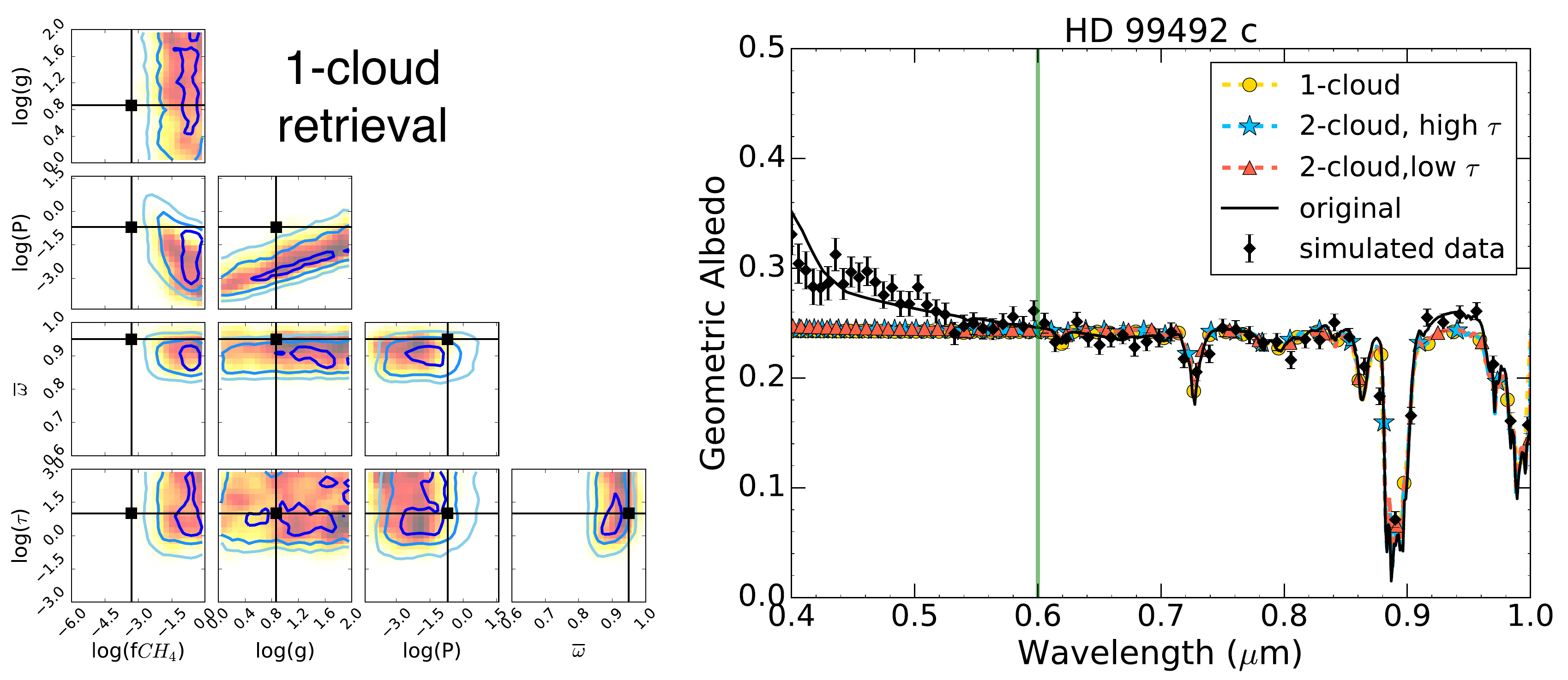}
\caption{Best-fit spectra and 2-D marginal posterior distributions for HD 99492 c (SNR=20, CL=25~nm), using a 1-cloud model.  The 2-cloud best fit parameters for the two modes are indicated in green in Figure~\ref{fig:hd9tri}. The black lines on the left plot show the 1-cloud parameter values that best match the ``theoretical model" on the right panel in Figure~\ref{fig:hd9cld}. \label{fig:hd9trispec}}
\end{figure*}

\begin {figure}
\centering
\hspace*{-0.3cm}\includegraphics*[scale=0.7,angle=0]{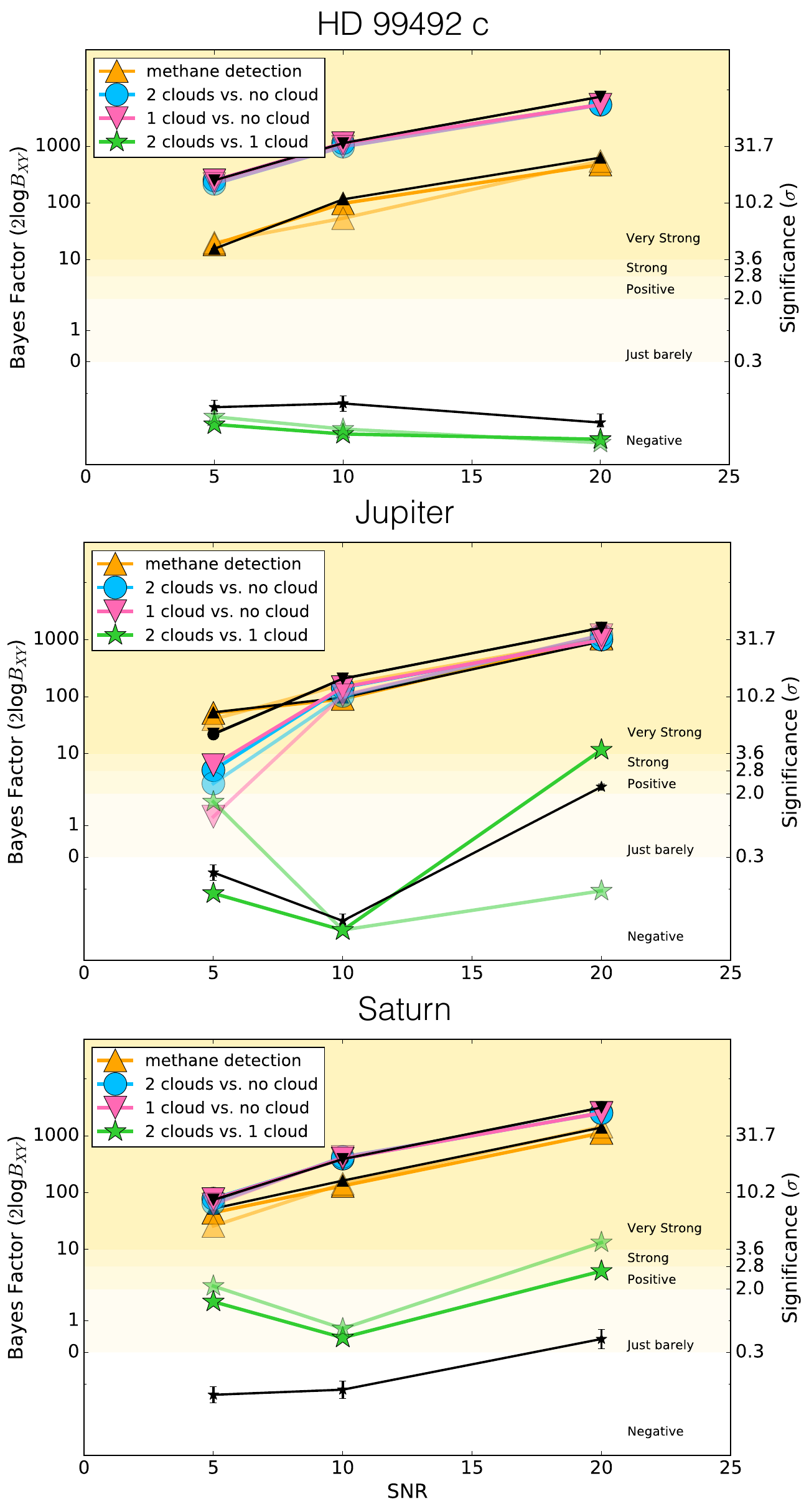}
\caption{Same as Figure~\ref{fig:cld2ev}, for the applications in Section~\ref{appl}.  The plots correspond to HD 99492 c, Jupiter, and Saturn, from top to bottom. As in the previous examples, the methane and cloud are clearly detected even with a SNR=5 dataset. \label{fig:tev}}
\end{figure}

\begin {figure*}
\centering
\includegraphics*[scale=0.3,angle=0]{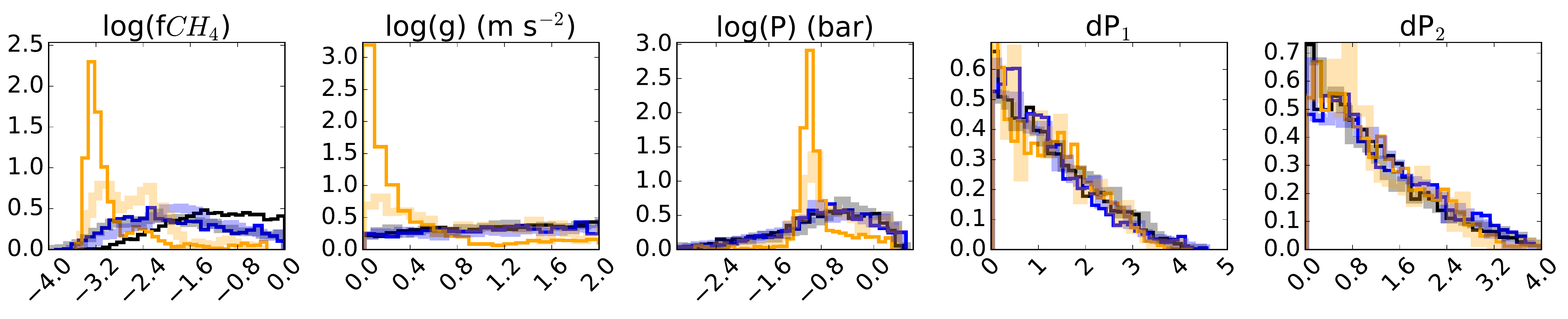}
\includegraphics*[scale=0.3,angle=0]{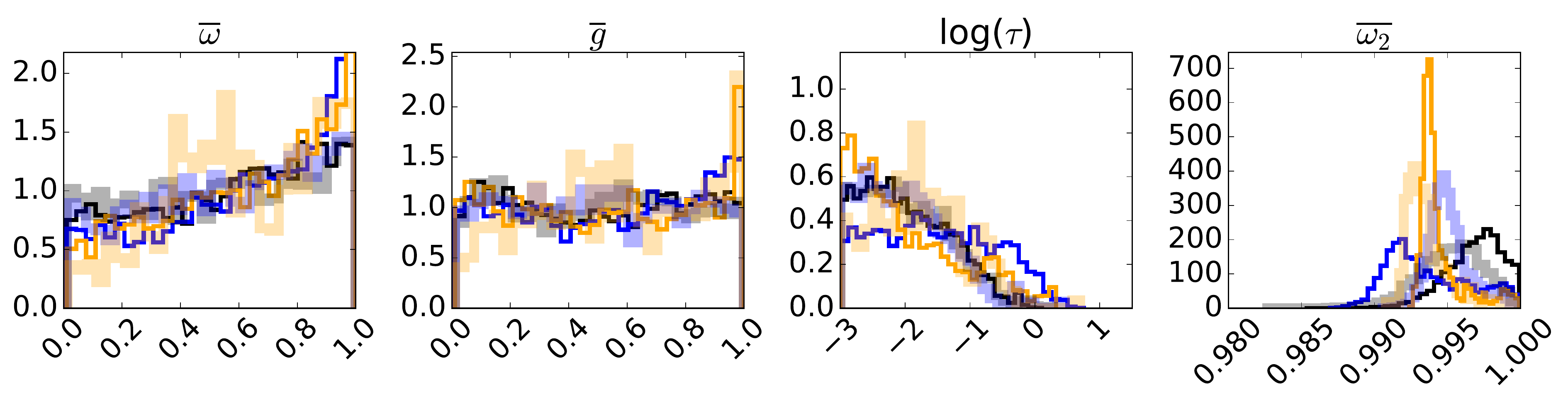}
\includegraphics*[scale=0.3,angle=0]{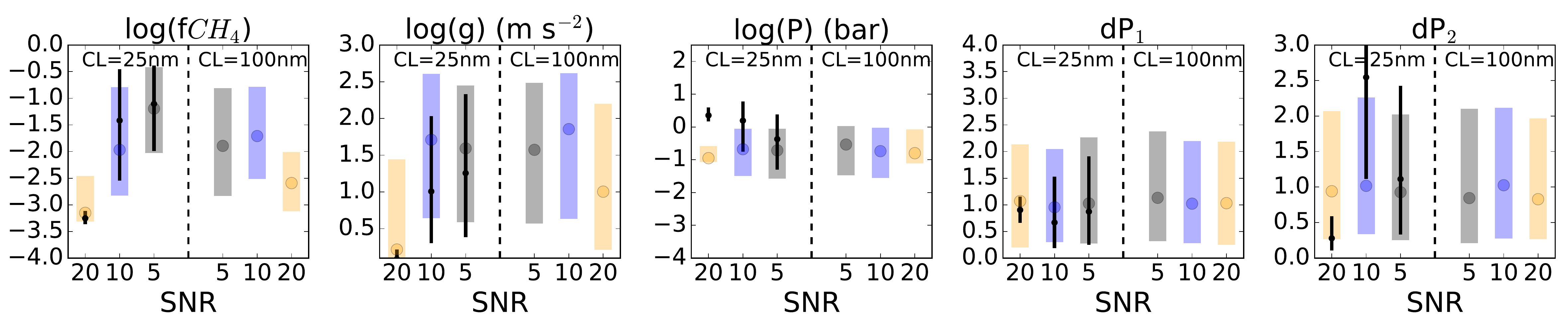}
\includegraphics*[scale=0.3,angle=0]{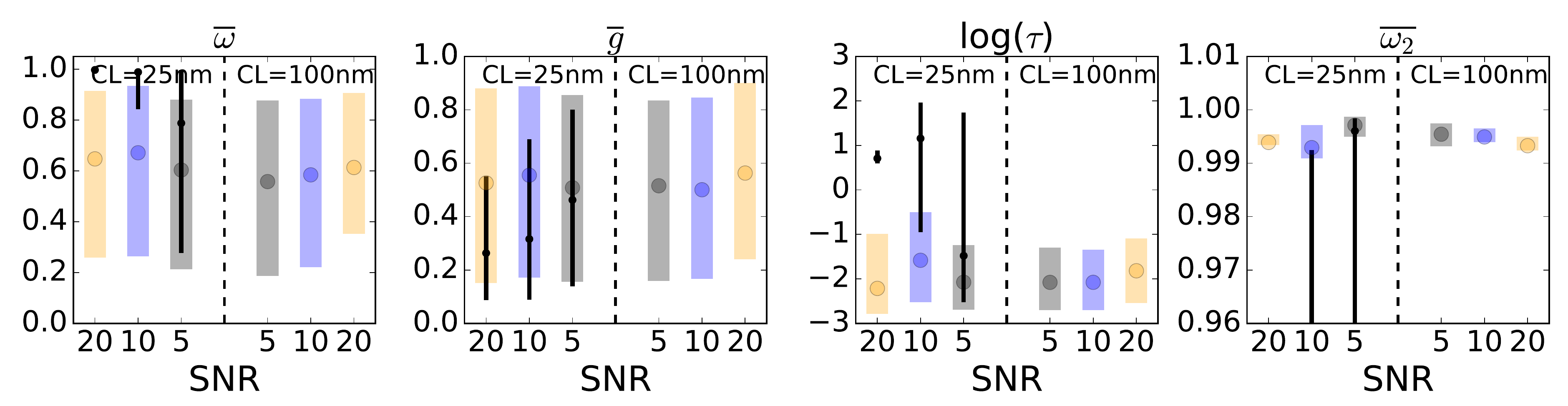}
\caption{Same as Figure~\ref{fig:hd9fit}, for the Jupiter albedo in Section~\ref{sjup}. \label{fig:jupfit}}
\end{figure*}

We also note that an optically thin top cloud favors a lower methane abundance, since now we integrate through the cloud, down to the bottom cloud, and thus see a greater column of atmosphere which can have a lower fractional CH$_4$ abundance. The position of the {\it best fit} parameter values for each mode was marked in green to emphasize that the best fit parameter {\it combination} is different from the set of median values of the marginal distributions, which are listed in Table~\ref{tab:hd9}. The range of spectra generated using random parameter sets from the posterior are shown in Figure~\ref{fig:tspec}.

In Figure~\ref{fig:hd9trispec} we show both the covariance plot for the retrieval using the 1-cloud model, as the more representative for the planet's vertical structure, and the best-fit spectra for the different models and modes. In the covariance plot the black lines show the parameter values that are closest to the theoretical planet structure. We note that this 1-cloud retrieval solution resembles the high-$\tau$ mode of the 2-cloud posterior, only with a tighter correlation between $P$ and $g$. In this case we find a lower bound for the pressure of the cloud surface, but a lack of constraints for $g$. Similar to the validation case, we can see that a tighter prior in $g$ would translate into better limits on $P$ (via correlation), and a narrower allowed range for {\it fCH4}. The best-fit spectra reveal the complete degeneracy of these solutions (red, blue and yellow lines overlapping). The differences between the retrieved and original spectra (black line) are due to a more comprehensive treatment of gas and cloud opacities in the original model. Additional constraints placed by available photometric points shortward of 0.6~$\mu$m will be investigated in future work.

The degeneracy between the best-fit solution given by the 2-cloud and 1-cloud models is also apparent in Figure~\ref{fig:hd9cld}, where the two cloud decks in the left panel overlap, within the error bars, and basically occupy the same vertical regions as the 1-cloud deck in the middle panel. This plot suggests that for a planet like HD 99492 c our simple cloud model can only provide a lower bound on the pressure at the top of the cloud deck (i.e. upper bound to the height above the surface) and a lower bound on the methane abundance (i.e. the methane abundance is inversely correlated to the cloud top pressure, such that the total CH$_4$ column is constant). Independent priors on the top cloud pressure (from equilibrium structure) and surface gravity (from radius and mass measurements) would help mitigate these uncertainties. 

Both Figure~\ref{fig:hd9cld} and \ref{fig:hd9trispec} show a retrieved CH$_4$ abundance that is significantly higher than the one used in the theoretical model. This is in contrast to the 1-cloud validation case, where the constraints on {\it fCH4} are much closer to the real value. This difference may be due to the fact that the forward model spectrum exhibits relatively few CH$_4$ bands compared to the previous test cases, with not enough constraints on continuum level, which sets the cloud top, methane absorption and atmospheric scale height determined by gravity. The cloud treatment in the inverse modeling is also very simplified. While the full theoretical model for HD 99492 c does include cloud optical depth variations with wavelength and depth in the atmosphere, these are not taken into account by the forward model in the retrieval. We note a similar bias toward high {\it fCH4} values in the case of Saturn below, which could be due to similar deficiencies in our simplified cloud model and will be investigated in future work.

As before, we show the Bayes factors between different model choices in the top panel of Figure~\ref{fig:tev}. The presence of methane and a cloud deck is confirmed at very high significance. The 2-cloud model is more disfavored relative to the 1-cloud model, likely due to the presence of additional unnecessary parameters. 

\begin {figure*}
\centering
\includegraphics*[scale=0.6,angle=0]{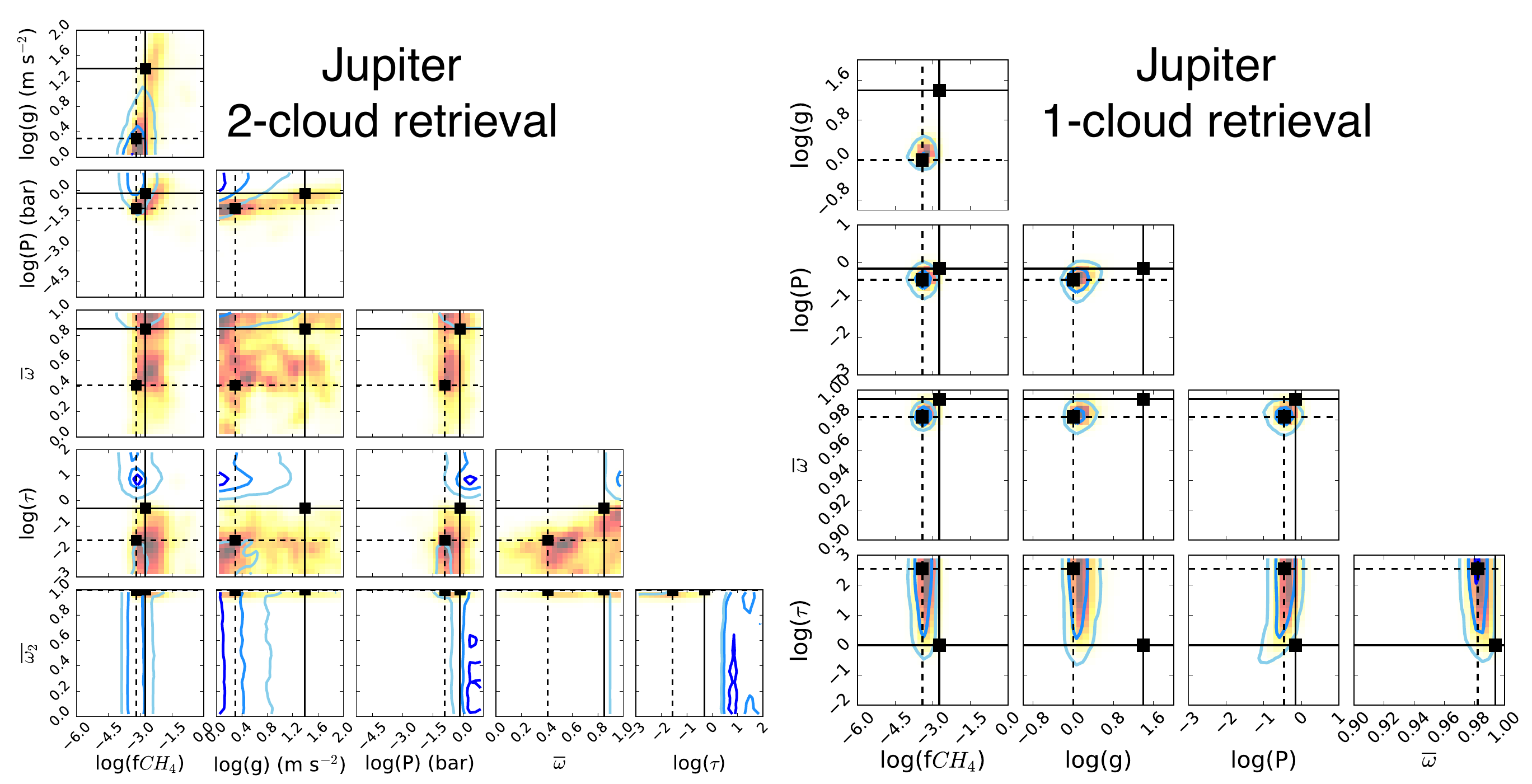}
\caption{ 2-D marginal posterior distributions for Jupiter (SNR=20, CL=25~nm), using a 2-cloud model (left) and a 1-cloud model (right). For the 2-cloud retrieval, the two posterior sampling methods lock onto different modes, one with low optical depth (MCMC, red colormap), and the other with high optical depth (nested sampling, blue contours). The best-fit solutions for both samplers, as well as for the 1-cloud model, are shown in Figure~\ref{fig:jupspec}. The dashed black lines show the {\it best fit} values, while the solid ones show the parameter values that best match the ``theoretical structure" of Jupiter shown in Figure~\ref{fig:jupcld}: $g=24.79$~m~s$^{-2}$, $f_{CH_4}=1.8\times 10^{-3}$, and $P=0.7$~bars. \label{fig:juptri}}
\end{figure*}

\subsection{Jupiter}
\label{sjup}

Arguably, a Jupiter-like planet is the closest real-world case to our 2-cloud forward model. We have simulated data for a Jupiter-like planet at 25 pc from the Sun using the observed Jupiter spectrum from \citet{Karkoschka:1994}. The results of our retrievals are shown in Figure~\ref{fig:jupfit}. This plot shows that the parameters that are best constrained by the data are {\it fCH4}, $P$, and $\bar{\omega}_2$. We note the narrowing of the distributions and therefore the tightening of the constraints for SNR=20 (orange lines), also shown by the size of the confidence intervals in the bottom plot. The derived CH$_4$ abundance is consistent with the generally adopted value of $(2.37\pm 0.57)\times10^{-3}$ (or -2.625 in log) in Jupiter \citep{Wang:2004}. However, the best constraint is only obtained at SNR=20 in our examples (see also Section~\ref{v2c}), suggesting that future observations should aim to achieve this SNR level. Also, the derived single scattering albedo of the lower cloud, $\bar{\omega}_2$, matches the observed value of 0.997 \citep[e.g., ][]{Sato:1979}. The mean values of these parameters are sensitive to the particular noise realization of each simulated dataset. Unconstrained parameters are $g$ and $\bar{g}$, and an upper limit is derived for $\tau$, showing that the upper cloud is likely optically thin, again consistent with Jupiter's observed stratospheric haze properties. The confidence intervals are summarized in Table~\ref{tab:jup}, and the range in spectra allowed by the posterior samples are shown in Figure~\ref{fig:tspec}.

\begin {figure}
\centering
\hspace*{-0.4cm}\includegraphics*[scale=0.35,angle=0]{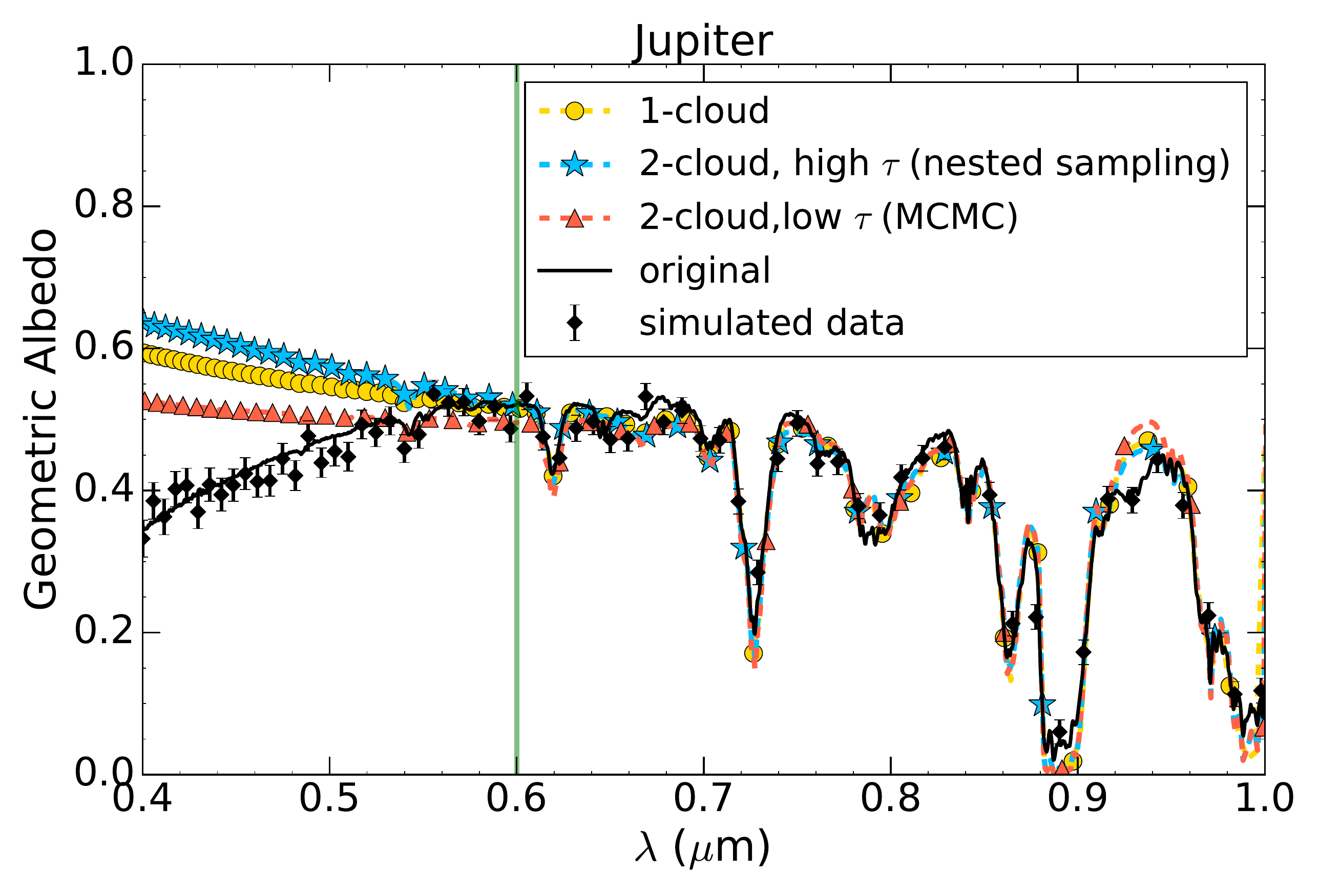}
\caption{Best-fit spectra for Jupiter (SNR=20, CL=25~nm), retrieved using the 2-cloud and 1-cloud models. The legend indicates that the low optical depth fit is favored by the MCMC method, while the high optical depth fit is favored by nested sampling (see also Figure~\ref{fig:juptri}). The vertical green line indicates that the retrieval is performed only on data between 0.6 and 1~$\mu$m.   \label{fig:jupspec}}
\end{figure}

\begin {figure}
\centering
\hspace*{-0.4cm}\includegraphics*[scale=0.4,angle=0,trim={10 0 0 0},clip]{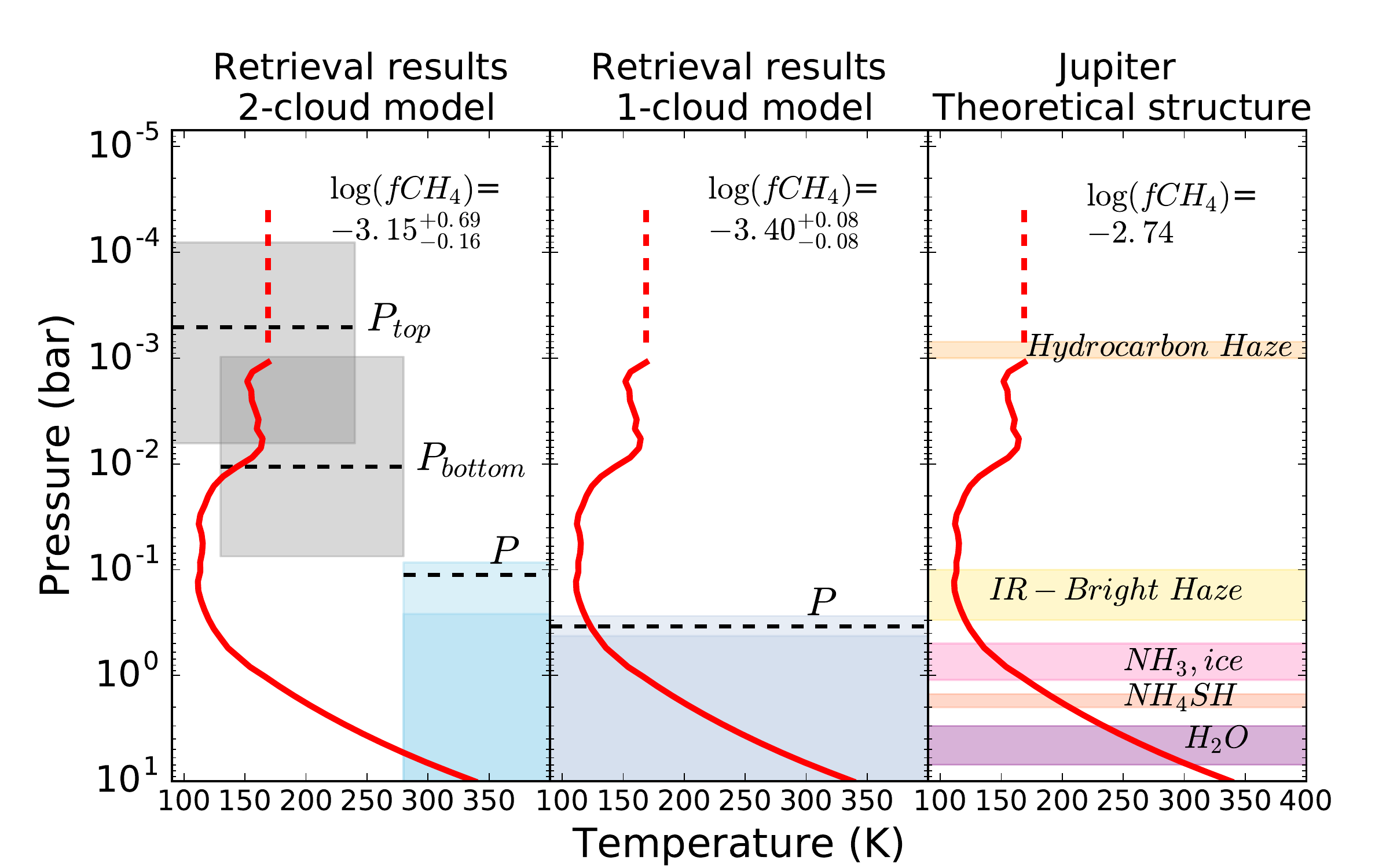}
\caption{Cloud structure for Jupiter, as retrieved using the 2-cloud model (left), and the 1-cloud model (right). The conventions are described in the Figure~\ref{fig:hd9cld} caption. The theoretical structure is shown in the right panel, with the cloud structure closely resembling available literature \citep[e.g.,][]{Simon-Miller:2001,Sato:2013}. The pressure-temperature profile is approximated as purely radiative in the top layers of the atmosphere (dashed red line).\label{fig:jupcld}}
\end{figure}

Although the MCMC algorithm strongly favors a single-mode posterior with an optically thin upper cloud, the nested sampling algorithm identifies two posterior modes, the second one having an optically thick upper cloud. This is reflected by the large confidence intervals shown in Figure~\ref{fig:jupfit} (black). The second, high optical depth mode, becomes favored by the nested sampling algorithm at SNR=20. Figure~\ref{fig:juptri} shows posterior covariance plots for some selected parameters for SNR=20, and noise correlation length 25~nm Jupiter data, using both the 2-cloud and 1-cloud models. The black solid lines indicate the parameter values that correspond to currently adopted values for Jupiter ($fCH_4=2.37\times10^{-3}$ and $g=24.79$~m~s$^{-2}$), while the dashed black lines show the best fit parameter values retrieved using the MCMC algorithm. The retrieved values for {\it fCH4}, top cloud pressure ($P$) and cloud albedo are close to the observed values. The constraints on {\it fCH4} and $P$ can be made even tighter by imposing better priors on surface gravity, following the correlation lines. The spectrum is not sensitive enough to the other model parameters, as shown by the large confidence regions. Therefore our initial guess or theoretical structure can lie far from the final best fit value.

It is apparent that the nested sampling (blue contours) favors a solution that resembles the 1-cloud model, with a deep, optically thick cloud and unphysically low gravity ($\sim 1$~m~s$^2$). Such low gravity solutions are also identified using the 2-cloud model. However, the 2-cloud model is still consistent with more realistic values of $g$, while the 1-cloud model is not. Such arguments can be used to favor one model over the other in the absence of quantitative Bayesian evidence. The correlations at the top of left panel in Figure~\ref{fig:juptri} show that a narrower allowed range in $g$ for known RV planets both constrain the methane abundance to match the real value and strongly disfavor the second, optically thick mode. The spectra corresponding to these best-fit solutions are shown in Figure~\ref{fig:jupspec}. This plot shows that the spectra are degenerate relative to these solutions at wavelengths between 0.6 and 1~$\mu$m, but physical arguments can be used to eliminate certain solutions. We note the need for wavelength-dependent continuum opacity, especially for using photometry data shortward of 0.6~$\mu$m. 

\begin {figure*}
\centering
\includegraphics*[scale=0.3,angle=0]{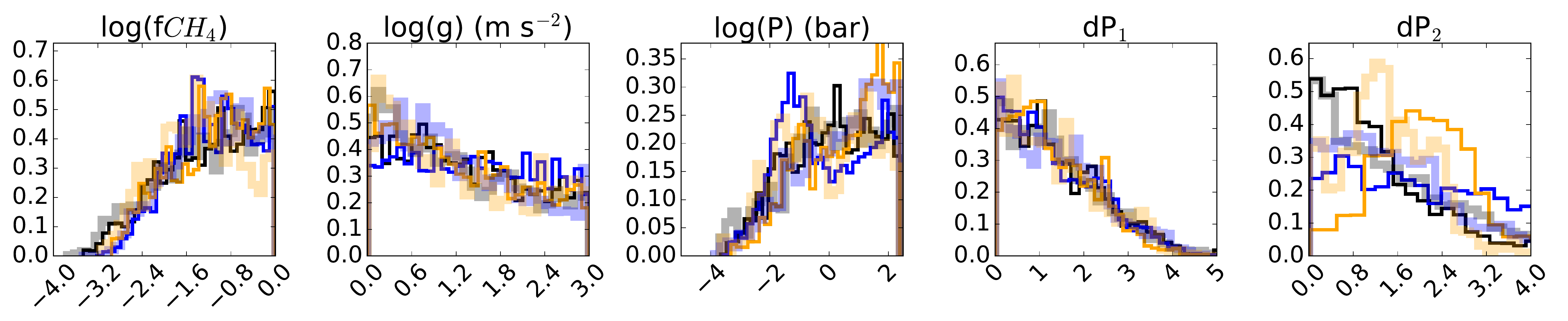}
\includegraphics*[scale=0.3,angle=0]{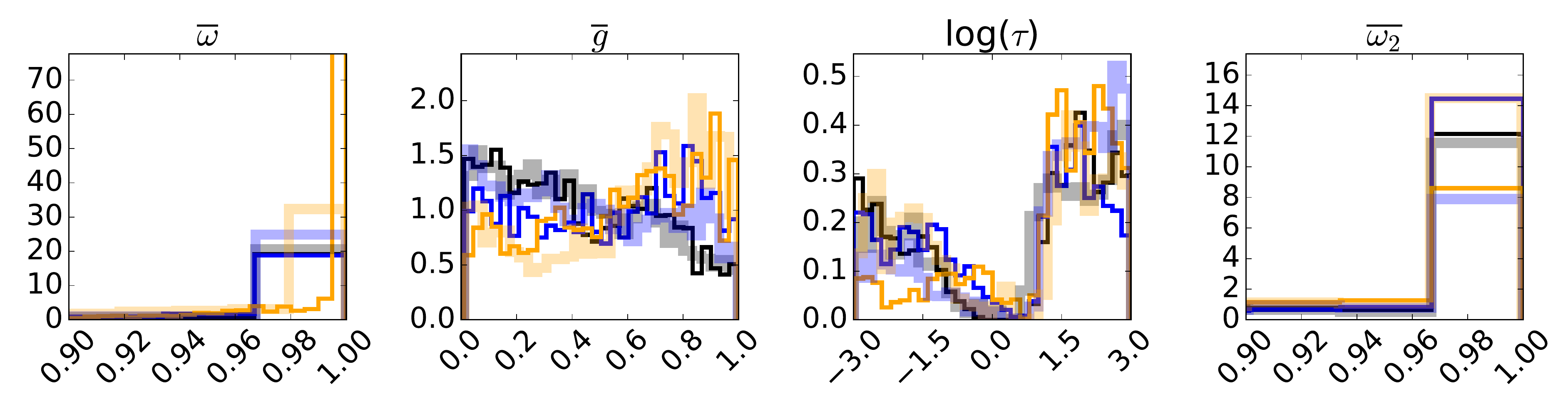}
\includegraphics*[scale=0.3,angle=0]{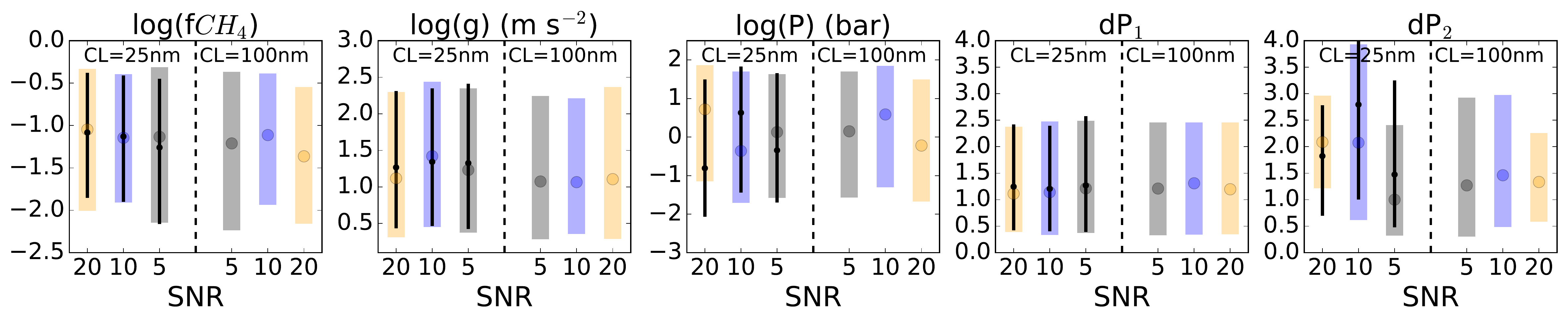}
\includegraphics*[scale=0.3,angle=0]{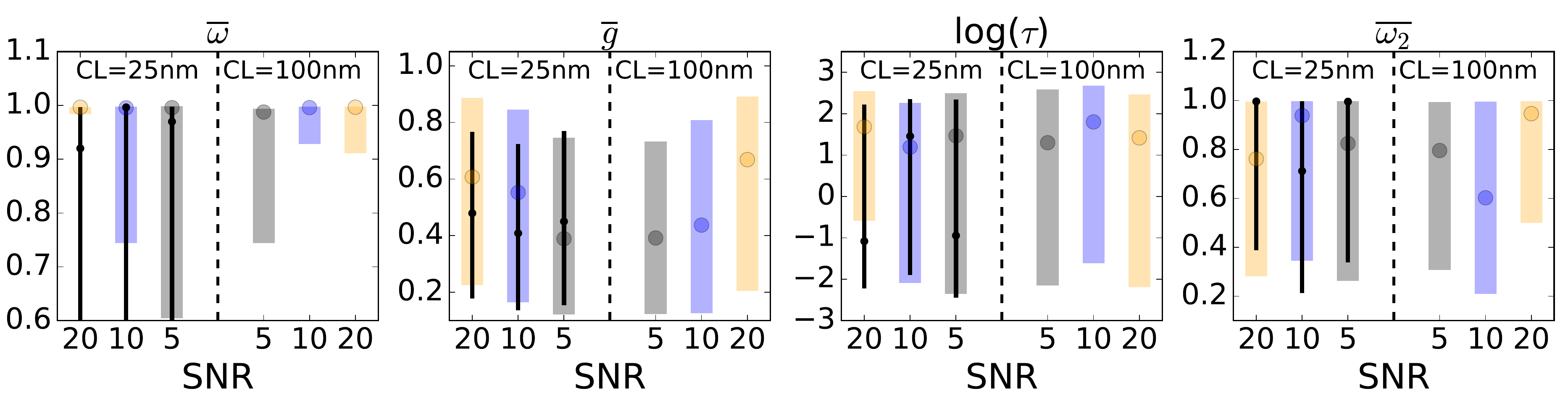}
\caption{Same as Figure~\ref{fig:hd9fit}, for the Saturn albedo in Section~\ref{ssat}. \label{fig:satfit}}
\end{figure*}

\begin {figure*}
\centering
\includegraphics*[scale=0.53,angle=0]{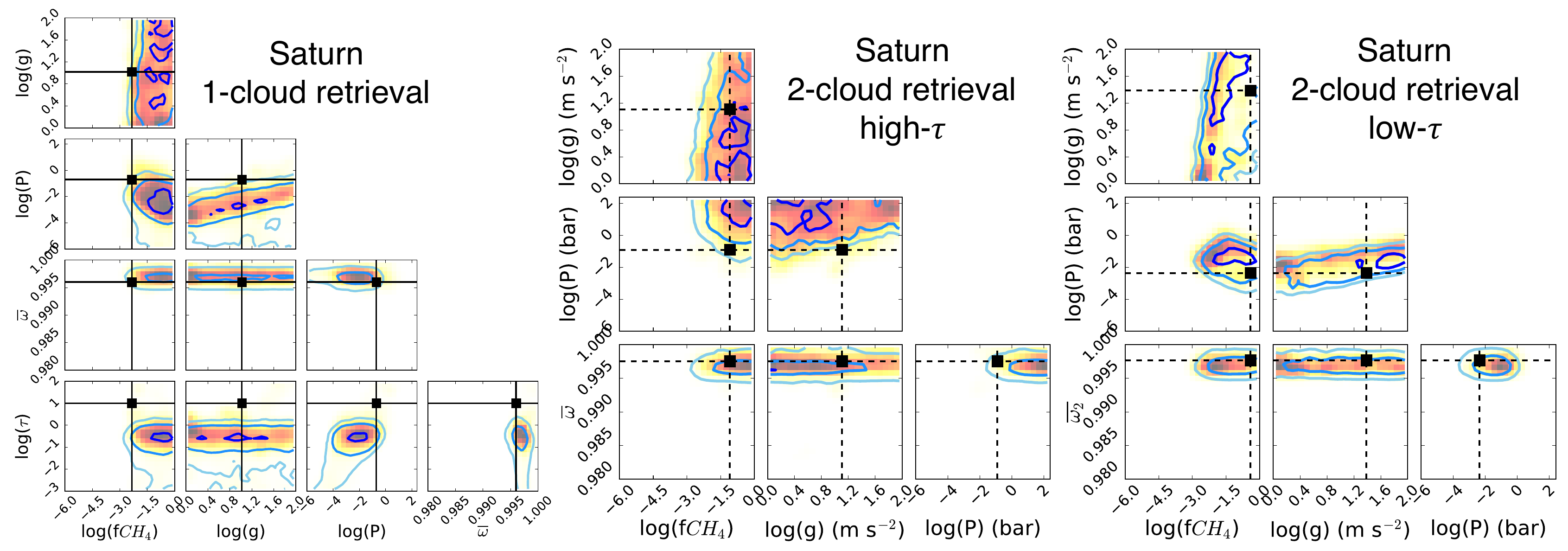}
\caption{2-D marginal posterior distributions for Saturn (SNR=20, CL=25~nm), using a 1-cloud model (left) and a 2-cloud model (middle and right). The posterior for the 2-cloud model is bi-modal, and the two modes are shown separately, for clarity. The dashed black lines mark the position of the {\it best fit solution} for each mode (corresponding to the spectra in Figure~\ref{fig:satspec}), while the black lines on the left plot show the 1-cloud parameter values that best match the ``theoretical model" on the right panel in Figure~\ref{fig:satcld} (e.g. known values for $g$ and {\it fCH4}). \label{fig:sattri}}
\end{figure*}

The Jupiter cloud structure as retrieved by our 2-cloud and 1-cloud models is compared to the theoretical vertical structure for Jupiter in Figure~\ref{fig:jupcld}. The cloud and haze layers shown in the right panel of Figure~\ref{fig:jupcld} approximately match the positions described elsewhere in the literature \citep[e.g.,][]{Simon-Miller:2001,Sato:2013}. The hazes are likely to have a wavelength-dependent continuum opacity, unlike our simple cloud model, and our notation was chosen to emphasize that the upper haze layer is likely absorbing and the lower haze/cloud layer is likely bright (reflective) at the wavelengths relevant in our study. We note that the upper cloud roughly matches the position of a hydrocarbon haze in the upper layers of the atmosphere, and the lower cloud deck overlaps with the bright haze and ammonia/water ice clouds in the deeper atmosphere. This deep cloud is also identified by the 1-cloud model retrieval, but without the opacity contribution of the upper haze/cloud, the retrieved suface gravity of the planet would be unphysically small ($g=1$~m s$^{-2}$, see Figure~\ref{fig:juptri}). 

The significance of the cloud and methane detection is shown in the middle panel of Figure~\ref{fig:tev}. The methane is detected at high significance for all SNR, while the cloud detection becomes {\it very strong} only when SNR$>$10. Due to the degeneracy of the solutions (see Figure~\ref{fig:jupspec}), the Bayes factor does not favor the 2-cloud vs.\ the 1-cloud model except at very high signal-to-noise. However, based on the previous arguments related to the surface gravity, it is reasonable to select the 2-cloud model in this case, and we expect a more clear distinction to appear once independent constraints on the surface gravity are provided.

We conclude that the two-layer cloud model is necessary for Jupiter, constraining the methane abundance to within factors of $\sim 20$ at SNR=5 and factors of $\sim 3$ at SNR=20, possibly much better when tighter limits on the surface gravity are available. The  single scattering albedo of the lower cloud is constrained within 0.5\% even at the lowest SNR. This gives us an indication for the composition of the lower cloud, since particles with high reflectivity are necessary to explain the large value of $\bar{\omega_2}$.

\subsection{Saturn}
\label{ssat}

Our third and final case study is Saturn, which falls between HD 99492 c and Jupiter in terms of retrieval results. We use again data from \citet{Karkoschka:1994} to generate simulated observations using the method in Section~\ref{noise}. The summary plots for the retrieval results are shown in Figure~\ref{fig:satfit}, with the confidence intervals listed in Table~\ref{tab:sat}. The posterior distribution for the 2-cloud retrieval is now clearly bimodal, with one mode corresponding to a low optical depth for the upper cloud, and the other to an optically thick upper cloud. The large confidence intervals plotted in the bottom panel of Figure~\ref{fig:satfit} are due to this bimodality. The range of the possible spectra with parameters drawn from the posterior are shown in the right panel of Figure~\ref{fig:tspec}.

\begin {figure}
\centering
\hspace*{-0.4cm}\includegraphics*[scale=0.4,angle=0]{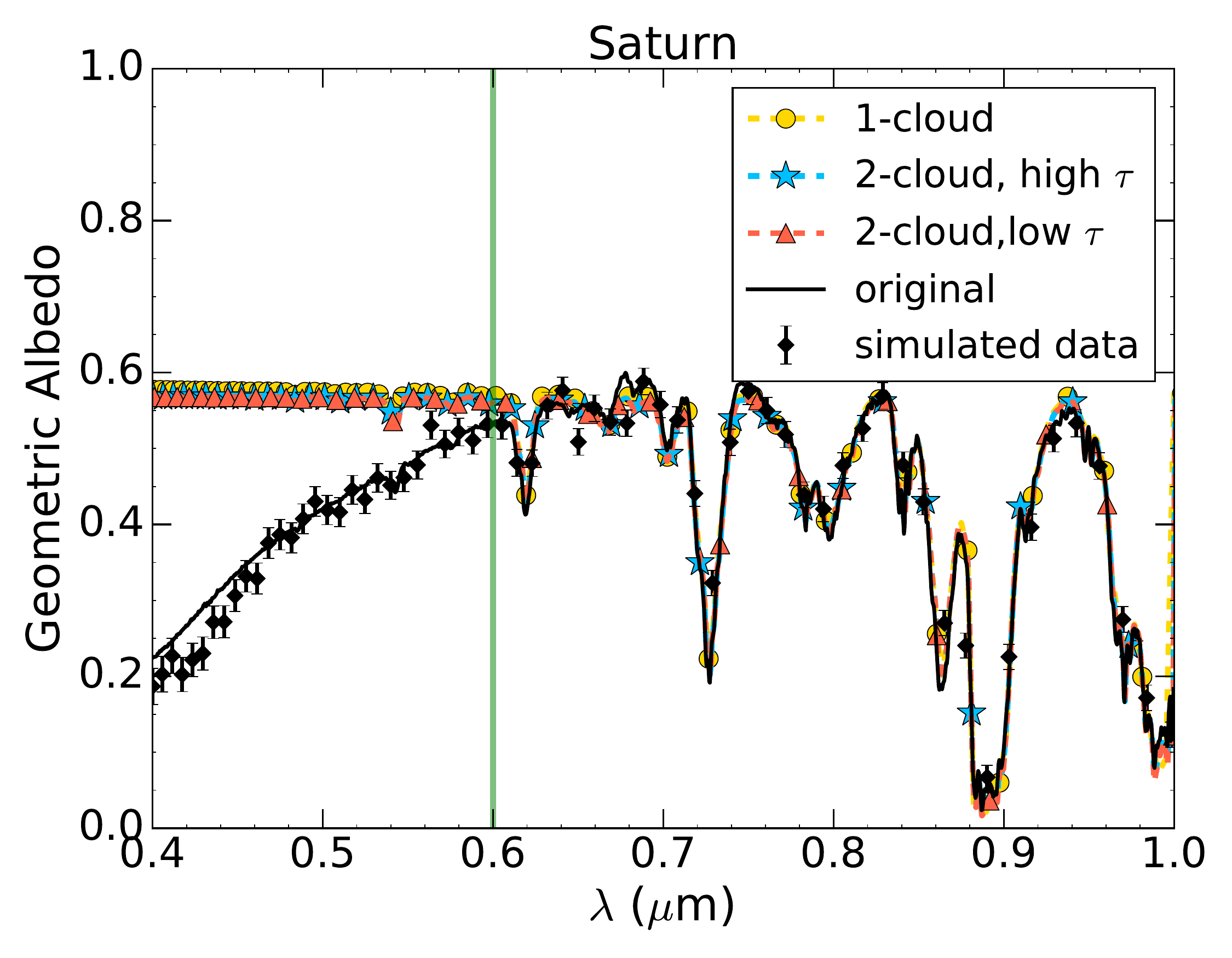}
\caption{Best-fit spectra for Saturn (SNR=20, CL=25~nm), retrieved using the 2-cloud and 1-cloud models. The 2-cloud posterior is bimodal, with the low optical depth and high optical depth best fit solutions shown separately (see also Figure~\ref{fig:sattri}).  \label{fig:satspec}}
\end{figure}

\begin {figure}
\centering
\hspace*{-0.4cm}\includegraphics*[scale=0.4,angle=0,trim={10 0 0 0},clip]{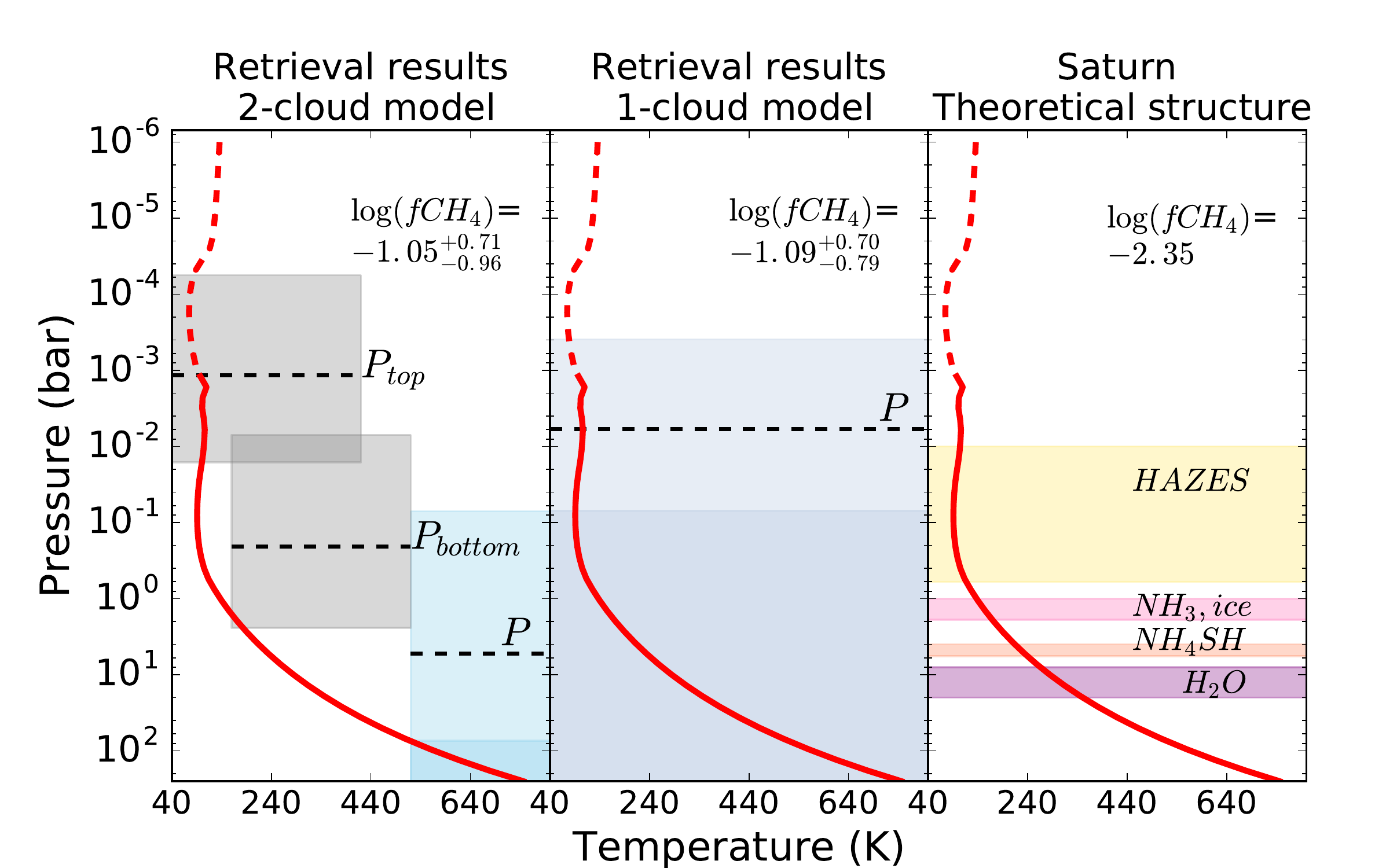}
\caption{Cloud structure for Saturn, as retrieved using the 2-cloud model (left), and the 1-cloud model (right). The conventions are described in the Figure~\ref{fig:hd9cld} and ~\ref{fig:jupcld} captions. The theoretical structure is shown in the right panel, with the cloud structure closely resembling available literature \citep[e.g.,][]{Roman:2013}. \label{fig:satcld}}
\end{figure}

For clarity, the two modes have been separated and the covariances of the most relevant parameters shown in Figure~\ref{fig:sattri} (middle and right panels). In the left panel of Figure~\ref{fig:sattri} we show the retrieved posterior distribution for the 1-cloud forward model, with the black lines indicating the parameter values that correspond to the currently adopted properties of Saturn ($fCH_4=4.5\times10^{-3}$ and $g=10.44$~m~s$^{-2}$). The dashed black lines in the middle and right panels show the {\it best fit} solutions for each of the two modes. As seen in the case of HD 99492 c, the mode with low optical depth constrains the albedo of the lower cloud ($\bar{\omega}_2$), while the optically thick mode constrains the albedo of the upper cloud ($\bar{\omega}$). However, in contrast to HD 99492 c, the 1-cloud retrieval mostly resembles the {\it low optical depth} mode of the 2-cloud retrieval. In this case, the reflecting surface ($P$) is found relatively high ($10^{-3}-1$~bar), with a position correlated with the methane abundance and $g$. The 1-cloud model also constrains the optical depth within a relatively narrow range of $\sim 0.1-1$.  The surface gravity $g$ is unconstrained by both the 1-cloud and 2-cloud retrievals, but independent constraints would translate into narrower confidence intervals for both $P$ and {\it fCH4}, as in the cases described above, especially considering the low optical depth mode. A more peaked distribution for {\it fCH4} is only obtained for the 2-cloud mode of low optical depth (right panel), while in the other two cases only lower limits can be inferred. The methane abundance is overall consistent with measured values, but biased towards higher values in the high optical depth mode, because the entire cloud structure is then obscuring most of the atmosphere.  

Figure~\ref{fig:satspec} shows the complete degeneracy between the 1-cloud retrieved solution and the two modes of the 2-cloud retrieval. Photometry shortward of 0.6~$\mu$m could be helpful for constraining haze properties. Based on these data, we cannot distinguish between the two possible modes, and the presence of the second cloud is not required. The retrieved cloud structure using the 1-cloud and 2-cloud models is presented in Figure~\ref{fig:satcld} and compared with the structure derived from the literature in the right panel \citep[e.g.,][]{Roman:2013}. The lack of evidence for a second cloud is also suggested by the overlap of the 2-cloud structure in the left panel, similar to the situation for HD 99492 c. By contrast, the cloud optical depth is low in this case, and therefore the transition from a clear to a cloudy atmosphere is very gradual. Overall, the retrieved cloud structure strongly overlaps with the theoretical structure, and all solutions are consistent with highly reflective layers present in the atmosphere. This is supported by the Bayes factors in the bottom panel of Figure~\ref{fig:tev}, where both methane and a cloud layer are detected with high significance for all SNR. The evidence for the second cloud is inconclusive, since these solutions are degenerate. We suggest that some evidence is provided by the tighter distribution in Figure~\ref{fig:sattri}, right panel vs. left panel, and a more relevant Bayes factor calculation would be between the 1-cloud model and each of the two modes of the 2-cloud model separately.

\section{SUMMARY AND CONCLUSIONS}
\label{sum}

We have used a Bayesian retrieval method to quantify the confidence intervals on the atmospheric methane abundance and cloud structures of extrasolar giant planets, using a simple atmospheric model with either 1 or 2 cloud decks. Our results should be viewed in the light of the limitations inherent to space coronagraph observations. Notably, we are trying to reproduce complex atmospheric structures by using simple 1-dimensional model approximations and low signal-to-noise, integrated light data. The $0.4-1~\mu$m and $0.6-1~\mu$m wavelength ranges used in the retrievals have also limited diagnostic power, but may be supplemented by other follow-up observations. Nevertheless we find that reflected light spectra of the quality expected from a space-based direct imaging exoplanet mission is sufficient to place interesting constraints on important planetary atmosphere characteristics, particularly methane mixing ratio and, in some cases, cloud albedo. In particular, the {\it presence} of clouds and/or methane absorption is detected at high significance even for a SNR of 5. However, higher SNRs, additional degeneracy-breaking constraints (e.g. on $g$), and even more sophisticated cloud models will be needed to determine accurate {\it abundances} and extracting useful information about mass-metallicity relationships. The retrieval methods presented are powerful for determining correlations among parameters and identifying which ones are unconstrained by the data, demonstrating the value in the synthetic datasets, even at low signal to noise ratios. We find that using both MCMC and nested sampling algorithms can provide us with better insights on the posterior probability distributions for the model parameters, especially in highly non-gaussian and multi-modal cases.

We found that our retrieval methods could reliably infer methane abundances to within factors of ten of the true value when the models are a good match for the data (such as the validation tests), and can accurately constrain cloud scattering properties in specific cases, thus providing a clue to the cloud composition. Gravity, however, is not well constrained by optical spectra in the presence of clouds. Observing planets with known masses therefore removes an important source of uncertainty and allows much greater precision in the inference of atmospheric abundances. Furthermore, cases in which the cloud model was inadequate are readily apparent in the retrieval output. These limitations are particularly apparent in our realistic test cases, where the posterior probability distribution is ofter bimodal, and only a lower limit is inferred for the methane abundance. This prompted us to calculate the Bayesian evidence for a set of models for each simulated spectrum. This is a method to quantify the significance associated with the methane and cloud detection, and the assumed cloud model (1-cloud vs. 2-cloud) in each case. Although time-consuming, this is a very powerful test that will become a necessity for interpreting future observations, as the complexity of our model atmospheres and understanding of planetary diversity is increasing. Our preliminary applications to realistic planets show that it is worthwhile to investigate different vertical cloud structures, such as the 1-cloud vs. the 2-cloud models. This can help us address degeneracies and identify unnecessary parameters. In summary, our first study on the characterization of extrasolar giant planets in reflected light found that retrieval methods using simple, gray cloud models can be applied to optical spectra of exoplanets to obtain insights on molecular abundances and cloud properties. We found that generally the retrieval results are equally sensitive to the particular noise realization as to the chosen spectral correlation length.

\subsection{Ongoing and Future Work}

For this initial study we made a number of simplifications to the analysis to make our task tractable and obtain a first look at parameter correlations. However future work should address these simplifications and their roles in the fidelity of the retrievals. Foremost among those that should be explored include: planetary radius uncertainty, thermal profile uncertainty, and orbital phase uncertainty. The second paper in this series (Nayak et al., submitted), addresses the radius and phase uncertainties. In addition the retrieval of more atmospheric abundances should be explored, particularly water and alkali gasses. We will also investigate the possibility of adopting a somewhat more general cloud model.

In this work we have focused on retrieving atmospheric parameters of giant planets, nevertheless the methods we are developing--and eventually the experience in applying them to real extrasolar planet spectra--will inform future efforts to characterize the atmospheres of lower mass planets. While detailed investigation of retrieval methods for such planets awaits future studies, we note several general conclusions. Planets with relatively flat spectra or few absorption features are, unsurprisingly, challenging. The methane-dominated spectra we studied here are well suited to retrieval methods as multiple bands of varying strength populate the optical, permitting constraints on both cloud top pressure and abundance when well resolved (e.g., Figure 3). This may not be the case for many potential terrestrial planet atmospheres leading to greater uncertainties in cloud top pressure and absorber column abundances. Furthermore lack of useful constraints on gravity, through mass determination, substantially increases the uncertainty in retrieved atmospheric abundances. Thus giant planets, even cloudless ones with steep Rayleigh scattering slopes, though not the pale blue dots we ultimately seek, do provide useful insights into the methods and limitations of our future characterization of such worlds.

\begin{deluxetable*}{llcccccc}
\tabletypesize{\scriptsize}
\tablecolumns{8}
\tablecaption{Retrieval verification results for the 1-cloud model.\label{tab:vres1}}
\tablehead{ \colhead{Parameter} & \colhead{Original} & \multicolumn{2}{c}{SNR=5} & \multicolumn{2}{c}{SNR=10} & \multicolumn{2}{c}{SNR = 20} \\
\colhead{} & \colhead{Value} & \colhead{CL\tablenotemark{a}=25nm} & \colhead{CL=100nm} & \colhead{CL=25nm} & \colhead{CL=100nm} & \colhead{CL=25nm} & \colhead{CL=100nm} }
\startdata
\sidehead{Cloud-free case}
\tableline
$\log(fCH_4)$ & -3.31 & $-3.22_{-0.22}^{+0.19}$ & $-2.92_{-0.24}^{+0.18}$ & $-3.42_{-0.11}^{+0.11}$ & $-3.20_{-0.10}^{+0.09}$ & $-3.27_{-0.03}^{+0.03}$ & $-3.20_{-0.03}^{+0.03}$\\
&& ($-3.21_{-0.20}^{+0.18}$)\tablenotemark{b} & & ($-3.42_{-0.10}^{+0.10}$) & & ($-3.27_{-0.03}^{+0.03}$) & \\
\\[-0.08in]
$\log(g)$ (m s$^{-2}$) & 0.86 & $0.84_{-0.39}^{+0.21}$ & $0.95_{-0.39}^{+0.22}$ & $0.90_{-0.22}^{+0.16}$ & $0.63_{-0.26}^{+0.28}$ & $0.85_{-0.04}^{+0.03}$ & $0.89_{-0.03}^{+0.03}$ \\
&& ($0.82_{-0.42}^{+0.22}$) & & ($0.93_{-0.21}^{+0.11}$) & & ($0.86_{-0.04}^{+0.03}$) & \\
\\[-0.08in]
$\log(P)$ (bar) & 1.00 & $-0.71_{-2.32}^{+1.74}$ & $-0.53_{-2.34}^{+1.61}$ & $-0.67_{-2.47}^{+1.82}$ & $-1.15_{-2.18}^{+1.94}$ & $-0.53_{-2.27}^{+1.55}$ & $-0.46_{-2.08}^{+1.51}$ \\
&& ($-0.60_{-2.32}^{+1.53}$) & & ($-0.83_{-2.41}^{+1.89}$) & & ($-0.45_{-2.14}^{+1.46}$) & \\
\\[-0.08in]
$\bar{\omega}$ & 0.50 & $0.51_{-0.32}^{+0.33}$ & $0.57_{-0.37}^{+0.31}$ & $0.57_{-0.36}^{+0.34}$ & $0.45_{-0.31}^{+0.34}$ & $0.52_{-0.35}^{+0.36}$ & $0.53_{-0.36}^{+0.35}$ \\
&& ($0.51_{-0.33}^{+0.32}$) & & ($0.57_{-0.37}^{+0.31}$) & & ($0.53_{-0.34}^{+0.32}$) & \\
\\[-0.08in]
$\bar{g}$ & 0.50 & $0.49_{-0.35}^{+0.36}$ & $0.47_{-0.31}^{+0.36}$ & $0.41_{-0.30}^{+0.35}$ & $0.59_{-0.37}^{+0.30}$ & $0.48_{-0.31}^{+0.35}$ & $0.50_{-0.34}^{+0.36}$ \\
&& ($0.50_{-0.32}^{+0.33}$) & & ($0.35_{-0.24}^{+0.38}$) & & ($0.50_{-0.32}^{+0.33}$) & \\
\\[-0.08in]
$\log(\tau)$ & -8.00 & $-7.01_{-2.01}^{+2.74}$ & $-6.81_{-2.19}^{+3.24}$ & $-4.81_{-2.73}^{+3.44}$ & $-4.34_{-2.32}^{+2.65}$ & $-7.35_{-1.76}^{+2.53}$ & $-7.64_{-1.69}^{+2.58}$ \\
&& ($-7.04_{-1.93}^{+2.92}$) & & ($-5.58_{-2.49}^{+3.56}$) & & ($-7.51_{-1.64}^{+2.49}$) & \\
\\[-0.08in]
\tableline
\sidehead{1-Cloud case}
\tableline
$\log(fCH_4)$ &  -3.31 &  $-3.54_{-0.31}^{+0.38}$ & $-3.47_{-0.32}^{+0.39}$ & $-3.27_{-0.22}^{+0.21}$ & $-1.42_{-0.82}^{+0.88}$ & $-3.31_{-0.22}^{+0.17}$ & $-2.73_{-0.27}^{+0.21}$ \\
& &  ($-3.52_{-0.32}^{+0.34}$) & & ($-3.25_{-0.20}^{+0.23}$) & & ($-3.13_{-0.11}^{+0.12}$) & \\
\\[-0.08in]
$\log(g)$ (m s$^{-2}$) &  0.86 &  $0.39_{-0.90}^{+0.85}$ & $0.19_{-0.81}^{+0.97}$ & $0.36_{-0.90}^{+0.91}$ & $1.08_{-1.04}^{+0.64}$ & $0.05_{-0.62}^{+0.50}$ & $1.31_{-1.47}^{+0.54}$ \\
& &  ($0.38_{-0.82}^{+0.88}$) & & ($0.41_{-0.91}^{+0.90}$) & & ($0.01_{-0.65}^{+0.67}$) & \\
\\[-0.08in]
$\log(P)$ (bar) &  -0.70 &  $-1.72_{-1.89}^{+1.36}$ & $-1.46_{-1.13}^{+1.14}$ & $-1.79_{-1.52}^{+1.40}$ & $-3.29_{-0.75}^{+1.22}$ & $-2.03_{-1.20}^{+1.02}$ & $-0.85_{-1.42}^{+0.84}$ \\
& &  ($-1.80_{-1.73}^{+1.51}$) & & ($-1.82_{-1.54}^{+1.33}$) & & ($-2.63_{-1.01}^{+0.98}$) & \\
\\[-0.08in]
$\bar{\omega}$ &  0.96 &  $0.90_{-0.05}^{+0.04}$ & $0.90_{-0.05}^{+0.03}$ & $0.92_{-0.03}^{+0.03}$ & $0.92_{-0.03}^{+0.03}$ & $0.95_{-0.03}^{+0.02}$ & $0.94_{-0.04}^{+0.02}$ \\
& &  ($0.90_{-0.04}^{+0.04}$) & & ($0.91_{-0.03}^{+0.03}$) & & ($0.92_{-0.03}^{+0.03}$) & \\
\\[-0.08in]
$\bar{g}$ &  0.85 &  $0.27_{-0.19}^{+0.38}$ & $0.35_{-0.24}^{+0.33}$ & $0.29_{-0.20}^{+0.39}$ & $0.27_{-0.19}^{+0.39}$ & $0.69_{-0.33}^{+0.24}$ & $0.52_{-0.35}^{+0.31}$ \\
& &  ($0.28_{-0.20}^{+0.38}$) & & ($0.26_{-0.18}^{+0.33}$) & & ($0.33_{-0.23}^{+0.29}$) & \\
\\[-0.08in]
$\log(\tau)$ &  0.00 &  $-1.31_{-2.20}^{+2.89}$ & $0.07_{-2.13}^{+1.85}$ & $-0.36_{-2.31}^{+2.36}$ & $-1.18_{-1.23}^{+2.43}$ & $-0.83_{-1.44}^{+2.49}$ & $0.73_{-1.38}^{+1.61}$ \\
& &  ($-1.40_{-2.05}^{+3.09}$) & & ($-0.48_{-2.33}^{+2.38}$) & & ($-1.45_{-1.18}^{+0.63}$) & \\
\\[-0.08in]
\enddata
\tablenotetext{a}{CL here is a shorthand notation for the spectral noise correlation length.}
\tablenotetext{b}{Numbers in parentheses show the nested sampling results.}
\end{deluxetable*}

\begin{deluxetable*}{llcccccc}
\tabletypesize{\scriptsize}
\tablecolumns{8}
\tablecaption{Retrieval verification results for the 2-cloud model.\label{tab:vres2}}
\tablehead{ \colhead{Parameter} & \colhead{Original} & \multicolumn{2}{c}{SNR=5} & \multicolumn{2}{c}{SNR=10} & \multicolumn{2}{c}{SNR = 20} \\
\colhead{} & \colhead{Value} & \colhead{CL=25nm} & \colhead{CL=100nm} & \colhead{CL=25nm} & \colhead{CL=100nm} & \colhead{CL=25nm} & \colhead{CL=100nm} }
\startdata
$\log(fCH_4)$ &  -2.74 &  $-1.95_{-0.67}^{+0.88}$ & $-2.54_{-0.53}^{+0.96}$ & $-1.79_{-0.65}^{+0.85}$ & $-1.90_{-0.60}^{+0.86}$ & $-2.66_{-0.19}^{+0.14}$ & $-2.65_{-0.17}^{+0.14}$ \\
& &  ($-1.35_{-0.95}^{+0.89}$) & & ($-1.37_{-0.86}^{+0.92}$) & & ($-2.65_{-0.21}^{+0.15}$) & \\
\\[-0.08in]
$\log(g)$ (m s$^{-2}$) &  1.39 &  $1.22_{-0.70}^{+0.55}$ & $1.21_{-0.70}^{+0.56}$ & $1.19_{-0.66}^{+0.54}$ & $1.28_{-0.68}^{+0.53}$ & $1.71_{-0.44}^{+0.22}$ & $1.62_{-0.39}^{+0.27}$ \\
& &  ($1.12_{-0.72}^{+0.60}$) & & ($1.07_{-0.69}^{+0.63}$) & & ($1.65_{-0.56}^{+0.24}$) & \\
\\[-0.08in]
$\log(P)$ (bar) &  -0.15 &  $-1.25_{-1.04}^{+0.84}$ & $-0.39_{-0.89}^{+0.54}$ & $-1.25_{-0.86}^{+0.70}$ & $-1.23_{-0.90}^{+0.78}$ & $0.06_{-0.32}^{+0.15}$ & $-0.04_{-0.27}^{+0.20}$ \\
& &  ($-0.72_{-1.07}^{+1.12}$) & & ($-0.90_{-0.85}^{+1.08}$) & & ($-0.07_{-0.37}^{+0.20}$) & \\
\\[-0.08in]
$dP_1$ (bar) &  0.54 &  $0.82_{-0.60}^{+1.05}$ & $1.21_{-0.85}^{+1.22}$ & $0.83_{-0.60}^{+1.03}$ & $0.87_{-0.63}^{+1.05}$ & $1.45_{-1.05}^{+1.23}$ & $1.44_{-0.99}^{+1.36}$ \\
& &  ($0.87_{-0.62}^{+0.97}$) & & ($0.87_{-0.60}^{+0.95}$) & & ($1.04_{-0.74}^{+1.38}$) & \\
\\[-0.08in]
$dP_2$ (bar) &  0.12 &  $0.89_{-0.66}^{+1.13}$ & $1.09_{-0.78}^{+1.12}$ & $0.76_{-0.57}^{+1.02}$ & $0.83_{-0.60}^{+0.91}$ & $1.28_{-0.88}^{+1.24}$ & $1.11_{-0.81}^{+1.28}$ \\
& &  ($1.04_{-0.75}^{+1.25}$) & & ($0.84_{-0.58}^{+0.94}$) & & ($1.49_{-1.00}^{+1.27}$) & \\
\\[-0.08in]
$\bar{\omega}$ &  0.85 &  $0.56_{-0.39}^{+0.32}$ & $0.62_{-0.42}^{+0.30}$ & $0.56_{-0.38}^{+0.31}$ & $0.54_{-0.37}^{+0.34}$ & $0.68_{-0.38}^{+0.23}$ & $0.69_{-0.35}^{+0.20}$ \\
& &  ($0.95_{-0.59}^{+0.03}$) & & ($0.80_{-0.54}^{+0.18}$) & & ($0.46_{-0.30}^{+0.33}$) & \\
\\[-0.08in]
$\bar{g}$ &  0.85 &  $0.48_{-0.31}^{+0.35}$ & $0.54_{-0.35}^{+0.32}$ & $0.55_{-0.34}^{+0.30}$ & $0.47_{-0.31}^{+0.34}$ & $0.60_{-0.40}^{+0.30}$ & $0.60_{-0.37}^{+0.27}$ \\
& &  ($0.39_{-0.27}^{+0.38}$) & & ($0.42_{-0.29}^{+0.37}$) & & ($0.55_{-0.35}^{+0.29}$) & \\
\\[-0.08in]
$\log(\tau)$ &  -0.30 &  $-1.85_{-0.79}^{+0.82}$ & $-1.67_{-0.88}^{+1.00}$ & $-2.06_{-0.65}^{+0.78}$ & $-1.99_{-0.73}^{+0.81}$ & $-1.02_{-0.70}^{+0.33}$ & $-1.00_{-1.04}^{+0.45}$ \\
& &  ($-0.63_{-1.70}^{+2.18}$) & & ($-1.43_{-1.08}^{+2.92}$) & & ($-1.01_{-0.60}^{+0.28}$) & \\
\\[-0.08in]
$\bar{\omega}_2$ &  0.997 &  $0.987_{-0.003}^{+0.004}$ & $0.991_{-0.003}^{+0.005}$ & $0.989_{-0.001}^{+0.002}$ & $0.988_{-0.001}^{+0.003}$ & $0.993_{-0.003}^{+0.003}$ & $0.993_{-0.003}^{+0.005}$ \\
& &  ($0.984_{-0.638}^{+0.005}$) & & ($0.989_{-0.564}^{+0.002}$) & & ($0.995_{-0.004}^{+0.003}$) & \\
\\[-0.08in]
\enddata
\end{deluxetable*}

\begin{deluxetable*}{lcccccc}
\tabletypesize{\scriptsize}
\tablecolumns{7}
\tablecaption{Retrieval results for HD 99492 c.\label{tab:hd9}}
\tablehead{ \colhead{Parameter} & \multicolumn{2}{c}{SNR=5} & \multicolumn{2}{c}{SNR=10} & \multicolumn{2}{c}{SNR = 20} \\
\colhead{} & \colhead{CL=25nm} & \colhead{CL=100nm} & \colhead{CL=25nm} & \colhead{CL=100nm} & \colhead{CL=25nm} & \colhead{CL=100nm} }
\startdata
$\log(fCH_4)$ &  $-1.76_{-1.29}^{+1.20}$ & $-1.68_{-1.12}^{+0.98}$ & $-1.37_{-1.00}^{+0.92}$ & $-1.24_{-1.09}^{+0.86}$ & $-1.13_{-0.73}^{+0.69}$ & $-1.25_{-0.80}^{+0.75}$ \\
& ($-1.85_{-1.18}^{+1.18}$) & & ($-1.48_{-0.95}^{+0.96}$) & & ($-1.14_{-0.94}^{+0.72}$) & \\
\\[-0.08in]
$\log(g)$ (m s$^{-2}$) & $0.55_{-1.01}^{+0.92}$ & $0.52_{-0.97}^{+0.99}$ & $0.41_{-0.85}^{+1.05}$ & $0.52_{-0.91}^{+0.93}$ & $0.51_{-0.86}^{+1.02}$ & $0.44_{-0.87}^{+0.93}$ \\
& ($1.51_{-0.95}^{+0.97}$) & & ($1.56_{-1.02}^{+0.95}$) & & ($1.71_{-1.10}^{+0.88}$) & \\
\\[-0.08in]
$\log(P)$ (bar) & $0.02_{-1.42}^{+1.13}$ & $0.08_{-1.42}^{+1.10}$ & $-0.12_{-1.43}^{+1.24}$ & $0.11_{-1.46}^{+1.06}$ & $-0.09_{-1.51}^{+1.13}$ & $-0.09_{-1.50}^{+1.26}$ \\
&  ($0.00_{-1.22}^{+1.05}$) & & ($-0.38_{-1.22}^{+1.28}$) & & ($-0.41_{-1.18}^{+1.24}$) & \\
\\[-0.08in]
$dP_1$ (bar) & $1.30_{-0.98}^{+1.35}$ & $1.26_{-0.94}^{+1.44}$ & $1.38_{-0.98}^{+1.33}$ & $1.58_{-1.19}^{+1.52}$ & $1.60_{-1.07}^{+1.26}$ & $1.59_{-1.17}^{+1.54}$ \\
&  ($1.03_{-0.72}^{+1.19}$) & & ($1.02_{-0.71}^{+1.12}$) & & ($1.15_{-0.80}^{+1.20}$) & \\
\\[-0.08in]
$dP_2$ (bar) & $1.24_{-0.93}^{+1.25}$ & $1.33_{-0.95}^{+1.43}$ & $0.79_{-0.56}^{+0.96}$ & $0.79_{-0.53}^{+0.83}$ & $0.63_{-0.44}^{+0.88}$ & $0.58_{-0.41}^{+0.64}$ \\
&  ($1.28_{-0.89}^{+1.47}$) & & ($0.91_{-0.62}^{+1.08}$) & & ($0.83_{-0.55}^{+0.85}$) & \\
\\[-0.08in]
$\bar{\omega}$ & $0.91_{-0.04}^{+0.04}$ & $0.91_{-0.04}^{+0.04}$ & $0.91_{-0.04}^{+0.03}$ & $0.90_{-0.04}^{+0.03}$ & $0.92_{-0.03}^{+0.02}$ & $0.92_{-0.03}^{+0.02}$ \\
&  ($0.89_{-0.49}^{+0.05}$) & & ($0.88_{-0.50}^{+0.05}$) & & ($0.87_{-0.46}^{+0.06}$) & \\
\\[-0.08in]
$\bar{g}$ & $0.31_{-0.23}^{+0.40}$ & $0.35_{-0.25}^{+0.41}$ & $0.30_{-0.23}^{+0.36}$ & $0.35_{-0.24}^{+0.35}$ & $0.46_{-0.25}^{+0.27}$ & $0.42_{-0.22}^{+0.29}$ \\
&  ($0.36_{-0.25}^{+0.37}$) & & ($0.38_{-0.26}^{+0.36}$) & & ($0.38_{-0.26}^{+0.33}$) & \\
\\[-0.08in]
$\log(\tau)$ & $1.49_{-1.02}^{+0.95}$ & $1.27_{-1.14}^{+1.10}$ & $2.00_{-0.88}^{+0.70}$ & $1.98_{-0.95}^{+0.75}$ & $2.18_{-0.89}^{+0.59}$ & $2.14_{-0.90}^{+0.60}$ \\
&  ($0.77_{-3.59}^{+1.38}$) & & ($0.77_{-3.75}^{+1.56}$) & & ($1.10_{-4.19}^{+1.35}$) & \\
\\[-0.08in]
$\bar{\omega}_2$ & $0.592_{-0.382}^{+0.354}$ & $0.644_{-0.441}^{+0.307}$ & $0.558_{-0.391}^{+0.348}$ & $0.562_{-0.358}^{+0.313}$ & $0.596_{-0.386}^{+0.293}$ & $0.542_{-0.382}^{+0.343}$ \\
&  ($0.880_{-0.580}^{+0.083}$) & & ($0.956_{-0.623}^{+0.006}$) & & ($0.878_{-0.559}^{+0.078}$) & \\
\\[-0.08in]
\enddata
\end{deluxetable*}

\begin{deluxetable*}{lcccccc}
\tabletypesize{\scriptsize}
\tablecolumns{7}
\tablecaption{Retrieval results for Jupiter.\label{tab:jup}}
\tablehead{ \colhead{Parameter} &  \multicolumn{2}{c}{SNR=5} & \multicolumn{2}{c}{SNR=10} & \multicolumn{2}{c}{SNR = 20} \\
\colhead{} &  \colhead{CL=25nm} & \colhead{CL=100nm} & \colhead{CL=25nm} & \colhead{CL=100nm} & \colhead{CL=25nm} & \colhead{CL=100nm} }
\startdata
$\log(fCH_4)$ & $-1.15_{-0.87}^{+0.74}$ & $-1.91_{-0.91}^{+1.13}$ & $-1.95_{-0.83}^{+1.14}$ & $-1.70_{-0.81}^{+0.88}$ & $-2.80_{-0.35}^{+0.48}$ & $-2.60_{-0.52}^{+0.61}$ \\
&  ($-1.10_{-0.89}^{+0.72}$) & & ($-1.42_{-1.13}^{+0.96}$) & & ($-3.25_{-0.11}^{+0.14}$) & \\
\\[-0.08in]
$\log(g)$ (m s$^{-2}$) & $1.63_{-1.04}^{+0.86}$ & $1.62_{-1.01}^{+0.88}$ & $1.74_{-1.08}^{+0.87}$ & $1.83_{-1.19}^{+0.79}$ & $0.76_{-0.62}^{+0.90}$ & $1.00_{-0.78}^{+1.19}$ \\
&  ($1.26_{-0.87}^{+1.07}$) & & ($1.01_{-0.70}^{+1.03}$) & & ($0.09_{-0.06}^{+0.13}$) & \\
\\[-0.08in]
$\log(P)$ (bar) & $-0.71_{-0.86}^{+0.67}$ & $-0.52_{-0.92}^{+0.56}$ & $-0.67_{-0.86}^{+0.62}$ & $-0.74_{-0.83}^{+0.71}$ & $-0.79_{-0.30}^{+0.53}$ & $-0.78_{-0.31}^{+0.70}$ \\
&  ($-0.37_{-0.93}^{+0.76}$) & & ($0.19_{-0.94}^{+0.59}$) & & ($0.35_{-0.18}^{+0.24}$) & \\
\\[-0.08in]
$dP_1$ (bar) & $1.06_{-0.77}^{+1.23}$ & $1.13_{-0.82}^{+1.32}$ & $0.95_{-0.65}^{+1.15}$ & $1.01_{-0.72}^{+1.16}$ & $0.62_{-0.46}^{+0.99}$ & $1.00_{-0.75}^{+1.20}$ \\
&  ($0.87_{-0.63}^{+1.04}$) & & ($0.67_{-0.48}^{+0.86}$) & & ($0.90_{-0.24}^{+0.25}$) & \\
\\[-0.08in]
$dP_2$ (bar) & $0.93_{-0.67}^{+1.09}$ & $0.88_{-0.66}^{+1.30}$ & $1.00_{-0.70}^{+1.23}$ & $1.05_{-0.77}^{+1.07}$ & $0.43_{-0.27}^{+0.71}$ & $0.83_{-0.57}^{+1.13}$ \\
&  ($1.11_{-0.78}^{+1.31}$) & & ($2.55_{-1.43}^{+1.14}$) & & ($0.28_{-0.17}^{+0.31}$) & \\
\\[-0.08in]
$\bar{\omega}$ & $0.60_{-0.39}^{+0.29}$ & $0.55_{-0.36}^{+0.32}$ & $0.67_{-0.43}^{+0.26}$ & $0.58_{-0.35}^{+0.30}$ & $0.84_{-0.33}^{+0.11}$ & $0.61_{-0.27}^{+0.29}$ \\
&  ($0.79_{-0.51}^{+0.21}$) & & ($0.99_{-0.15}^{+0.00}$) & & ($1.00_{-0.00}^{+0.00}$) & \\
\\[-0.08in]
$\bar{g}$ & $0.52_{-0.36}^{+0.34}$ & $0.50_{-0.36}^{+0.33}$ & $0.53_{-0.37}^{+0.35}$ & $0.49_{-0.33}^{+0.34}$ & $0.88_{-0.48}^{+0.11}$ & $0.56_{-0.33}^{+0.33}$ \\
&  ($0.46_{-0.32}^{+0.34}$) & & ($0.32_{-0.23}^{+0.37}$) & & ($0.26_{-0.18}^{+0.29}$) & \\
\\[-0.08in]
$\log(\tau)$ & $-2.04_{-0.64}^{+0.81}$ & $-2.12_{-0.60}^{+0.78}$ & $-1.59_{-0.93}^{+1.08}$ & $-2.08_{-0.63}^{+0.75}$ & $-1.12_{-0.95}^{+0.62}$ & $-1.83_{-0.72}^{+0.78}$ \\
&  ($-1.48_{-1.05}^{+3.22}$) & & ($1.16_{-2.11}^{+0.81}$) & & ($0.71_{-0.11}^{+0.18}$) & \\
\\[-0.08in]
$\bar{\omega}_2$ & $0.997_{-0.002}^{+0.002}$ & $0.995_{-0.002}^{+0.002}$ & $0.993_{-0.002}^{+0.004}$ & $0.995_{-0.001}^{+0.002}$ & $0.995_{-0.001}^{+0.001}$ & $0.993_{-0.001}^{+0.002}$ \\
&  ($0.996_{-0.494}^{+0.002}$) & & ($0.645_{-0.435}^{+0.348}$) & & ($0.379_{-0.254}^{+0.313}$) & \\
\\[-0.08in]
\enddata
\end{deluxetable*}

\begin{deluxetable*}{lcccccc}
\tabletypesize{\scriptsize}
\tablecolumns{7}
\tablecaption{Retrieval results for Saturn.\label{tab:sat}}
\tablehead{ \colhead{Parameter} & \multicolumn{2}{c}{SNR=5} & \multicolumn{2}{c}{SNR=10} & \multicolumn{2}{c}{SNR = 20} \\
\colhead{} & \colhead{CL=25nm} & \colhead{CL=100nm} & \colhead{CL=25nm} & \colhead{CL=100nm} & \colhead{CL=25nm} & \colhead{CL=100nm} }
\startdata
$\log(fCH_4)$ & $-1.15_{-0.99}^{+0.83}$ & $-1.20_{-1.00}^{+0.85}$ & $-1.14_{-0.76}^{+0.77}$ & $-1.10_{-0.86}^{+0.69}$ & $-1.29_{-0.63}^{+0.73}$ & $-1.37_{-0.83}^{+0.83}$ \\
&  ($-1.26_{-0.90}^{+0.81}$) & & ($-1.13_{-0.77}^{+0.72}$) & & ($-1.08_{-0.77}^{+0.70}$) & \\
\\[-0.08in]
$\log(g)$ (m s$^{-2}$) & $1.24_{-0.86}^{+1.18}$ & $1.13_{-0.82}^{+1.11}$ & $1.43_{-0.98}^{+1.00}$ & $1.07_{-0.72}^{+1.14}$ & $1.18_{-0.80}^{+1.14}$ & $1.14_{-0.84}^{+1.23}$ \\
&  ($1.33_{-0.90}^{+1.09}$) & & ($1.34_{-0.88}^{+1.01}$) & & ($1.27_{-0.83}^{+1.04}$) & \\
\\[-0.08in]
$\log(P)$ (bar) & $0.12_{-1.70}^{+1.51}$ & $0.19_{-1.86}^{+1.53}$ & $-0.32_{-1.37}^{+2.03}$ & $0.60_{-1.90}^{+1.23}$ & $-0.37_{-1.21}^{+2.11}$ & $-0.19_{-1.49}^{+1.75}$ \\
&  ($-0.34_{-1.35}^{+2.00}$) & & ($0.63_{-2.07}^{+1.20}$) & & ($-0.81_{-1.26}^{+2.31}$) & \\
\\[-0.08in]
$dP_1$ (bar) & $1.20_{-0.85}^{+1.32}$ & $1.22_{-0.88}^{+1.23}$ & $1.18_{-0.81}^{+1.28}$ & $1.31_{-0.97}^{+1.16}$ & $1.21_{-0.90}^{+1.24}$ & $1.22_{-0.85}^{+1.23}$ \\
&  ($1.27_{-0.88}^{+1.31}$) & & ($1.21_{-0.80}^{+1.19}$) & & ($1.25_{-0.82}^{+1.17}$) & \\
\\[-0.08in]
$dP_2$ (bar) & $0.99_{-0.68}^{+1.43}$ & $1.22_{-0.90}^{+1.68}$ & $2.08_{-1.46}^{+1.85}$ & $1.47_{-0.98}^{+1.60}$ & $1.75_{-0.91}^{+0.96}$ & $1.32_{-0.70}^{+0.94}$ \\
&  ($1.47_{-1.00}^{+1.78}$) & & ($2.79_{-1.79}^{+1.96}$) & & ($1.82_{-1.13}^{+0.96}$) & \\
\\[-0.08in]
$\bar{\omega}$ & $1.00_{-0.36}^{+0.00}$ & $0.99_{-0.25}^{+0.01}$ & $1.00_{-0.25}^{+0.00}$ & $1.00_{-0.06}^{+0.00}$ & $1.00_{-0.05}^{+0.00}$ & $1.00_{-0.08}^{+0.00}$ \\
&  ($0.97_{-0.59}^{+0.03}$) & & ($1.00_{-0.41}^{+0.00}$) & & ($0.92_{-0.54}^{+0.08}$) & \\
\\[-0.08in]
$\bar{g}$ & $0.39_{-0.27}^{+0.36}$ & $0.39_{-0.27}^{+0.34}$ & $0.55_{-0.38}^{+0.30}$ & $0.46_{-0.31}^{+0.35}$ & $0.67_{-0.41}^{+0.25}$ & $0.67_{-0.46}^{+0.22}$ \\
&  ($0.45_{-0.30}^{+0.32}$) & & ($0.41_{-0.27}^{+0.32}$) & & ($0.48_{-0.30}^{+0.29}$) & \\
\\[-0.08in]
$\log(\tau)$ & $1.49_{-3.77}^{+1.01}$ & $1.28_{-3.41}^{+1.30}$ & $1.20_{-3.33}^{+1.09}$ & $1.80_{-3.38}^{+0.88}$ & $1.28_{-2.83}^{+1.05}$ & $1.42_{-3.56}^{+1.04}$ \\
&  ($-0.94_{-1.50}^{+3.28}$) & & ($1.46_{-3.35}^{+0.89}$) & & ($-1.08_{-1.14}^{+3.31}$) & \\
\\[-0.08in]
$\bar{\omega}_2$ & $0.812_{-0.552}^{+0.186}$ & $0.806_{-0.507}^{+0.188}$ & $0.946_{-0.593}^{+0.052}$ & $0.610_{-0.405}^{+0.386}$ & $0.968_{-0.598}^{+0.029}$ & $0.949_{-0.449}^{+0.048}$ \\
&  ($0.996_{-0.658}^{+0.003}$) & & ($0.712_{-0.500}^{+0.285}$) & & ($0.996_{-0.608}^{+0.001}$) & \\
\\[-0.08in]
\enddata
\end{deluxetable*}

\clearpage
\appendix
\section{Sampling Methods and Evidence Calculation}

In Bayesian inference, the allowed ranges of model parameters are given by the posterior probability distribution of the parameter vector $\bm{\theta}$,

\begin{equation}
\mathcal{P}(\bm{\theta}) = \frac{\mathcal{L}(\bm{\theta})\pi(\bm{\theta})}{\mathcal{Z}},
\end{equation}
where $\mathcal{P}(\bm{\theta})\equiv \mathrm{Pr}(\bm{\theta}\mid \mathcal{D}, \mathcal{M})$, $\mathcal{L}(\bm{\theta})\equiv\mathrm{Pr}(\mathcal{D}\mid \bm{\theta}, \mathcal{M})$ is the likelihood, $\pi(\bm{\theta})\equiv\mathrm{Pr}(\bm{\theta}\mid \mathcal{M})$ is the prior on model parameters, and $\mathcal{Z}\equiv\mathrm{Pr}(\mathcal{D}\mid \mathcal{M})$ is the Bayesian evidence. Here $\mathcal{D}$ and $\mathcal{M}$ denote the data and the model, respectively. Normalization of the posterior distribution requires that

\begin{equation}
\label{eq:ev}
\mathcal{Z}=\int \mathcal{L}(\bm{\theta})\pi(\bm{\theta}) d\bm{\theta}.
\end{equation}

The calculation of $\mathcal{Z}$ is not necessary for parameter estimation, and best-fit parameter values with associated confidence intervals are obtained from the un-normalized $\mathcal{P}(\bm{\theta})$. In general, the posterior $\mathcal{P}(\bm{\theta})$ is difficult or impossible to calculate analytically, and in practice the shape of this distribution is approximated by taking a large number of samples. The methods described below are optimized to sample more efficiently the regions of parameter space where $\mathcal{L}(\bm{\theta})$ is large, such that a good approximation of $\mathcal{P}(\bm{\theta})$ is obtained with a minimum number of samples. The Bayesian evidence $\mathcal{Z}$ is by definition model-dependent, and provides the information necessary for model selection. The evaluation of this multi-dimensional integral is also often difficult, and addressed by various approximations (Section~\ref{evidence}).

\subsection{Model selection}
\label{evidence}

In Bayesian inference, the probability associated with a given model $\mathcal{M}$, given the data, is defined as $\mathrm{Pr}(\mathcal{M}\mid \mathcal{D})=\mathrm{Pr}(\mathcal{D}\mid \mathcal{M})\mathrm{Pr}(\mathcal{M})=\mathcal{Z}\mathrm{Pr}(\mathcal{M})$. In our calculations of Bayesian evidence we have employed the approximations described below. 

In the Laplace-Metropolis approximation \citep{Lopes:2004}, $\mathcal{Z}$ is computed using the covariance matrix $\bm{C}$ of the posterior, or the minimum volume ellipsoid enclosing the posterior distribution

\begin{equation}
\mathcal{Z}\simeq \mathcal{L}_{max}(\bm{\theta})(2\pi)^{n/2}\sqrt{\det\bm{C}},
\end{equation}
where n is the dimension of the parameter space, and $\mathcal{L}_{max}(\bm{\theta})$ is the maximum likelihood value. This approximation clearly breaks down when the posterior is multi-modal.

The BIC estimate is a result obtained in the asymptotic limit for distributions in the exponential family, and gives the largest penalty to models with a large number of parameters. In this approximation

\begin{equation}
\ln \mathcal{Z}\simeq \ln \mathcal{L}_{max}(\bm{\theta})-\frac{n}{2}\ln N_D,
\end{equation}
where $N_D$ is the number of data points. In most cases, this offers a simple, order-of magnitude estimate for $\mathcal{Z}$.

Finally, the NLA computes the evidence using the equality

\begin{equation}
\label{eq:wein}
\begin{split}
\frac{1}{\mathcal{Z}} &= \int \frac{\mathcal{P}(\bm{\theta})}{\mathcal{L}(\bm{\theta})}d\bm{\theta}\\
&=\int_{Y_0}^{Y_N} M(Y)dY + M(Y_0)Y_0,
\end{split}
\end{equation}
where the last term contains a Lebesgue integral with $Y=\mathcal{L}(\bm{\theta})^{-1}$ and measure $M(y)$

\begin{equation}
M(y)=\int_{Y(\bm{\theta})>y} \mathcal{P}(\bm{\theta}) d\bm{\theta}.
\end{equation}
 
This conversion to a Lebesgue integral has the clear advantage of replacing the n-dimensional integral by a 1-dimensional one. This approach is also used by the nested sampling algorithm (Section~\ref{mn}) where $\mathcal{Z}$ is computed as 

\begin{equation}
\label{eq:leb}
\mathcal{Z}=\int_{0}^{1}\mathcal{L}(X)dX; \mbox{   } X(\lambda)=\int_{\mathcal{L}(\bm{\theta})>\lambda} \pi(\bm{\theta})d\bm{\theta}.
\end{equation}

Since the final MCMC sample is distributed as the posterior probability $\mathcal{P}(\bm{\theta})$, in Equation~\ref{eq:wein}, $M$ can be approximated as $M(Y_i)\approx \frac{1}{N}\sum_{j=1}^{N}\bm{1}_{{Y_j>Y_i}}$ for each $\mathcal{L}_i$, where $\bm{1}$ is the indicator function. With this approximation we have

\begin{equation}
\mathcal{Z}\approx \left(\frac{1}{N}\sum_{j}\frac{1}{\mathcal{L}_j}\right)^{-1},
\end{equation}
which is also known as the harmonic mean estimator (HME). This disadvantages of this estimator are well known in the literature \citep[e.g., ][]{Raftery:2007,Calderhead:2009}. Due to the presence of $1/\mathcal{L}_j$ terms this method is unstable for very small likelihood values that dominate the sum. The proposed solution is to restrict the integration space only to well-sampled regions of high likelihood. Therefore this method suffers from problems intrinsic to MCMC sampling. In addition, \citet{Calderhead:2009} show that even in well-behaved scenarios, the HME can produce biased (lower) results. To avoid these issues, the nested sampled approach (Equation~\ref{eq:leb}) is the preferred alternative to thermodynamic integration.

The $B_{XY}$ factor can also be estimated directly using the reverse jump MCMC \citep[e.g., ][]{Lopes:2004}, or the Savage-Dickie density ratio \citep[e.g., ][]{Trotta:2007}. The reverse jump MCMC is essentially a chain moving between different models, and can be either slow to converge or inaccurate for a small number of samples. The last method can provide high accuracy for nested models, as long as the parameter priors are separable, which is not generally true for our atmospheric models. 

To draw the analogy with the frequentist approach, the Bayes factor for nested models can be shown to satisfy the relation \citep{Trotta:2008,Sellke:2001}

\begin{equation}
B_XY\leq -\frac{1}{e \mathsf{p}\ln\mathsf{p}},
\end{equation}
where $e = \exp(1)$, and $\mathsf{p}$ is the p-value. Equivalently, this probability can be expressed as the number of standard deviations from the mean $x\sigma$, assuming a Gaussian distribution, $\mathsf{p}=\mathrm{erf}(x/\sqrt{2})$. This upper bound is the significance $\sigma$ value we refer to in our model comparison examples. 

\subsection{Markov chain Monte Carlo}
\label{emcee}

MCMC methods are widely used in investigating multi-dimensional, non-gaussian and highly correlated posteriors, since they don't require any {\it a priori} assumption about the shape of the posterior probability distribution. The most common form is the Metropolis-Hastings algorithm, where the chain is created as a random walk towards the region of maximum likelihood. Each sample is generated from a proposal distribution centered on the current point, and accepted with a probability $\mathrm{pr}=\min(1,\mathcal{L}(\bm{\theta'})/\mathcal{L}(\bm{\theta}))$. If the new sample is rejected, the position of the chain remains unchanged. The chain is initialized by a first guess $\bm{\theta_0}$, and after a burn-in period reaches a stationary state where the sample distribution reflects the shape of the posterior (more samples are drawn from high-likelihood regions). The un-normalized posterior distribution is simply the histograms of all the MCMC samples after the burn-in stage, and the marginal probability distributions for all parameters can be derived from it. Although much more efficient than just a simple Monte Carlo technique, MCMC still has a series of drawbacks: the convergence is not easily testable and can require a very large number of samples; due to its Markov chain nature, it is not easily parallelizable in this form; can be sensitive to the initial guess and get stuck in local minima; sample correlation can affect the final distribution. 

The affine-invariant MCMC ensemble sampler proposed by \citet{Goodman:2010} solves some of these problems. In this paper we use the version of this algorithm {\it emcee} implemented in Python by \citet{Foreman-Mackey:2013}\footnote{\url{http://dan.iel.fm/emcee/}}. This algorithm uses multiple chains, or ``walkers" run in parallel for a faster exploration of the parameter space. The $K$ chains are initialized in a n-dimensional Gaussian distribution around the initial guess. At each step, the position of a given walker $W_i$ is determined by randomly choosing a different walker from the set $W_j$ and generating the new position $W_j+Z(W_i-W_j)$, with $Z$ is distributed as 

\begin{equation}
Z\sim \frac{1}{\sqrt{z}}, \mbox{ for } z\in\left[\frac{1}{a},a\right] \mbox{ and } 0 \mbox{ otherwise}, 
\end{equation}
where $a=2$ is the scale parameter. This new position is accepted with the probability $\mathrm{pr}=\min(1,Z^{K-1}\mathcal{L}(\bm{\theta'_i})/\mathcal{L}(\bm{\theta_i}))$. Alternate sets of walkers can be updated in parallel, greatly enhancing computing time. This method produces more independent (uncorrelated) samples than the traditional MCMC. Essentially, with a few hundred walkers each iteration can be considered a snapshot of the full posterior, after the burn-in time. The multiple walkers can also more easily spread out to explore the parameter space, such that a large number of iterations is not necessary. We adopted this method for speed, reliability, and ease of implementation for retrieving model parameters. However, it does not provide a direct estimate of $\mathcal{Z}$ and we have to resort to the approximations presented in Section~\ref{evidence}.

\subsection{Multimodal nested sampling}
\label{mn}

The multimodal nested sampling method was devised by \citet{Skilling:2004}, further refined by \citet{Shaw:2007, Feroz:2008}, and implemented into the {\it MultiNest} package by \citet{Feroz:2009}\footnote{\url{https://ccpforge.cse.rl.ac.uk/gf/project/multinest/}}, with an easy-to-use Python wrapper \citep{Buchner:2014}\footnote{\url{https://github.com/JohannesBuchner/PyMultiNest}}. It was initially designed as a tool for more reliable Bayesian evidence calculation, but was also found to provide low-noise estimates of the posterior distribution, and thus constraints on the model parameters.

Nested sampling starts with $N$ ``live points" uniformly spaced across the entire initial prior volume, mapped into a unit hypercube. At every iteration $i$, the ``live points" with the lowest likelihood value $\mathcal{L}_i$ are iteratively replaced by requiring that new ones have $\mathcal{L}>\mathcal{L}_i$. In order to ensure that last condition is satisfied, the iso-likelihood contour is approximated by a set of (possibly overlapping) ellipsoids containing the active points, and new samples are drawn from within this new volume until one is found that satisfies $\mathcal{L'}>\mathcal{L}_i$. This new point then replaces the one with $\mathcal{L}_i$ in the set of active points. The volume occupied by the points with $\mathcal{L}_i>\mathcal{L}_{i-1}$ at iteration $i$ is a random variable that can be approximated by its expectation value as $\ln X_i\approx -(i\pm \sqrt{i})/N$ \citep{Feroz:2008} and used in the evaluation of the Bayesian evidence $\mathcal{Z}$ as a 1-dimensional integral (Equation~\ref{eq:leb}): 

\begin{equation}
\mathcal{Z}=\sum_{i=1}^{M}\mathcal{L}_iw_i + \bar{\mathcal{L}}X_M, 
\end{equation}
where the last term represents the contribution of the current set of active points, and $w_i$ are the weights for the trapezoidal rule $w_i=\frac{1}{2}(X_{i-1}-X_{i+1})$.

The error in $\mathcal{Z}$ is estimated \citep{Skilling:2004} as $\sqrt{H/N}$, where

\begin{equation}
H\approx\sum_{i=1}^{M}\frac{\mathcal{L}_iw_i}{\mathcal{Z}}\ln\frac{\mathcal{L}_i}{\mathcal{Z}},
\end{equation}
and $M$ is the number of iterations. The posterior distribution is approximated by the total set of active and discarded points and their weights $p_i=\mathcal{L}_iw_i/\mathcal{Z}$, where $w_i$ is calculated as above for the set of discarded points, and as $w_i=X_M/N$ for the current set of active points. The mean and covariance of the parameters are then

\begin{equation}
\bar{\bm{\theta}}=\sum_{i=1}^{M+N}p_i\bm{\theta_i},
\end{equation}

\begin{equation}
\bm{C}=\sum_{i=1}^{M+N}p_i(\bm{\theta}_i-\bar{\bm{\theta}})(\bm{\theta}_i-\bar{\bm{\theta}})^T,
\end{equation}

In addition to providing the Bayesian evidence as a by-product, {\it MultiNest} also employs a well-defined convergence criterion that can significantly reduce the number of required posterior samples, and therefore the running time. Convergence is achieved when the estimated change in likelihood $\Delta \mathcal{Z}_i=\max(\mathcal{L}_i)X_i$ is less than a user-specified tolerance. Generally, the number of likelihood evaluations until convergence grows exponentially with the number of dimensions of the parameter space. This makes the algorithm unfeasible for a large number of dimensions ($\gtrsim 10$). However, at avery step new samples can be drawn in parallel, significantly increasing computational speed. In practice, we find that {\it MultiNest} can be run for a much shorter time than {\it emcee} to converge, mainly because {\it emcee} does not have a self-stopping criterion and is left to run long enough to cover the entire parameter space and obtain sufficient independent samples. Similar to MCMC, in some cases the acceptance rate is low for {\it MultiNest}, and therefore convergence is also slow. 

\section*{Acknowledgments}

This research was supported by the {\it WFIRST} Preparatory Science Program and the JPL Exoplanet Exploration Program Office. Resources supporting this work were provided by the NASA High-End Computing (HEC) Program through the NASA Advanced Supercomputing (NAS) Division at Ames Research Center. The research was carried out in part at the Jet Propulsion Laboratory, California Institute of Technology, under a contract with the National Aeronautics and Space Administration. We thank the anonymous referee for the constructive comments that helped improve this paper. RL would like to thank the other co-authors for paying her a living wage for the duration of the project, and Mom for endless moral support.

\bibliography{references}

\end{document}